\documentclass[
  aps,
  prx,
  reprint,
  superscriptaddress,
  nofootinbib,
  floatfix
]{revtex4-2}
\usepackage[utf8]{inputenc}
\usepackage[T1]{fontenc}
\usepackage{amsmath}
\usepackage{amssymb}
\usepackage{physics}
\usepackage{mathtools}
\usepackage{array}
\usepackage[dvipsnames]{xcolor}
    \definecolor{darkgreen}{rgb}{0,0.5,0}
    \definecolor{darkblue}{rgb}{0,0,0.6}
    \definecolor{purple}{rgb}{0.4,.2,0.7}
	\definecolor{linkblue}{rgb}{0.0,0.0,0.6}
	\definecolor{citepurple}{rgb}{0.35,0.2,0.55}
\usepackage[hyperfootnotes = false, 
	colorlinks = true,
	linkcolor = darkblue, 
	citecolor = purple, 
	urlcolor = linkblue]{hyperref}
\usepackage{relsize}
\usepackage{CJKutf8}

\newcommand{\ii}{\mathrm{i}}
\newcommand{\ee}{e}
\newcommand{\sgn}{\operatorname{sgn}}

\newcommand{\eff}{{\rm{eff}}}
\newcommand{\gR}{{\rm{R}}}
\newcommand{\gA}{{\rm{A}}}
\newcommand{\gK}{{\rm{K}}}

\newcommand{\gW}{{\rm{W}}}

\newcommand{\sys}{{\rm{S}}}
\newcommand{\bath}{{\rm{B}}}
\newcommand{\sysb}{{\rm{SB}}}
\newcommand{\lind}{{\rm{L}}}

\begin{document}

\title{Inducing, and enhancing, many-body quantum chaos by continuous monitoring}

\author{Xianlong Liu ({\begin{CJK}{UTF8}{gbsn}刘显龙\end{CJK}})}
\email{xianlong\_liu@sjtu.edu.cn}
\affiliation{Shanghai Center for Complex Physics, School of Physics and Astronomy,
    Shanghai Jiao Tong University, Shanghai 200240, China
}

\author{Jie-ping Zheng ({\begin{CJK*}{UTF8}{gbsn}郑杰平\end{CJK*}})}
\email{jpzheng@sjtu.edu.cn}
\affiliation{Shanghai Center for Complex Physics, School of Physics and Astronomy, 
    Shanghai Jiao Tong University, Shanghai 200240, China
}

\author{Antonio M. Garc\'ia-Garc\'ia}
\email{amgg@sjtu.edu.cn}
\affiliation{Shanghai Center for Complex Physics, School of Physics and Astronomy,
	Shanghai Jiao Tong University, Shanghai 200240, China
}

\begin{abstract}
It is intuitively expected, and supported by earlier studies, that many-body quantum chaos is suppressed, or even destroyed, by dissipative effects induced by continuous monitoring. We show here that this is not always the case. For this purpose, we study the quenched dynamics of a continuously monitored Sachdev-Ye-Kitaev (SYK) model, described by the Lindblad formalism, coupled to a thermal environment modeled by another SYK maintained at constant temperature. We find that the combined effect of monitoring and the thermal bath drives the system toward a non-thermal steady state independently of the initial conditions. The corresponding retarded Green's function exhibits two stages of exponential decay, with rates that depend non-monotonously on the thermal bath coupling and the monitoring strength. In the limit of weak coupling, the late time decay of the Green's function, computed analytically, is closely related to that of the thermal bath. Strikingly, we identify a range of parameters in which continuous monitoring, despite being a source of decoherence, induces or enhances quantum chaotic dynamics suppressed by the thermal bath. For instance, in the limit of weak coupling to the thermal bath, the Lyapunov exponent increases sharply when monitoring is turned on. For intermediate values of the thermal bath coupling, the Lyapunov exponent exhibits re-entrant behavior: it vanishes at zero or sufficiently weak monitoring strength, and becomes positive again as the monitoring strength is increased. Our results offer intriguing insights on the mechanisms leading to quantum scrambling which paves the way to its experimental control and consequently to a performance enhancement of quantum information devices.
\end{abstract}

\maketitle

Developments in quantum information and quantum computation have boosted dramatically the interest in the dynamics of open, also termed ``monitored'' in this context, many-body quantum systems~\cite{Fazio2024lectures}, because in realistic settings the extraction of quantum information necessarily involves a series of quantum measurements or monitoring. For that purpose, the process of measurement is modeled by protocols such as  quantum jumps~\cite{warren1986,zoller1987,gleyzes2007,minev2019,plenio1998,molmer93,dalibard1992,dum1992,daley2014} or quantum state diffusion~\cite{wiseman1993,collett1987,wiseman2014,fuwa2015,cao2019a,alberton2021a,carisch2023,ladewig2022}. These protocols, especially the quantum state diffusion, tend to minimize~\cite{misra1977,itano1990,raizen2001} the impact of measurements, that are expected to suppress quantum coherence. 
This monitored dynamics is equivalent \cite{molmer1993,dum1992,dalibard1992} to the coupling of the system to a Markovian bath.
A full description of the effect of the bath {\it a la} Caldeira-Leggett~\cite{caldeira1981} is challenging because of the technical complexity of modeling the environment analytically in a many-body context. Purely numerical approaches are limited to small number of particles. 
A notable exception is the case of the Sachdev-Ye-Kitaev (SYK) model~\cite{kitaev2015,french1970,bohigas1971,sachdev1993,benet2001,maldacena2016,garcia2022d}, consisting of~$N$ Majorana fermions with random infinite range interactions in Fock space, where the effect of a thermal bath, given by another much larger SYK model, has been studied quantitatively using different approximation schemes. Assuming that the bath has a much larger number of degrees of freedom than that of the system, the back-reaction of system dynamics to the bath becomes negligible, and the computation of Lyapunov exponent~\cite{chen2017a,zhang2019}, which characterizes quantum chaotic dynamics in the semi-classical condition, becomes feasible.
The full quenched dynamics in SYK, resulting in a final thermal state, was investigated~\cite{eberlein2017} by numerically solving the Kadanoff-Baym (KB) equations \cite{stefanucci2025,kamenev2023,eberlein2017,zhang2019,almheiri2019} for non-equilibrium Green's functions. The process of black hole evaporation has been modeled~\cite{almheiri2019} by an SYK model which is abruptly coupled to a thermal bath at a different temperature. 
A similar calculation for two weakly coupled SYK in contact with a bath was proposed~\cite{milekhin2019,milekhin2024} as a toy model for the formation of a wormhole in real time. 
More recently, a similar problem, the dynamics of a driven SYK characterized by a periodic in time coupling constant has been investigated in both a single SYK~\cite{knap2020} and a two-site weakly coupled SYK models~\cite{berenguer2024} dual to a kicked traversable wormhole.

Another tractable approach is to adopt a Markovian bath in the weak coupling limit through the so-called Lindblad formalism~\cite{lindblad1976,gorini1976}, where the dynamics can be cast in terms of a non-Hermitian Liouvillian depending only on the system degrees of freedom.
Interestingly, as mentioned earlier, the same formalism describes the process of continuous measurement ~\cite{molmer93,dalibard1992}, where the jump operators in the Liouvillian are interpreted as operators representing the observable being monitored. As in the previous case, applying path integral techniques in the Lindbladian SYK model~\cite{sa2022,kulkarni2022,garcia2022e}, an SYK system coupled to a Markovian environment, enables the computation of the retarded Green's function and the Lyapunov exponent, which characterizes the early-time exponential growth of out-of-time-ordered correlation functions (OTOC)~\cite{garcia2024}.
It was found that the typical relaxation time, captured by the decay of retarded Green's functions, has an intriguing non-monotonic dependence on the bath/monitoring strength.  
By contrast, increasing the environment/measurement strength gradually reduces the Lyapunov exponent, which vanishes at a critical monitoring coupling~\cite{garcia2024}.
This behavior is expected since an environment typically induces decoherence and disturbs quantum evolution, while a positive Lyapunov exponent characterizes quantum chaotic dynamics for short times of the order but shorter than the Ehrenfest/scrambling time.

Likewise, a general feature of the Lindblad formalism~\cite{lindblad1976,gorini1976} is that, except for non-interacting or integrable many-body systems~\cite{breuer2002,garcia2025}, the system is generically driven to a infinite-temperature steady state, for a broad class of jump operators and coupling strengths, reflecting the quantum white noise character of generic Markovian dissipation.
This heating effect is in most cases a problem as it narrows the situations in which the formalism is applicable to model realistic systems. Therefore, it would be of broad interest identifying situations in which a continuous monitoring of the dynamics would lead to a steady state which is not necessarily at infinite temperature. 

In this paper, we address these two problems, how to avoid an infinite temperature steady state and the mentioned suppression of quantum effects by continuous monitoring, by studying the quenched dynamics of a continuously monitored Majorana SYK, modeled by the Lindblad formalism, which is coupled to a large thermal bath described by another SYK with a much larger number of degrees of freedom. 
We show that the steady state is in general non-thermal, with two typical decay rates. Moreover, in certain circumstances, counter-intuitively, the continuous monitoring induces many-body quantum chaos or makes its features more robust. A cartoon summarizing our model and the main results of the paper are depicted in Fig.~\ref{fig:quenched_Lindblad_illustration}. We start with the definition of the model.      

\section{The model and Kadanoff-Baym equations }
\label{sec:large_N_equations}

The dynamics of a quantum system subjected to continuous monitoring is described by a Lindblad equation
\begin{equation} \label{eq:Lindblad_equation}
	\frac{\dd{\rho}}{\dd{t}} = 
	-\ii [H, \rho(t)] + \sum_{k} \Bigl( L_k \rho(t) L_k^\dagger - \frac{1}{2} \{L_k^\dagger L_k, \rho(t)\} \Bigr) ,
\end{equation}
where the Liouvillian $\mathcal{L}(\rho)$ is given by the right hand side of the equation. Here $\{\cdot, \cdot\}$ denotes the anti-commutator, and $L_{k}$'s are the quantum jump operators. Lindblad dynamics admits steady states satisfying $\mathcal{L}(\rho) = 0$. In generic many body systems, except for certain integrable cases~\cite{breuer2002,garcia2025}, Lindblad evolution drives the system toward the infinite-temperature state $\rho = \mathbb{I} / \mathrm{Tr}(\mathbb{I})$, and Hermitian jump operators provide a prominent example of this generic behavior.
\begin{figure}[tbp]
		\includegraphics[width=\columnwidth]{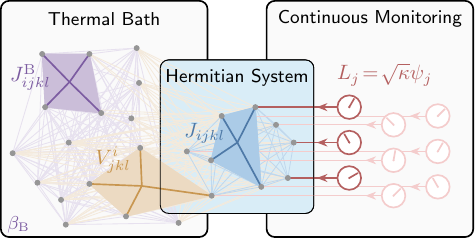}
		\includegraphics[width=\columnwidth]{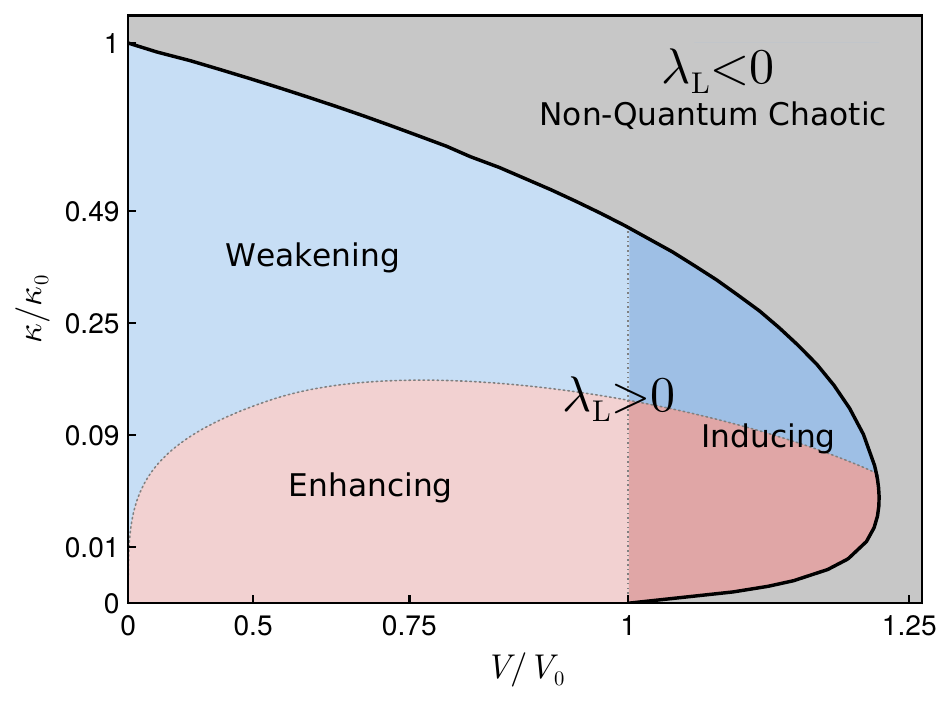}
		\caption{\textbf{Top:} Illustration of the model  Eq.~(\ref{eq:Lindblad_equation}) that consists of a Hermitian SYK ($q=4$) with $N$ Majorana fermions coupled to a thermal bath, another SYK ($q_{\bath} =4$) with $N_{\bath} \gg N$ Majoranas at inverse temperature $\beta_{\bath}$ with coupling~$V$, $(f_{\sys}, f_{\bath}) = (1, 3)$ to the Hermitian SYK. The Hermitian SYK is also subjected to continuous measurements described by the jump operators $L_j = \sqrt{\kappa} \psi_j$ ($j=1, \dots, N$). \textbf{Bottom:} Illustration of the Lyapunov exponent $\lambda_{\mathrm{L}}$, that characterizes quantum chaotic dynamics, in the parameter space of the coupling strengths, $\kappa$ (monitoring) and $V$ (thermal bath with bath inverse temperature $\beta_{\rm{B}}=1000$). The values $V_0 \approx 0.94$ and $\kappa_0 \approx 0.33$ indicate the critical values at which $\lambda_{\rm{L}}$ vanishes when the Markovian or thermal bath is absent, respectively. The horizontal axis is shown on a square scale ($x \mapsto x^2$), while the $y$-axis is shown on a square root scale ($y \mapsto \sqrt{y}$). The ``Inducing'' (dark red and dark blue), ``Enhancing'' (light blue), and ``Weakening'' (light red) regions indicate positive Lyapunov exponents $\lambda_{\rm L} > 0$ such that the system is quantum chaotic. By ``Inducing'' we mean that the Lindblad dynamics induces quantum chaos for $V \geq V_0$. By ``Enhancing'' (``Weakening'') we mean that when $V$ is fixed, the increase of monitoring rate $\kappa$ increases (decreases) the Lyapunov exponent, and hence enhances (weakens) quantum chaos. The ``Inducing'' region is divided into two parts: for a fixed $V$ with $V \geq V_0$, $\lambda_{\rm L}$ increases as $\kappa$ increases in the dark red region until it reaches a maximum, then it decreases, until it vanishes, in the dark blue region. The ``Non-quantum chaotic'' (light gray) region corresponds to $\lambda_{\rm L} < 0$.}
		\label{fig:quenched_Lindblad_illustration}
\end{figure}

We search for a broader set of steady states by coupling the system above to a (non-Markovian) thermal bath. We refer to the resulting time evolution as {\it monitored dissipative dynamics} where dissipative stands for the role of the thermal bath. To be specific, we investigate the SYK model~\cite{kitaev2015,maldacena2016,french1970,bohigas1971,sachdev1993,benet2001,garcia2016} described by the Hamiltonian 
\begin{equation}
    H^{{\rm SYK}}[\psi ; q, J, N] = 
	\ii^{\frac{q}{2}} 
	\sum_{1 \leq j_1 < \dots < j_{q} \leq N} J_{j_1 \dots j_{q}} \psi_{j_1} \dots \psi_{j_q} ,
\end{equation}
with $\psi_j$, Majorana fermions satisfying $\{\psi_i, \psi_j\} = \delta_{ij}$. The random couplings $J_{j_1 \dots j_q}$ are drawn from a Gaussian distribution with
\begin{equation}
    \langle J_{j_1 \dots j_q} \rangle = 0 , \qquad  
    \langle J_{j_1 \dots j_q}^2 \rangle = \frac{(q-1)! J^2}{N^{q-1}}.
\end{equation}
We will consider the Lindblad dynamics Eq.~\eqref{eq:Lindblad_equation} where the Hamiltonian is given by \cite{almheiri2019}
\begin{equation}
	\begin{split}
    H & = H_{\sys}^{\rm{SYK}}[\psi ; q, J, N] + H_{\bath}^{\rm{SYK}}[\chi ; q_{\rm{B}}, J_{\rm{B}}, N_{\rm{B}}] \\
	& \quad + \theta(t) H_{\sysb}[\psi, \chi ; f_{\rm{S}}, f_{\rm{B}}, V] .
	\end{split}
\end{equation}
Here $H_{\sys}$ is the system SYK Hamiltonian with $N$ Majorana fermions $\psi_j$, and $H_{\bath}$ is the thermal bath (non-Markovian reservoir) SYK Hamiltonian with $N_{\bath}$ Majorana fermions $\chi$. The interaction Hamiltonian $H_{\sysb}$ between the system and the thermal bath is given by
\begin{equation}
    H_{\sysb} = 
	\sum_{\substack{1 \leq j_1 < \dots < j_{f_\sys} \leq N \\ 
	1 \leq i_1 < \dots < i_{f_\bath} \leq N_\bath}
	}
	V^{i_1 \dots i_{f_\bath}}_{j_1 \dots j_{f_\sys}} \psi_{j_1} \dots \psi_{j_{f_\sys}} \chi_{i_1} \dots \chi_{i_{f_\bath}} ,
\end{equation}
where
\begin{equation}
    \gamma = f_\sys f_\bath + \frac{f_\sys(f_\sys -1)}{2} + \frac{f_\bath (f_\bath - 1)}{2} , 
\end{equation}
such that the interaction Hamiltonian is Hermitian. The random couplings $V^{i_1 \dots i_{f_\bath}}_{j_1 \dots j_{f_\sys}}$ are drawn from a Gaussian distribution with
\begin{equation}
    \langle V^{i_1 \dots i_{f_\bath}}_{j_1 \dots j_{f_\sys}} \rangle = 0 \, , \quad
    \langle (V^{i_1 \dots i_{f_\bath}}_{j_1 \dots j_{f_\sys}})^2 \rangle = \frac{(f_\sys-1)! f_\bath! V^2}{N^{f_\sys - 1} N_\bath^{f_\bath}} \, .
\end{equation}
We will assume $N_\bath \gg N$ so that the back-reaction of the system on the bath can be neglected, and the bath is characterized by an inverse temperature $\beta_\bath$. At last, the Lindblad jump operators $L_j$ are 
\begin{equation}
    L_j = \sqrt{\kappa} \, \psi_j , \quad j = 1, \dots, N .
\end{equation}

An illustration for our setup is shown in the top panel of Fig.~\ref{fig:quenched_Lindblad_illustration}. In the absence of the thermal bath ($V=0$), as was proved in~\cite{garcia2024,sa2022,kulkarni2022}, the steady state for the pure Lindblad dynamics is a thermal state at infinite temperature. We would like to see the effects of the additional bath at inverse temperature $\beta_\bath$ on the steady state of the system.

\subsection{Kadanoff-Baym equations}
As a first step, we set up the Schwinger-Keldysh path-integral along the contour $z \in \mathcal{C} = \mathcal{C}^+ \cup \, \mathcal{C}^-$ and obtain the saddle-point KB equations by using the standard procedure of performing disorder average over couplings, introducing the bi-local collective fields 
\begin{align}
	G(z_1, z_2) & = - \frac{\ii}{N} \sum_{j=1}^{N} \psi_j(z_1) \psi_j(z_2) , \\
	G_{\bath}(z_1, z_2) & = - \frac{\ii}{N_{\bath}} \sum_{j=1}^{N_{\bath}} \chi_j(z_1) \chi_j(z_2) ,
\end{align}
for the system and the bath Majorana fermions along the Schwinger-Keldysh contour and the corresponding Lagrange multipliers (self-energies) $\Sigma$, $\Sigma'$ that enforce these relations. In a second step, we integrate out the original fermionic fields $\psi_i(z)$. The resulting partition function is given by
\begin{equation}
	Z = \int \mathcal{D} G \mathcal{D} G_{\bath} \mathcal{D} \Sigma \mathcal{D} \Sigma' \, \ee^{\ii S[G, G_{\bath}, \Sigma, \Sigma']} .
\end{equation}
A saddle point analysis of this action results in the mentioned KB equations whose solution and analysis are our main focus of study.

A precise definition of the action is found in the Supplementary Materials. Here, we present the two-time KB equations for the system dynamics. The Green's function $G^{\alpha \beta}(t_1, t_2)$ defined on the Schwinger-Keldysh contour has four components, since each time argument may lie on either the forward $\mathcal{C}^+$ or backward $\mathcal{C}^-$ branch, with $\alpha, \beta \in \{+, -\}$. However, due to the contour ordering identity $G^{++} + G^{--} = G^{+-} + G^{-+}$, and the relation $[G^{++}(t_1, t_2)]^{\dagger} = G^{--}(t_1, t_2)$, $[G^{+-}(t_1, t_2)]^{\dagger} = G^{-+}(t_1, t_2)$ for Majorana fermions, only one Green's functions are independent. A convenient choice is to take the ``greater'' or ``lesser'' Green's functions, $G^> \equiv G^{-+}$ and $G^< \equiv G^{+-}$, whose dynamics is governed by the KB equations \cite{stefanucci2025,kamenev2023,eberlein2017,zhang2019,almheiri2019}:
\begin{align} 
\label{eq:KB_equation_1}
\ii \partial_{t_1} G^{\gtrless}(t_1, t_2) & = 
\bigl( \Sigma^{\gR} * G^{\gtrless} + \Sigma^{\gtrless} * G^{\gA}\bigr)(t_1, t_2) , \\
\label{eq:KB_equation_2}
- \ii \partial_{t_2} G^{\gtrless}(t_1, t_2) & = 
\bigl( G^{\gR} * \Sigma^{\gtrless} + G^{\gtrless} * \Sigma^{\gA} \bigr)(t_1, t_2) ,
\end{align}
where $(A * B) (t_1, t_2) = \int_{-\infty}^{\infty} A(t_1, t) B(t, t_2) \dd{t}$ denotes the convolution. The retarded and advanced Green's function $G^{\gR}$ and $G^{\gA}$ are defined as
\begin{align}
	\label{eq:GR_def}
    G^{\gR}(t_1, t_2) & = \theta(t_1 - t_2) \big[G^>(t_1, t_2) - G^<(t_1, t_2) \big] , \\
    G^{\gA}(t_1, t_2) & = \theta(t_2 - t_1) \big[G^<(t_1, t_2) - G^>(t_1, t_2)] .
\end{align}
The self-energies $\Sigma^{\gtrless}$ consist of three parts: the system SYK interaction, the system interaction with the thermal bath, and the monitored dynamics described by the Lindblad formalism:
\begin{equation}
    \Sigma^{\gtrless} = \Sigma^{\gtrless}_{\sys} + \Sigma^{\gtrless}_{\bath} + \Sigma^{\gtrless}_{\lind} .
\end{equation}
Each term is given by
\begin{align}
\label{eq:Sigma_S}
\Sigma^{\gtrless}_{\sys}(t_1, t_2) & = - \ii^{q} J^2 \big[G^{\gtrless}(t_1, t_2)\big]^{q-1} , \\
\label{eq:Sigma_B}
\Sigma^{\gtrless}_{\bath}(t_1, t_2) & = 
- \ii^{f_{\sys} + f_{\bath}} V^2 \theta(t_1) \theta(t_2) \notag \\ 
& \quad\quad \times \big[G^{\gtrless}(t_1, t_2)\big]^{f_{\sys} - 1} \big[G_{\bath}^{\gtrless}(t_1, t_2)\big]^{f_{\bath}} ,
\\
\label{eq:Sigma_L}
\Sigma^{\gtrless}_{\lind}(t_1, t_2) & = \mp \ii \, \kappa \, \delta(t_1 - t_2) .
\end{align}
These general two-time KB equations do not assume that the Green's function depends only on the time difference, i.e., time-translation symmetry, and therefore offer a platform to study the general non-equilibrium dynamics of quantum many-body systems without any assumption about whether the system is in thermal equilibrium or the fluctuation-dissipation theorem applies. Due to the relation $N_{\bath} \gg N$, the back-reaction of the system on the thermal bath is parametrically suppressed. As a result, the bath dynamics for the Green's function $G_{\bath}$ is governed by the standard SYK saddle-point SD equations~\cite{maldacena2016}. We therefore take the thermal bath SYK to remain in a thermal state with inverse temperature $\beta_{\bath} \gg 1$, such that it lies in the conformal regime where the SYK model admits a gravity dual.
As stated earlier, the pure Lindblad dynamics heats up the system~\cite{garcia2023} until it reaches infinite temperature. Meanwhile, the quench dynamics after the abrupt coupling to the thermal bath~\cite{eberlein2017,almheiri2019,zhang2019} results in a steady state characterized by the bath temperature $T_{\bath} = 1/\beta_{\bath}$. We now study the dynamics resulting from the competition between these two effects that, as mentioned earlier, we termed {\it monitored dissipative dynamics}. For that purpose, we solve the two-time KB equations for the Green's functions with different initial conditions.

\section{Monitored Dissipative Dynamics: insensitivity to initial conditions and non-thermal steady state}
\label{sec:solutions_Kadanoff_Baym_equations}
\begin{figure}[tbp]
    \includegraphics[width=\columnwidth]{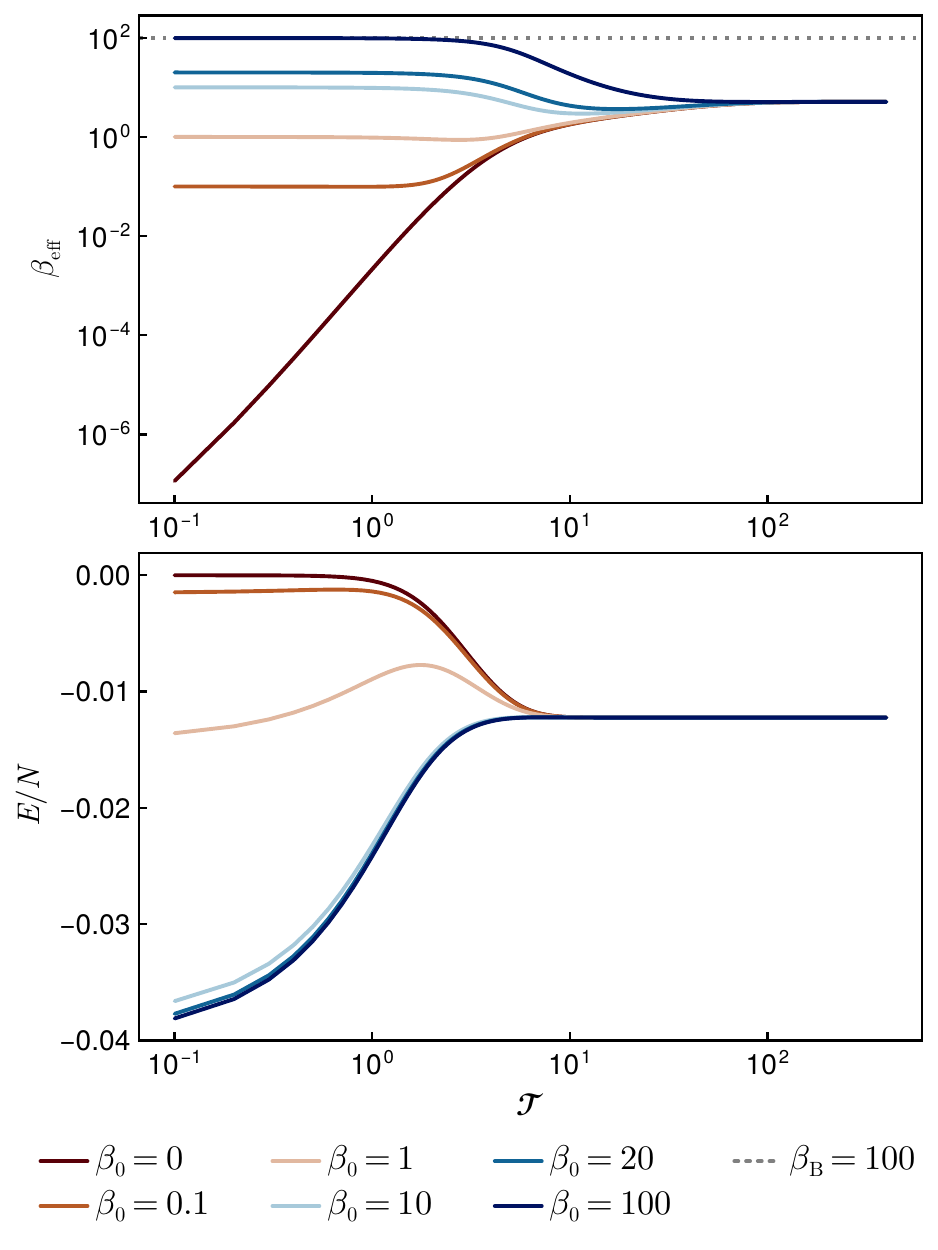}
    \caption{The effective temperature $\beta_{\rm eff}$ Eq.~(\ref{eq:betaeff}) and the energy density $E/N$ Eq.~(\ref{eq:ET}) as a function of the center-of-mass time $\cal{T}$ obtained from the solution of the KB equations for the monitored dissipative dynamics of the SYK model with parameters $q = 4$, $(f_{\sys}, f_{\bath}) = (1, 3)$, $V = 0.5$ and $\kappa = 0.1$. The thermal bath is chosen as another SYK with $J_{\bath} = J = 1$, $q = 4$ at finite inverse temperature $\beta_{\bath} = 100$. The initial conditions are thermal states at various inverse temperatures $\beta_{0} \in \{0, 0.1, 1, 10, 20, 100\}$. The equations are solved in an $8001 \times 8001$ grid with the initial conditions encoded in the third quadruple $(t_1, t_2) \in [-T_{\max}, 0) \times [-T_{\max}, 0)$, with size $4000 \times 4000$. The time separation is set to be $\Delta t = 0.1$ such that $T_{\max} = 400$. We observe that different initial states lead to the same final steady state solution at long time, indicating that the steady state is insensitive to the initial conditions.}
    \label{fig:SYK4_quenched_Lindblad_KB_solutions}
\end{figure}
\begin{figure}[tbp]
    \includegraphics[width=\columnwidth]{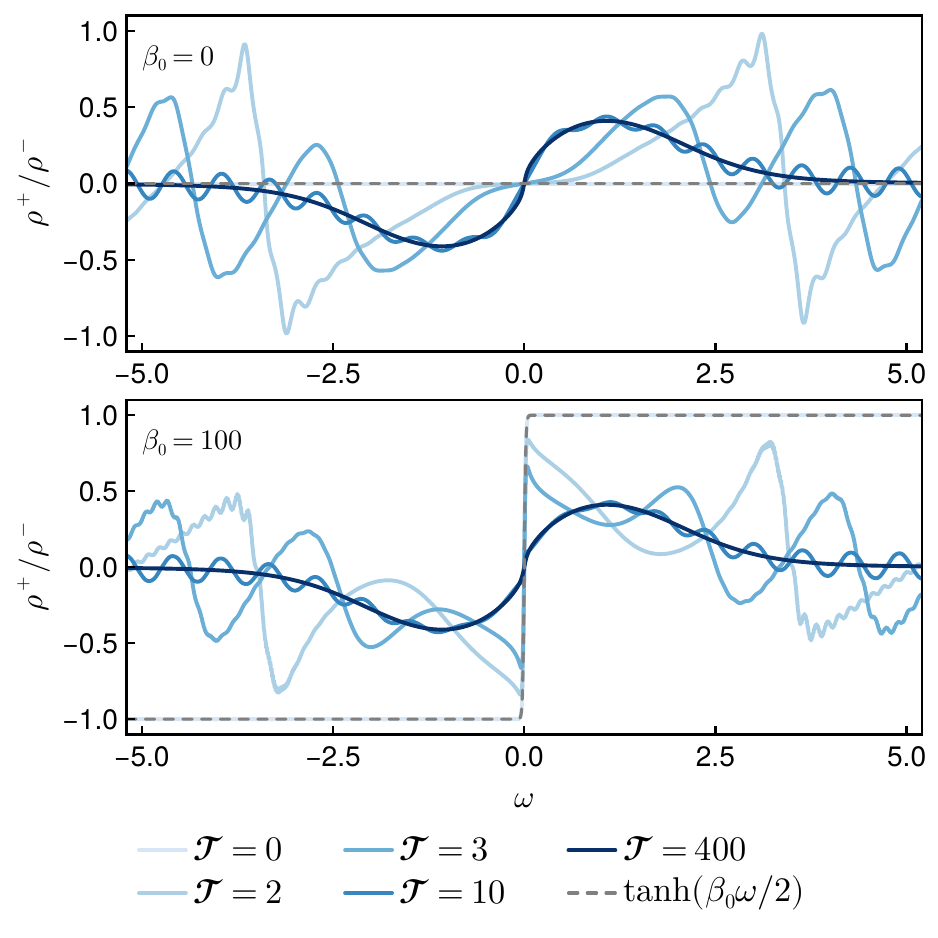}
    \caption{The function $\rho^+/\rho^-$ Eq.~(\ref{eq:rhopm}) at different center-of-mass time $\mathcal{T}$ for different initial states. \textbf{Top:} Thermal initial state with inverse temperature $\beta_0=0$. \textbf{Bottom:} Thermal initial state with inverse temperature $\beta_0=100$.}
    \label{fig:SYK4_quenched_Lindblad_KB_solutions_rhop_over_rhom}
\end{figure}

\begin{figure}[tbp]
    \includegraphics[width=\columnwidth]{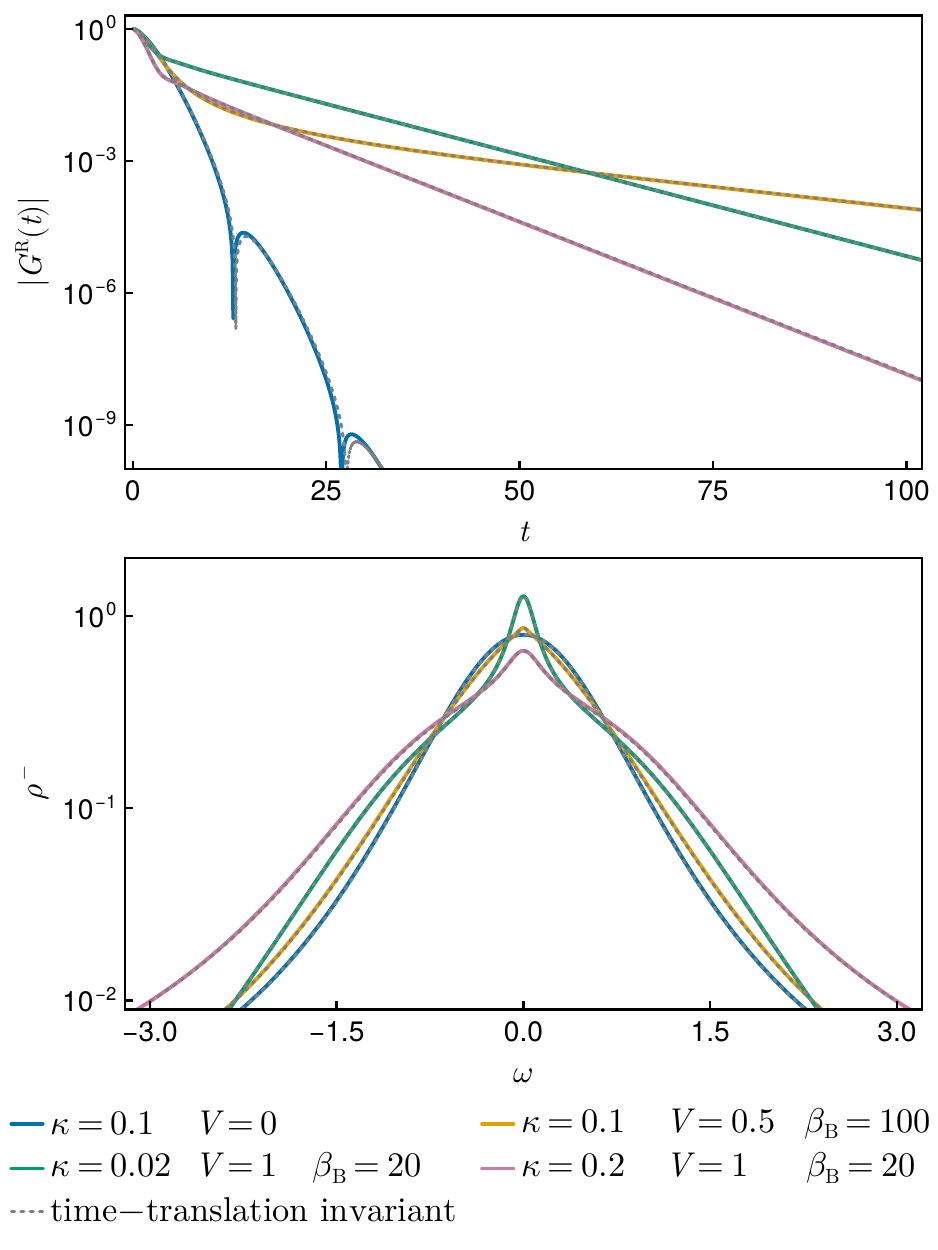}
    \caption{Comparison between the steady state solution of the KB equations and the time-translational invariant SD equations with different $V$, $\beta_{\bath}$ and $\kappa$. \textbf{Top:} Comparison for the magnitude of the retarded Green's function Eq.~\eqref{eq:GR_def}. \textbf{Bottom:} Comparison for the spectral function Eq.~\eqref{eq:rhopm}. Other parameters are fixed as $q=q_{\bath}=4$, $(f_{\sys}, f_{\bath}) = (1, 3)$ and $J = J_{\bath}=1$, see Eqs.~\eqref{eq:Sigma_S} and \eqref{eq:Sigma_B}. Excellent agreement is observed in all cases.}
    \label{fig:steady_state_comparison_KB_vs_SD}
\end{figure}
Firstly, we study whether the KB equations for the monitored dissipative dynamics admit steady state solutions, and, if so, whether the steady state is independent of the chosen initial condition. Secondly, we shall find out whether this steady state, and more generally the late time solutions are consistent with those of a system close to thermal equilibrium. In order to proceed, we need to compute different Green's functions by solving the full KB equations in the two-time space $(t_1, t_2) \in \mathbb{R}^2$ with different initial conditions. To compute physical observables at fixed center-of-mass time $\mathcal{T} \equiv (t_1 + t_2) / 2$, we consider the ``corner slice'' Green's function \cite{almheiri2019}
\begin{equation}
	G^{\gtrless}_{{\rm{cor}}}(\mathcal{T}, t) \equiv \theta(t) G^{\gtrless}(\mathcal{T} - t, \mathcal{T}) + \theta(- t) G^{\gtrless}(\mathcal{T}, \mathcal{T} + t) ,
\end{equation}
and perform Fourier transform on the Green's functions only at the relative time $t \equiv t_1 - t_2$ (Wigner transform):
\begin{equation}
	G^{\gtrless}_{{\rm{cor}}}(\mathcal{T}, \omega) = \int_{-\infty}^{\infty} G^{\gtrless}_{{\rm{cor}}}(\mathcal{T}, t) \ee^{\ii \omega t} \dd{t} \! .
\end{equation}
If the spectral function, 
\begin{equation}
	\rho^{\pm}(\mathcal{T}, \omega) \equiv - \frac{1}{2\pi \ii} \big[G^>_{\rm{cor}}(\mathcal{T}, \omega) \pm G^<_{\rm{cor}}(\mathcal{T}, \omega) \big] , \label{eq:rhopm}
\end{equation}
based on the calculation of this Green's function, obeys the fluctuation-dissipation theorem
\begin{equation}
	\frac{\rho^+(\mathcal{T, \omega})}{\rho^-(\mathcal{T, \omega})} = \tanh(\frac{\beta \omega}{2}) ,
\end{equation}
it would mean that the system is close to thermal equilibrium at late times $\mathcal{T}$.
In order to investigate whether the monitored dissipative dynamics leads to a steady state, we define the effective temperature as \cite{zhang2019}
\begin{equation}
	\beta_{\eff}(\mathcal{T}) = \left. 2 \pdv{}{\omega} \frac{\rho^+(\mathcal{T, \omega})}{\rho^-(\mathcal{T}, \omega)} \right|_{\omega = 0} , \label{eq:betaeff}
\end{equation}
and the system energy density $E(\mathcal{T})/N$, defined as the instantaneous expectation value of the system Hamiltonian per fermion. In terms of the Green's function, 
\begin{equation}
	\frac{E(\mathcal{T})}{N} = - \frac{\ii^{q+1} J^2}{q} \int_{-\infty}^{\infty} \sgn(t) \, G^>_{{\rm{cor}}}(\mathcal{T}, t)^q \dd{t} \! . \label{eq:ET}
\end{equation}
We present results for these observables in Fig.~\ref{fig:SYK4_quenched_Lindblad_KB_solutions} for different initial states given by thermal equilibrium states for the Hermitian SYK, with $q=4$, initial inverse temperatures $\beta_0 \in \{0, 0.1, 1, 10, 20, 100\}$, $J = 1$, $(f_{\sys}, f_{\bath})=(1,3)$, $V = 0.5$, and $\kappa = 0.1$. The thermal bath is taken as another SYK with $q = 4$, $J_{\bath}=1$ at inverse temperature $\beta_{\bath} = 100$. We find that all these different initial states lead to the same energy density and effective temperature at long times. This indicates that the steady state is independent of the initial conditions, and is only determined by the system-bath couplings $V$, bath temperature $1/\beta_{\bath}$ and the Lindblad coupling $\kappa$. 
Therefore, to study the steady state, we can consider the long time limit $\mathcal{T} \rightarrow \infty$. In that limit, the KB equations reduce to time-translational invariant equations that only depend on the relative time $t$, which we will consider in the next section. In Fig.~\ref{fig:steady_state_comparison_KB_vs_SD}, we present a comparison between the KB equation steady state solutions and those corresponding to the time-translational invariant SD equations which reveals an excellent agreement. In Fig.~\ref{fig:SYK4_quenched_Lindblad_KB_solutions_rhop_over_rhom}, we present the ratio $\rho^+ / \rho^-$ at different center-of-mass-times $\mathcal{T}$ with initial condition $\beta_0 = 0$ (Top) and $\beta_{0} = 100$ (Bottom), which exhibit distinct intermediate-time dynamics while converging to the same final steady state.

At the same time, results depicted in Fig.~\ref{fig:SYK4_quenched_Lindblad_KB_solutions_rhop_over_rhom} for the ratio $\rho^+ / \rho^-$ at different center-of-mass times $\mathcal{T}$  indicate that the steady state is not a thermal state, due to the violation of the fluctuation-dissipation theorem. This is because, in the $\mathcal{T} \gg 1$ limit, $\rho^+ / \rho^-$  does not agree with the fluctuation-dissipation theorem prediction $\tanh(\beta \omega / 2)$. Indeed, the behavior of $\rho^+ / \rho^-$ is completely different: its maximum value is less than 1, and asymptotes to 0 as $|\omega| \rightarrow \infty$.

Once we have established that the monitored dissipative dynamics is characterized by a non-thermal steady state which is independent of the initial conditions, we focus on its characterization by the study of retarded Green's functions, and out-of-time ordered correlation functions, specifically the Lyapunov exponent, that describes early time quantum chaotic features \cite{larkin1969,kitaev2015,maldacena2016}. 
\begin{figure*}
    \includegraphics[width=\textwidth]{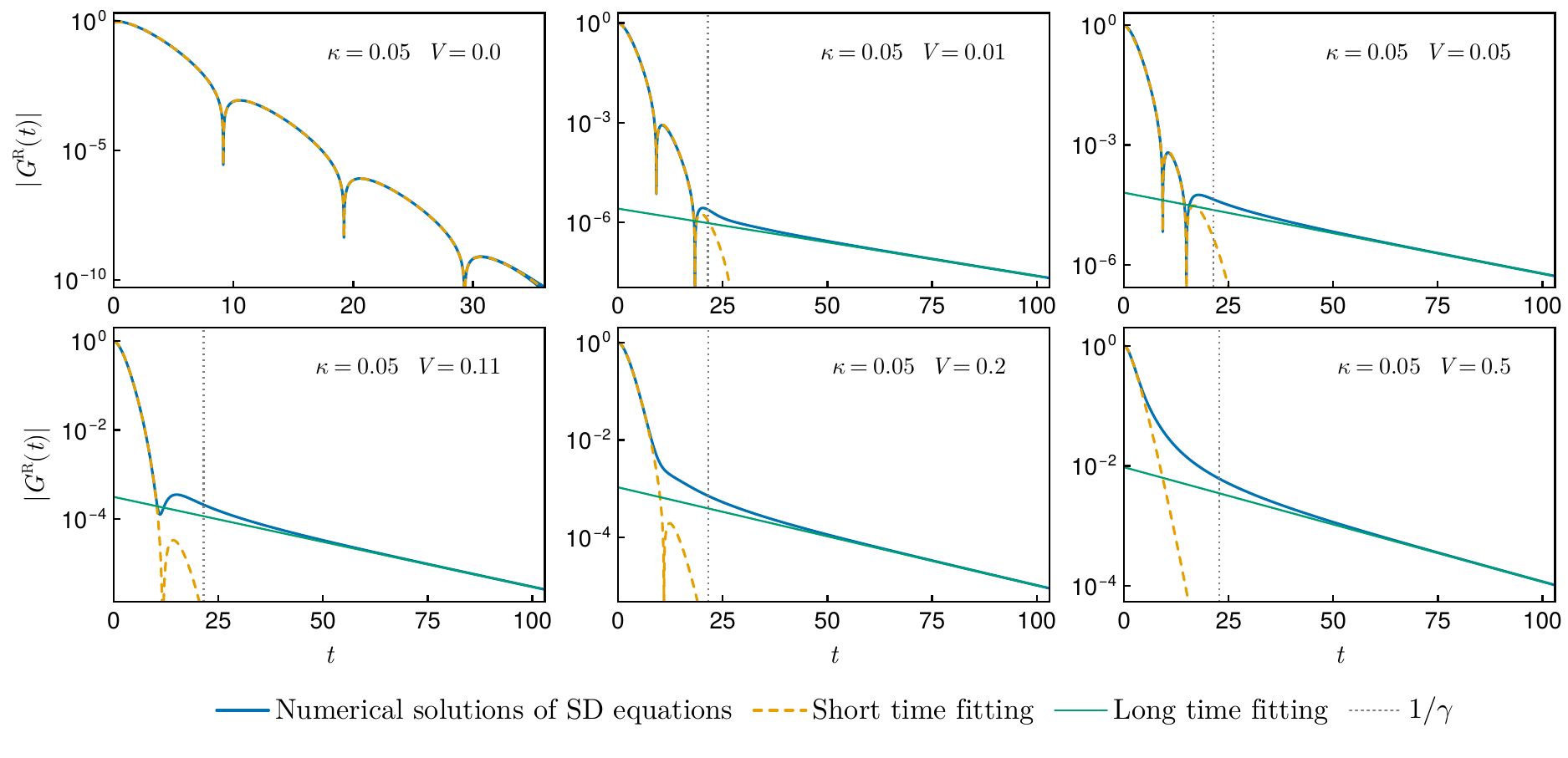}
    \caption{Fitting of $|G^{\gR}(t)|$ with Eq.~(\ref{eq:fitting}) with respect to $t$ for $\kappa = 0.05$ and different values of $V$. The exponential decay occurs in two stages. The early time (short time fitting) decay depends on both the monitoring and the bath strengths while the late time decay (log time fitting) is mostly dependent on the thermal bath only. $1/\gamma$ stands for the typical long-time decay.}
    \label{fig:GRt_fitting_kappa0.05_with_different_V_fS1_fB3}
\end{figure*}

\section{Approach to the steady state: decays(s) of the retarded Green's function}
In the previous section we showed that there exist unique, in general non-thermal, steady states which are independent of the initial conditions. For that reason, from now on we work on the long time limit $\mathcal{T} \equiv (t_1+t_2)/2 \rightarrow \infty$ which simplifies substantially the calculations. In order to characterize this steady state, we first study the retarded Green's function. Specifically, we consider the $q=4$ case with $J = 1$. The thermal bath is also an SYK model at inverse temperature $\beta_{\bath} = 100$ with $q = 4$ and $J_{\bath} = J = 1$. The system-bath interaction form is chosen as $(f_{\sys}, f_{\bath}) = (1, 3)$ and we compute the retarded Green's function by solving the simplified KB equations numerically in the parameter space $(V, \kappa)$. 

In the long time limit, we have the time-translational invariant saddle-point SD equations
\begin{equation}
    \ii \partial_t G^{\alpha \beta}(t) \!=\! s_{\alpha} \delta_{\alpha \beta} \delta(t) \!+\! \sum_{\gamma = \pm} s_{\gamma}\! \int_{-\infty}^{\infty} \!\!\Sigma^{\alpha \gamma}(t - t^\prime) G^{\gamma \beta}(t^\prime) \dd t^\prime \!, 
\end{equation}
with $\alpha, \beta \in \{+, -\}$ labeling the branch of the Schwinger-Keldysh contour, and $s_{+} = 1$, $s_{-} = -1$. The self-energies are the time-translational invariant reductions of Eqs.~\eqref{eq:Sigma_S}, \eqref{eq:Sigma_B} and \eqref{eq:Sigma_L}. These equations can be solved numerically in a similar manner to the SD equations for the Hermitian Majorana SYK model at finite temperature~\cite{maldacena2016}, with an important difference. In the Hermitian case, one may impose the Kubo-Martin-Schwinger (KMS) condition $G^>(\omega) = - \ee^{\beta \omega} G^<(\omega)$, together with the relation $G^>(t) = -G^<(-t)$ for Majorana fermions. As a result, in frequency space, it suffices to solve for either the real or the imaginary part of the Green's function~$G^>$. In the present setting, however, the absence of KMS symmetry requires solving for both the real and the imaginary parts of the Green's function.
Nevertheless, we are able to obtain unique solutions once the parameters and interaction couplings are fixed, and find excellent agreement with the long time results of the KB equations illustrated in Fig.~\ref{fig:steady_state_comparison_KB_vs_SD}.

Since one of our main goals is to show that the addition of a thermal bath deviates the steady state from the infinite-temperature steady state that characterizes the continuously monitored SYK, we start by reviewing the retarded Green's function $G^{\gR}$ in the $V = 0$ limit~\cite{garcia2023,sa2022,kulkarni2022}. In the long time limit, $G^{\gR}(t) = \theta(t) (G^>(t) - G^<(t))$. As is depicted in Fig.~\ref{fig:GRt_fitting_kappa0.05_with_different_V_fS1_fB3}, the Green's function decays exponentially to zero at a rate $\Gamma$ which for $\kappa\to 0$ agrees with the decay rate for the SYK model at infinite temperature. For $\kappa \ll 1$, the decay rate $\Gamma$ increases linearly and at $\kappa_{\rm{c}} \approx 0.15$ experiences a sharp drop until reaches a local minimum followed by a smooth monotonic increases for larger values of $\kappa$. Physically, the sharp drop is interpreted \cite{mori2024} as a transition from a (Hermitian) system-dominated dynamics to a monitoring-dominated dynamics, which admits a spectral characterization based on the role played by exceptional points after becoming real eigenvalues of the spectrum of the vectorized Liouvillian \cite{garcia2025a}. 

For $\kappa < \kappa_{\rm{c}}$~\cite{garcia2023}, the observed superimposed oscillations are well described by $|G^{\gR}(t)| = A \ee^{- \Gamma t} |\sin(\Omega t + \phi)|$ with the frequency $\Omega\sim\sqrt{|\kappa-\kappa_{\rm{c}}|  }$. By contrast, for $\kappa = 0$, the steady state is a thermal state with the same temperature as the bath~\cite{eberlein2017,almheiri2019,zhang2019}. The decay of $G^{\gR}$ is still exponential. In the conformal limit corresponding to low bath temperatures, $|G^{\gR}(t)| = B \exp(- \gamma_{\rm{B}}t)$ for $t \gg 1$, where $\gamma_{\rm B} = 2\pi / (q_{\rm B}\beta_{\rm B})$~\cite{maldacena2016}.

\begin{figure}[tbp]
    \includegraphics[width=\columnwidth]{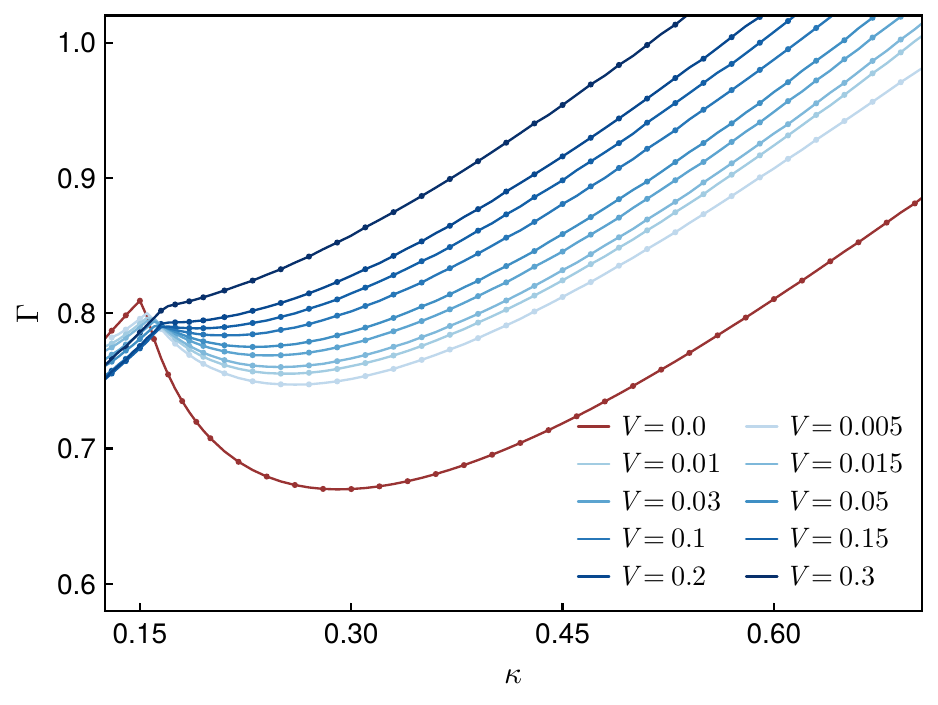}
    \caption{The dissipation rate $\Gamma$, using the fitting function Eq.~(\ref{eq:fitting}), as a function of $\kappa$ for different values of $V$ in the case $(f_{\sys}, f_{\bath})=(1,3)$. We do not present results for very small $\kappa < 0.1$ because the fitting interval is too short to get robust results. We note that even a very small $V = 0.005$ induces a large change with respect to the $V = 0$ limit which indicates that the steady state is quite different for any finite coupling to the thermal bath.}
    \label{fig:Gamma_vs_kappa_with_different_V_fS1_fB3}
\end{figure}

Results for $|G^{\gR}(t)|$, depicted in Fig.~\ref{fig:GRt_fitting_kappa0.05_with_different_V_fS1_fB3} for small $\kappa = 0.05$ and $V \lesssim 0.5$, reveal a more intricate structure than in the previous limiting cases. For intermediate times, we observe an exponential decay with oscillations, as in the pure monitored dynamics case ($V=0$, $\kappa < \kappa_{\rm{c}}$) \cite{garcia2023}, but the pattern of oscillations is more complicated. By contrast, for longer times, oscillations vanish and the decay rate becomes substantially smaller. Results depicted Fig.~\ref{fig:GRt_fitting_kappa0.05_with_different_V_fS1_fB3} illustrate that this simple fitting function  
\begin{equation}
	\bigl|G^{\gR}(t)\bigr| =
	\begin{dcases}
		A \ee^{- \Gamma t} |\!\sin(\Omega_1 t + \phi_1) \sin(\Omega_2 t + \phi_2)|, & \!\! t < t_{\mathrm{th}}, \\
		B \ee^{- \gamma t}, & \!\! t > t_{\mathrm{th}},
	\end{dcases}\label{eq:fitting}
\end{equation}
where $t_{\mathrm{th}}$ stands for a crossover time between the two decays, agrees well with the numerical results. From the fittings, we obtain the dependence of $\Gamma$, $\gamma$ and $\Omega$'s on $V$, $\kappa$. For small $V$ and $\kappa \to 0$, the decay rate~$\Gamma$, see Fig.~\ref{fig:Gamma_vs_kappa_with_different_V_fS1_fB3}, deviates from the infinite temperature SYK result. For larger $\kappa$, and still small $V$, the transition from system-dominated to bath-dominated still occurs but it is gradually smeared out as $V$ increases until the dependence of $\Gamma$ on $\kappa$ becomes monotonous. Importantly, $\Gamma$ only controls the initial decay of $G^{\gR}(t)$ for intermediate times. 

\begin{figure}[tbp]
    \includegraphics[width=\columnwidth]{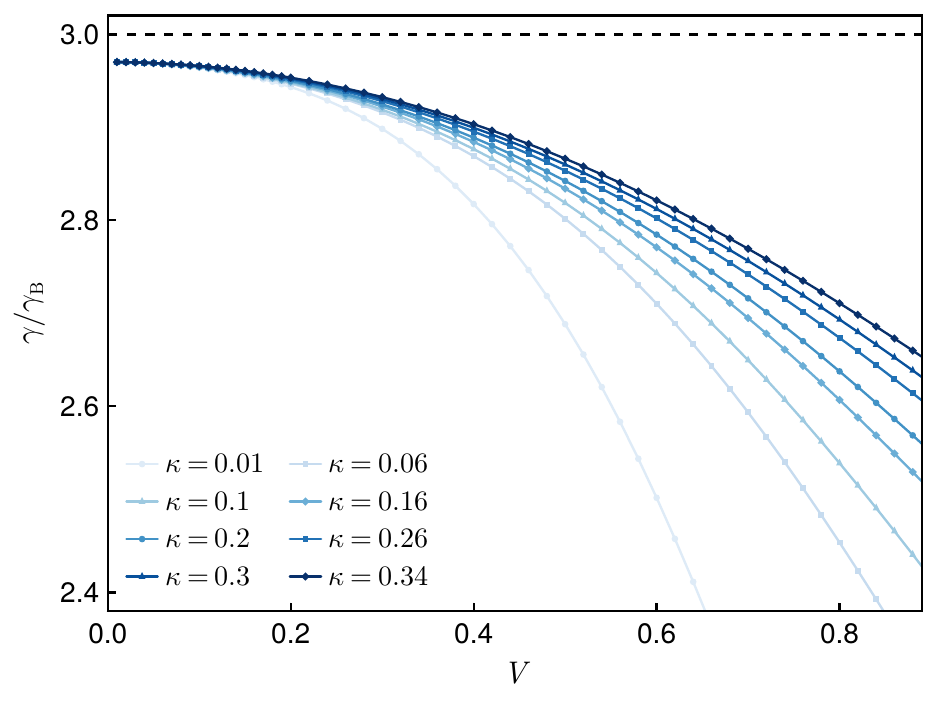}
    \includegraphics[width=\columnwidth]{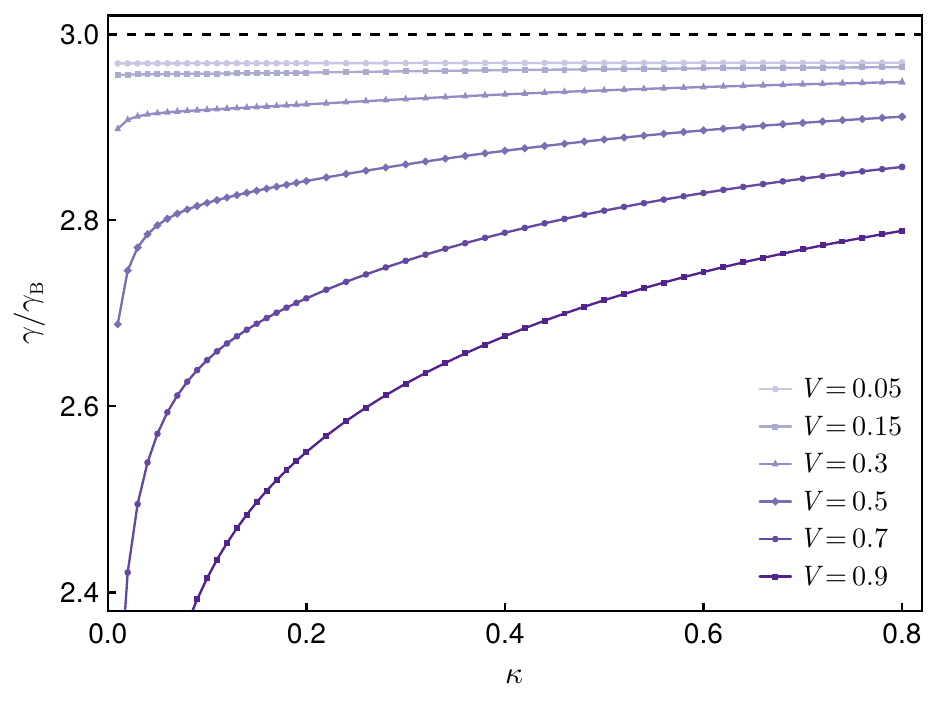}
	\caption{\textbf{Top:} Long time decay rate $\gamma / \gamma_{{\rm B}}$ versus $V$ for several values of $\kappa$ obtained by fitting the solution of the saddle point equation with Eq.~(\ref{eq:fitting}). \textbf{Bottom:} $\gamma / \gamma_{{\rm B}}$ versus $\kappa$ for several values of $V$. We take $\beta_{\rm B}=100$, $q=4$ and $J=1$. $\gamma$ is obtained by fitting with Eq.~(\ref{eq:fitting}). The top dashed line is the analytical prediction $\gamma \approx (q-1)\gamma_{\rm B}$ in the weak coupling limit.}
	\label{fig:gamma_ratio_vs_V_beta}
\end{figure}
In the long time limit, $t>t_\mathrm{th}$, results depicted in Fig.~\ref{fig:gamma_ratio_vs_V_beta} for the dependence of the ratio $\gamma/\gamma_{\mathrm{B}}$ on $V$ and $\kappa$, obtained by varying one parameter while keeping the other fixed, shows that for a fixed $V$, $\gamma$ increases monotonically as $\kappa$ increases, while for a fixed $\kappa$, $\gamma$ decreases monotonically as $V$ increases. This suggests that the continuous monitoring heats up the system while the thermal bath cools it down though we stress that the steady state is never thermal. In particular, we provide evidence below that the long time decay rate in the weak coupling limit, in both $\kappa$ and $V$, is given by
\begin{equation} 
\gamma = (q-1) \gamma_{\mathrm{B}} + \delta \gamma, 
\end{equation}
where $\delta \gamma$ is a small negative correction whose magnitude decreases as $\beta_{\mathrm{B}}$ increases. Since the system-bath interaction is of type $(1, q-1)$, the retarded system-bath self-energy~$\Sigma_{\rm B}^{\rm R}$ then has a decay rate $(q-1) \gamma_{\rm{B}}$. The long time decay rate is therefore a bath-induced effect, and is given by the $\gamma = - \Im \omega_*$, whereby $\omega_*$ is the pole of the retarded Green's function $G^{\gR}(\omega)$ in the lower half plane that is closest to the real axis.

We now show analytically why $\gamma$ depends predominantly on $\beta_{\bath}$, with sub-leading corrections from the coupling strengths $\kappa$ and $V$. From the Schwinger-Dyson equations we have
\begin{equation}
    G^{\rm R}(\omega) = \frac{1}{D(\omega)} , 
    \quad
    D(\omega) \equiv \omega - \Sigma_{\rm S}^{\rm R}(\omega) - \Sigma_{\rm B}^{\rm R}(\omega) + \ii \kappa ,
\end{equation}
where $\Sigma_{\rm S}^{\rm R}$ is the self-energy corresponding to the SYK interaction. Since $\Sigma_{\bath}^{\gR}(t)$ decays exponentially with rate $(q-1)\gamma_{\bath}$, its Fourier transform $\Sigma_{\bath}^{\gR}(\omega)$ has a pole at $\omega_0=-\ii(q-1)\gamma_{\bath}$. When $V \ll J$, $\Sigma_{\bath}^{\gR}(\omega)$ is negligible across a wide range of $\omega$ in the complex plane except in the vicinity of its pole $\omega_0$, whereby it overwhelms other terms. This implies that there exists a zero $\omega_*$ of $D(\omega)$ near $\omega_0$, owing to the analyticity of $D(\omega)$ away from its poles. The simple pole $\omega_*$ satisfies the necessary condition $D(\omega_*) = 0$ and $D'(\omega_*) \neq 0$, giving an expansion of the retarded Green's function as 
\begin{equation}
    G^{\rm R}(\omega) = \frac{1/D'(\omega_*)}{\omega - \omega_*} + \dots
\end{equation}
To compute the corrections, we first expand the simple pole by $\omega_* = \omega_0 + \delta \omega$, where $\delta \omega$ is a small correction compared to $\omega_0$. We then expand $D(\omega)$ near $\omega_0$ to solve for $\delta \omega$. Substituting the expansion of $\Sigma_{\bath}^{\gR}$ into the expression for $D(\omega)$, noticing that the derivative of $\Sigma_{\bath}^{\gR}$ at $\omega_*$ is dominant over the other terms, and keeping terms up to order $\mathcal{O}(\delta \omega)$, we obtain both the correction $\delta \omega$ and the residue,
\begin{align}
    \delta \omega & = \frac{A}{D_{\rm reg}(\omega_0)} , \\
    \frac{1}{D'(\omega_*)} & = \frac{A}{[D_{\rm reg}(\omega_0)]^2} + \mathcal{O}(\delta \omega^4) ,
\end{align}
where the regular part is given by,
\begin{equation}
	D_{\rm reg}(\omega_0) = \omega_0 - \Sigma_{\rm S}^{\rm R}(\omega_0) - B + \ii \kappa .
\end{equation}
For computing the residue, we plug $D_{\rm reg}$ into the solution for~$\delta \omega$. Here $A \propto - V^2 \beta_{\bath}^{-2(1 - \Delta)}$ and $B \propto - \ii V^2 \beta_{\bath}^{-(1 - 2\Delta)}$, see Eqs.~(S27) and (S28) of the Supplementary Materials for precise definitions of $A, B$. The correction to the long time decay rate therefore is given by $\delta \gamma = - \Im \delta \omega$. Due to the non-negligible system self-energy $\Sigma_{\rm S}^{\rm R}(\omega_0)$, one cannot explicitly compute the correction $\delta \omega$. However, we can still discuss its qualitative behavior in the long time decay rate. First of all, it explains that $\delta \gamma$ is negative, such that the corrections decrease the decay rate. Secondly, as $\kappa$ increases, the magnitude of $\delta \gamma$ decreases, hence $\gamma$ increases. Thirdly, since~$A$ is proportional to $V^2$, as $V$ increases, the magnitude of $\delta \gamma$ increases, and therefore $\gamma$ decreases with~$V$. These properties are in  agreement with the numerical results depicted in Fig.~\ref{fig:gamma_ratio_vs_V_beta}. Finally, for larger $\beta_{\rm B}$ (lower temperature), we have a smaller $A$, consequently the corrections are smaller than in the high temperature limit, which agrees with the numerical results for $\beta_{\rm B}=1000$ presented in the Supplementary Material. 
 
\section{Quantum chaos induced by continuous monitoring}
In order to further characterize the quantum dynamics, we compute the Lyapunov exponent $\lambda_{\mathrm{L}}$, defined as the rate of exponential growth of out-of-time-ordered correlation function (OTOC) before the scrambling time $t_{\rm scr} \sim \lambda_{\mathrm{L}}^{-1} \log{N}$. Its positivity is a distinct feature of many-body quantum chaotic dynamics~\cite{larkin1969,maldacena2016}. In the low temperature limit of the Hermitian SKY model, $\lambda_{\mathrm{L}}$ saturates a bound on chaos~\cite{maldacena2015} that indicates the existence of a gravity dual. The procedure to compute it in our case is similar to that of the Hermitian SYK model, but differs in one important respect. In the Hermitian SYK~\cite{maldacena2016,bermudez2019}, at a given temperature, the OTOC is {\it regularized} by splitting the thermal loop into two real-time folds separated by intervals of imaginary time. This regularization inherently requires a well-defined thermal equilibrium state. In our case, however, as shown in previous sections, the steady state is not thermal. Therefore, to compute the OTOCs, we do not perform any regularization, namely, in the relevant two folded contours, all imaginary times are set to zero while keeping the contour ordering fixed. Fortunately, it has been shown, in SYK~\cite{bermudez2019} and melonic quantum mechanics~\cite{Ferrari:2019ogc}, that while the OTOC depends on the chosen regularization, the Lyapunov exponent is independent of it so a comparison with systems at thermal equilibrium is meaningful. We present here a brief derivation and leave the full details of the calculation to the Supplementary Materials.

OTOCs are defined as the four-point correlation functions with non-trivial operator ordering, naturally formulated on the Schwinger-Keldysh contour~$\mathcal{C}$ which involves two real-time folds. In the bi-local collective field description, they can be expressed in terms of two-point functions $G$, with time arguments placed on different folds in an out-of-real-time ordered manner. Consider time argument $z = t + \ii \tau \in \mathcal{C}$, where $t = \mathrm{Re}(z)$ ($\tau = \mathrm{Im}(z) \leq 0$) denotes the real (imaginary) time. Due to time translational invariance, we may take one of the bi-local fields as~$G(\ii \tau, 0)$, and the other as~$G(z_1, z_2)$, with the following relation enforcing the contour ordering
\begin{equation} \label{eq:OTOC_ordering}
	\tau_1 < \tau < \tau_2 < 0 .
\end{equation}
Performing the contour ordering, the four-point function becomes
\begin{equation}
	\langle \operatorname{T}_{\mathcal{C}}\! G(z_1 , z_2) G(\ii \tau , 0) \rangle \!=\! - \frac{1}{N^2} \!\!\! \sum_{i,j=1}^{N} 
	\!\! \langle \psi_i(t_1) \psi_j(0) \psi_i(t_2) \psi_i(0) \rangle ,
\end{equation}
which is manifestly in an out-of-(real)-time ordered manner.
Expanding the four-point function up to order $1/N$, we have the connected OTOC
\begin{equation} \label{eq:OTOC_def}
    \frac{1}{N} C(t_1, t_2) = \langle \operatorname{T}_{\mathcal{C}}\! G(z_1, z_2) G(\ii \tau, 0) \rangle \!-\!
	G_0(z_1, z_2) G_0(\ii \tau, 0) , 
\end{equation}
where $G_0$ denotes the large~$N$ saddle point solution that was numerically evaluated in the last section, and~$C$ accounts for the $1/N$-corrections. In the long time limit $t_1, t_2 \gg 0$, which governs the onset for quantum chaos~\cite{Ferrari:2019ogc}, the connected OTOC satisfies the recursion relation
\begin{equation}
    C(t_1, t_2) = \int \mathcal{K}(t_1, t_2, t_3, t_4) C(t_3, t_4) \dd t_3 \dd t_4 ,
\end{equation}
with the kernel
\begin{equation}
	\mathcal{K}(t_1, t_2, t_3, t_4) = - 3 J^2 G_0^{\gR}(t_{13}) G_0^{\gR}(t_{24}) \bigl[G_0^{\gW}(t_{34})\bigr]^2 ,
\end{equation}
where $t_{ij} \equiv t_i - t_j$, and $G_0^{\gW}$ is the Wightman function
\begin{equation}
	G_0^{\gW}(t_1, t_2) = G_0\bigl(t_1 - t_2 + \ii (\tau_1 - \tau_2)\bigr) .
\end{equation}
\begin{figure}[tbp]
    \includegraphics[width=\columnwidth]{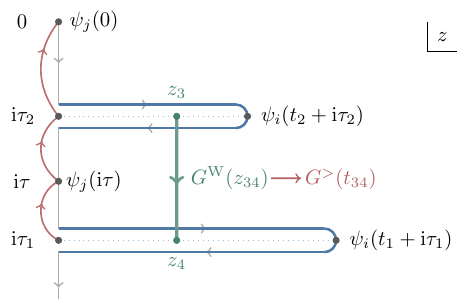}
    \caption{Schwinger-Keldysh contour of OTOC with contour ordering $\tau_1 < \tau < \tau_2$. The red arcs represent taking the limit sequentially: first $\tau_1 \rightarrow \tau^-$, then $\tau \rightarrow \tau_2^-$, and finally $\tau_2^- \rightarrow 0^-$, which is equivalent to taking all Euclidean time arguments to 0 while keeping the contour ordering relation fixed. The Wightman function $G^{\gW}$ colored in dark green with non-zero imaginary time separation consequently reduces to the greater Green's function $G^>$, giving rise to the un-regularized OTOC.}
    \label{fig:OTOC_contours}
\end{figure}

The regularized OTOCs are obtained by setting $\tau_1 = 3 \beta/4$, $\tau = \beta/2$, and $\tau_2 = \beta/4$ with $\beta$ the inverse temperature of the system. In the absence of thermal steady states, which in our case means the absence of the KMS condition, we instead compute the un-regularized OTOC by sending all imaginary time arguments $\tau_i$ to 0 while keeping the contour ordering fixed. This gives a valid definition of OTOCs. Indeed, we have numerically verified that, in agreement with the results of Ref.~\cite{bermudez2019}, the un-regularized OTOCs of Hermitian SYK at finite temperature produce the same Lyapunov exponents extracted from the regularized ones. An illustration for our setup, and the mentioned limiting process, is exhibited in Fig.~\ref{fig:OTOC_contours}. For the un-regularized OTOCs, the Wightman function reduces to the greater Green's function
\begin{equation}
\lim_{\tau_2^{} \rightarrow  0_{}^-} \, \lim_{\tau_1 \rightarrow \tau_2^-} G_0^{\gW}(t_1, t_2) = G_0^{>}(t_{12}) .
\end{equation}
The limits are taken sequentially, with the ordering $\tau_1 < \tau_2$ kept fixed. This reduction, together with the fact that the greater Green's function $G^>$ can be expressed by retarded Green's function $G^{\rm R}$, implies that the kernel~$\mathcal{K}$ is completely determined by $G^{\rm R}_0$. We may extract the Lyapunov exponent $\lambda_{\rm L}$ from the recursion relations of the connected OTOCs. Specifically, making an exponential growth ansatz in the center-of-mass frame
\begin{equation} \label{eq:OTOC_ansatz}
    C(\mathcal{T}, t) = \ee^{\lambda_{\mathrm{L}} \mathcal{T}} f(t),
\end{equation}
the center-of-mass time $\mathcal{T} = (t_1 + t_2) / 2$ then decouples from the recursion relation, leaving an eigenvalue problem for $f(t)$ with $t = t_1 - t_2$:
\begin{equation}
    f(t) = \int_{-\infty}^{\infty} M(t, t'; \lambda_\mathrm{L}) f(t') \dd{t'} \! .
\end{equation}
The reduced kernel is
\begin{equation} \label{eq:reduced_kernel_M_i}
\begin{split}
    M(t, t' ; \lambda_\mathrm{L}) = & 3 J^2 \bigl[G^{>}(t')\bigr]^{2}  \\
	& \times
	\int_{-\infty}^{\infty} G^{\gR}(u_+) G^{\gR}(u_-)
    \ee^{-\lambda_{\mathrm{L}} u} \dd{u} \! ,
\end{split}
\end{equation}
where $u_{\pm} \equiv u \pm (t - t')/2$. To be consistent with the exponential growth ansatz Eq.~ \eqref{eq:OTOC_ansatz}, the Lyapunov exponent~$\lambda_\mathrm{L}$ must correspond to the value for which the kernel matrix $M(t, t'; \lambda_{\mathrm{L}})$ has its largest eigenvalue to be~$1$. Numerically, this can be done by discretizing times such that $M$ becomes a discretized matrix. Due to the high sparsity of the discretized kernel matrix, one may employ  Krylov subspace method to search for the largest eigenvalue, and then perform a binary search for $\lambda_\mathrm{L}$. We have verified the uniqueness of the Lyapunov exponent by varying $\lambda_{\rm L}$ over a large range. The growth ansatz~\eqref{eq:OTOC_ansatz} itself implies that a positive $\lambda_{\rm L}$ leads to exponential growth of the connected four-point function $C$ and hence quantum chaos, while for $\lambda_{\rm L} < 0$, $C$ vanishes in the long time limit. The positiveness of $\lambda_{\rm L}$ can therefore serve as a diagnostic for quantum chaos~\cite{garcia2024}. In addition, the alternative ordering $\tau_2 < \tau < \tau_1 < 0$ gives rise to the other independent OTOC,~$C'$. In general, the recursion relation for~$C$ and~$C'$ are coupled through off-diagonal kernels. However, for the Majorana SYK model, the off-diagonal kernels vanish, hence the recursion relations become diagonal, and $C$ and $C'$ each satisfy a closed recursion relation independently. Nevertheless, we have checked that it gives the same Lyapunov exponent. For more details, we refer to the Supplementary Materials. We now present the main results for the Lyapunov exponent.
\begin{figure}[tbp]
    \includegraphics[width=\columnwidth]{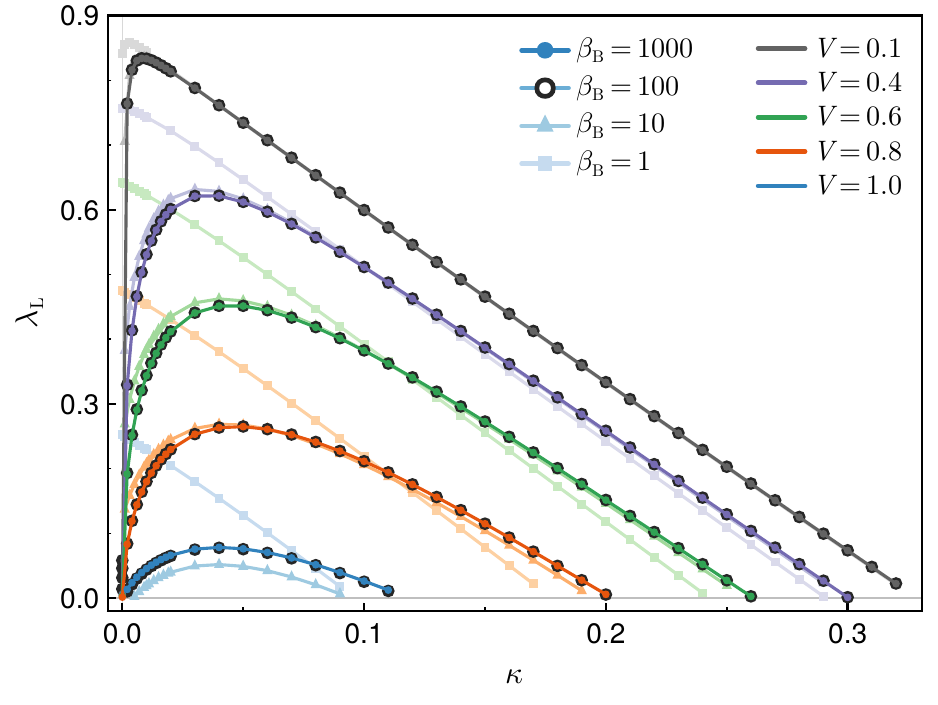}
    \includegraphics[width=\columnwidth]{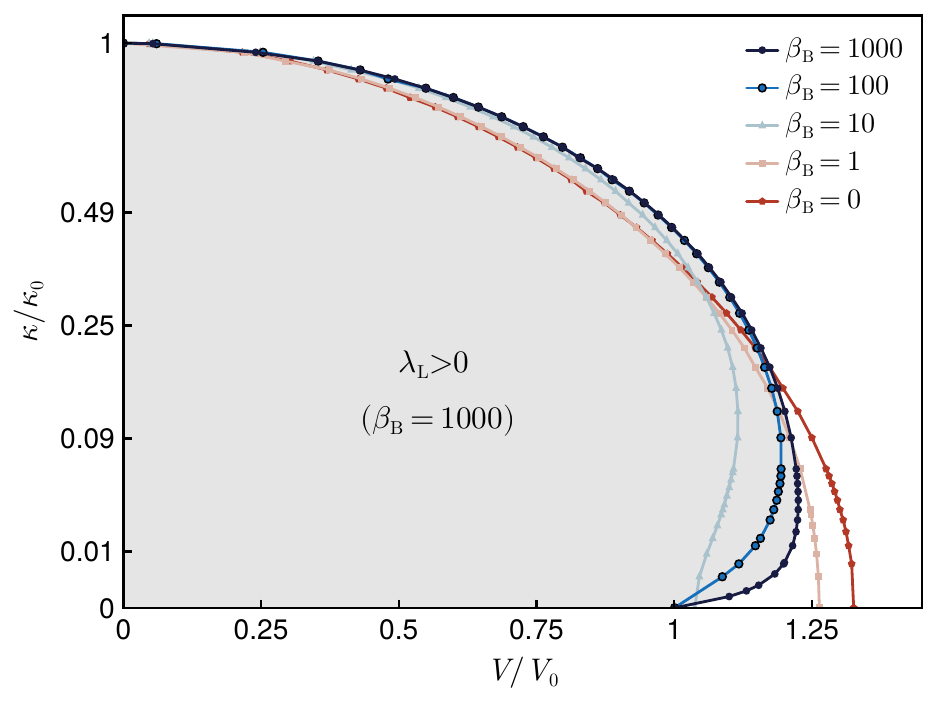}
    \caption{\textbf{Top:} The Lyapunov exponent $\lambda_{\mathrm{L}}$ versus $\kappa$, the monitoring strength, for a fixed $V$ with different $\beta_{\bath}$. Different colors denote different values of the coupling $V$, while different shades and markers denote different values of the bath inverse temperature $\beta_{\bath}$. \textbf{Bottom:} Critical lines at which $\lambda_{\mathrm{L}}=0$ for different $\beta_{\bath}$. The grey area is the corresponding quantum chaotic region ($\lambda_{\rm L}>0$) for $\beta_{\bath} =1000$. The $y$-axis is shown with a power transform $y \mapsto \sqrt{y}$ to enhance re-entrant behavior near $\kappa=0$.}
    \label{fig:Lyapunov_vs_kappa_different_betaB}
\end{figure}

Our first goal is to provide further evidence that our model deviates from the infinite-temperature limit typical of continuously monitored systems modeled by the Lindblad formalism. We recall that for $V = 0$ and $\kappa \to 0$ but finite \cite{garcia2024}, $\lambda_{\mathrm{L}}$ agrees with the infinite temperature Hermitian SYK model result \cite{maldacena2016} which is proportional to the coupling $J$, and therefore of order one, while in the low temperature limit of the Hermitian SYK, $\lambda_{\mathrm{L}} = 2\pi T k_{\mathrm{B}}/\hbar$ is small. 
In Fig.~\ref{fig:Lyapunov_vs_kappa_different_betaB} (Top), we depict results for $\lambda_{\mathrm{L}}$ as a function of $\kappa$ for  various $V$, where different shades and markers with the same color denote different values of $\beta_{\bath}$ for a fixed $V$. We observe that the effect of a finite $V$, for any $\beta_{\bath}$, is to monotonically decrease  $\lambda_{\mathrm{L}}$, indicating that the non-thermal steady state studied in the previous section is qualitatively different from an infinite temperature state. 

A natural question to ask, in light of these results, is whether the Lyapunov exponent is controlled by the early or by the late time decay of the Green's function. In the weak coupling limit, which is our major focus of interest, we have checked that the Lyapunov exponent is mostly given by the early exponential decay which combines features of both the monitoring, described by the Lindblad formalism, and the thermal bath. This is not surprising as for sufficiently large $V$, the Lyapunov exponents vanish if continuous monitoring is not turned on. 
We have also checked that our results are consistent with previous findings in two limiting cases. For $V=0$ and $\kappa \neq 0$, pure Lindblad dynamics, the results agrees those those of Ref.~\cite{garcia2024}: $\lambda_{\mathrm{L}}$ decreases monotonically as $\kappa$ increases and vanishes at $\kappa_0 \approx 0.3306$. For $\kappa = 0$ and $V \neq 0$, $\lambda_{\mathrm{L}}$ exhibits agreement with that of Ref.~\cite{zhang2019}, and vanishes exactly at $V_0 = 3^{-3/8} \sqrt{2} \approx 0.9367$.

A standout feature of Fig.~\ref{fig:Lyapunov_vs_kappa_different_betaB} is that the dependence of~$\lambda_{\mathrm{L}}$ on $\kappa$ is non-monotonous. Especially for large~$\beta_{\mathrm{B}}$ and small~$V$, there is a sharp increase of $\lambda_{\mathrm{L}}$ for small $\kappa$, until it reaches a local maximum for intermediate values of $\kappa$. For larger values of $\kappa$, $\lambda_{\mathrm{L}}$ decreases until it vanishes. In the small $\beta_{\mathrm{B}}$ limit, the effect is almost absent because the system is already at high temperature before any continuous monitoring. Likewise, for sufficiently high~$\beta_{\mathrm{B}}$, results are independent on $\beta_{\mathrm{B}}$, which indicates that once the thermal bath is close to the conformal limit, the impact of the monitoring coupling $\kappa$ is similar because the effective temperature is already close to zero.

The increase of $\lambda_{\mathrm{L}}$ with $\kappa$, for small $V$, reveals that, counterintuitively, continuous monitoring can enhance quantum chaos.  
This increase eventually terminates because strong continuous monitoring suppresses quantum effects such as scrambling that cause the exponential growth of the OTOC. The addition of a thermal bath reveals more clearly the dual effect of continuous monitoring on quantum chaotic features: it can make quantum chaos more robust by the heating effect that increase the Lyapunov exponent, but at the same time, its dissipative character will weaken it as well. Which feature prevails depends on the parameters of the model, but overall the combined effect of monitoring and a thermal bath provides a much richer quantum chaotic dynamics than anticipated. 

Another interesting feature of the role of continuous monitoring on quantum chaos is depicted in Fig.~\ref{fig:Lyapunov_vs_kappa_different_betaB} (Bottom) that identifies the quantum chaotic regions in $\kappa, V$ space, namely, the regions with $\lambda_{\mathrm{L}}> 0$ for different $\beta_{\mathrm{B}}$.  Intriguingly, we observe a re-entrant behavior only for low bath temperatures, indicating that, in a narrow range of $V$, the quantum chaotic domain satisfying $\lambda_{\mathrm{L}} > 0$ requires a minimum non-zero value of $\kappa$. Therefore, rather unexpectedly, many-body quantum chaos, defined by a positive Lyapunov exponent, can be induced by continuous monitoring. We believe that this mechanism is not specific to the SYK model, or the specific choice of jump operators, but rather to the large coupling of a quantum chaotic system to a thermal cold bath. This setting is quite generic so we believe our results are potentially of interest for the design of quantum devices where it is necessary to dial the quantum chaotic properties of a many-body quantum system for the optimization of quantum coherence.   
We also note that in the limit of large dissipation rate $\kappa$ ($\kappa \gtrsim 0.1$), the critical lines for large $\beta_{\mathrm{B}}$ are almost indistinguishable for the reasons explained above. However, for small $\kappa$, there are deviations, and the critical line for smaller $\beta_{\bath}$ is bounded by the one with larger $\beta_{\bath}$. This is an expected behavior, since in the small $\beta_{\bath}$ limit, the Lyapunov exponent is much larger and therefore it can stay positive for larger values of $\kappa$.

Finally, we discuss the bath dominated region corresponding to sufficiently large $V, \kappa$  where the Lyapunov exponent is not positive. We first stress that in all cases we still find solutions with a negative Lyapunov exponent. Physically, this corresponds to a regime where quantum effects are exponentially suppressed, due to the monitoring, or the effect of the thermal bath if $V \gtrsim 1$, so the dynamics is no longer quantum chaotic. Although it is tempting to term it a quantum-classical transition, we consider that further research is needed for a complete characterization of this region.

\section{Conclusions}
We have studied the impact of continuous monitoring on the dynamics of quantum many-body systems coupled to a thermal bath.
For technical reasons, we have focused our analysis in the SYK model but we think that this setting combining monitoring and a thermal bath will lead to similar results for other models. 
We have first shown that the quenched dynamics governed by the two-times KB equations have unique steady state solutions that do not depend on the initial state and that these solutions are in general non-thermal and different from those of the monitored SYK model with no coupling to a thermal bath where the steady state is at infinite temperature.  

The approach to the steady state, as described by the retarded Green's function, also shows qualitative differences with the infinite temperature case: there are two exponential decays. For late times, we show analytically that the decay rate is controlled by an effective temperature closely related to the bath temperature. For intermediate times, the decay is also exponential but at a different rate. Oscillations occur in the limit of weak monitoring limit which is qualitatively similar to the behavior of the SYK if no thermal bath is present. 

We have also studied quantum chaotic properties by computing the Lyapunov exponent. Two features stand out: in the weak monitoring limit, turning on the thermal bath, at sufficiently low temperature, induces a sharp decrease of the Lyapunov exponent which is a strong indication of both that continuous monitoring can enhance quantum chaos and that the steady state of the system is qualitatively different from the infinite-temperature limit expected in systems governed by the Lindblad formalism. 
More interestingly, we have identified a range of parameters for which the Lyapunov exponent vanishes if there is no continuous monitoring but, 
after we turning it on, the Lyapunov exponent becomes eventually finite if the coupling is not too strong. Therefore, unexpectedly, continuous monitoring can induce many-body quantum chaos. This is one of the main results of the paper.

Moreover, for any finite monitoring strength, the Lyapunov exponent vanishes at a certain coupling $V$ of the thermal bath which illustrates the expected suppression of quantum effects due to dissipation and monitoring. 
  
Although for technical reasons we have restricted our analysis to the SYK model, and a single choice of jump operators, we think our results are of general applicability though a full confirmation is still challenging due to the technical difficulties of simulating strongly interacting systems coupled to two baths where, unlike the SYK model, it is not possible to express the action in terms of bi-linear fields and solve it in the saddle-point limit.  
Both the re-entrant behavior of the Lyapunov exponent and its enhancement by increasing the monitoring strength are intriguing features that offer insights about how to dial quantum scrambling in a many-body quantum systems which is of direct practical relevance in quantum information applications.   

\section*{Acknowledgments}
We were partially supported by the National Science Foundation of China
(NSFC), Individual Grant No.12374138.
\bibliographystyle{apsrev4-2}
\bibliography{cooling}

\begin{thebibliography}{56}%
\makeatletter
\providecommand \@ifxundefined [1]{%
 \@ifx{#1\undefined}
}%
\providecommand \@ifnum [1]{%
 \ifnum #1\expandafter \@firstoftwo
 \else \expandafter \@secondoftwo
 \fi
}%
\providecommand \@ifx [1]{%
 \ifx #1\expandafter \@firstoftwo
 \else \expandafter \@secondoftwo
 \fi
}%
\providecommand \natexlab [1]{#1}%
\providecommand \enquote  [1]{``#1''}%
\providecommand \bibnamefont  [1]{#1}%
\providecommand \bibfnamefont [1]{#1}%
\providecommand \citenamefont [1]{#1}%
\providecommand \href@noop [0]{\@secondoftwo}%
\providecommand \href [0]{\begingroup \@sanitize@url \@href}%
\providecommand \@href[1]{\@@startlink{#1}\@@href}%
\providecommand \@@href[1]{\endgroup#1\@@endlink}%
\providecommand \@sanitize@url [0]{\catcode `\\12\catcode `\$12\catcode
  `\&12\catcode `\#12\catcode `\^12\catcode `\_12\catcode `\%12\relax}%
\providecommand \@@startlink[1]{}%
\providecommand \@@endlink[0]{}%
\providecommand \url  [0]{\begingroup\@sanitize@url \@url }%
\providecommand \@url [1]{\endgroup\@href {#1}{\urlprefix }}%
\providecommand \urlprefix  [0]{URL }%
\providecommand \Eprint [0]{\href }%
\providecommand \doibase [0]{https://doi.org/}%
\providecommand \selectlanguage [0]{\@gobble}%
\providecommand \bibinfo  [0]{\@secondoftwo}%
\providecommand \bibfield  [0]{\@secondoftwo}%
\providecommand \translation [1]{[#1]}%
\providecommand \BibitemOpen [0]{}%
\providecommand \bibitemStop [0]{}%
\providecommand \bibitemNoStop [0]{.\EOS\space}%
\providecommand \EOS [0]{\spacefactor3000\relax}%
\providecommand \BibitemShut  [1]{\csname bibitem#1\endcsname}%
\let\auto@bib@innerbib\@empty
\bibitem [{\citenamefont {Fazio}\ \emph {et~al.}(2024)\citenamefont {Fazio},
  \citenamefont {Keeling}, \citenamefont {Mazza},\ and\ \citenamefont
  {Schirò}}]{Fazio2024lectures}%
  \BibitemOpen
  \bibfield  {author} {\bibinfo {author} {\bibfnamefont {R.}~\bibnamefont
  {Fazio}}, \bibinfo {author} {\bibfnamefont {J.}~\bibnamefont {Keeling}},
  \bibinfo {author} {\bibfnamefont {L.}~\bibnamefont {Mazza}},\ and\ \bibinfo
  {author} {\bibfnamefont {M.}~\bibnamefont {Schirò}},\ }\href
  {https://arxiv.org/abs/2409.10300} {\bibinfo {title} {Many-body open quantum
  systems}} (\bibinfo {year} {2024}),\ \Eprint
  {https://arxiv.org/abs/2409.10300} {arXiv:2409.10300 [quant-ph]} \BibitemShut
  {NoStop}%
\bibitem [{\citenamefont {Nagourney}\ \emph {et~al.}(1986)\citenamefont
  {Nagourney}, \citenamefont {Sandberg},\ and\ \citenamefont
  {Dehmelt}}]{warren1986}%
  \BibitemOpen
  \bibfield  {author} {\bibinfo {author} {\bibfnamefont {W.}~\bibnamefont
  {Nagourney}}, \bibinfo {author} {\bibfnamefont {J.}~\bibnamefont
  {Sandberg}},\ and\ \bibinfo {author} {\bibfnamefont {H.}~\bibnamefont
  {Dehmelt}},\ }\href {https://doi.org/10.1103/PhysRevLett.56.2797} {\bibfield
  {journal} {\bibinfo  {journal} {Phys. Rev. Lett.}\ }\textbf {\bibinfo
  {volume} {56}},\ \bibinfo {pages} {2797} (\bibinfo {year}
  {1986})}\BibitemShut {NoStop}%
\bibitem [{\citenamefont {Zoller}\ \emph {et~al.}(1987)\citenamefont {Zoller},
  \citenamefont {Marte},\ and\ \citenamefont {Walls}}]{zoller1987}%
  \BibitemOpen
  \bibfield  {author} {\bibinfo {author} {\bibfnamefont {P.}~\bibnamefont
  {Zoller}}, \bibinfo {author} {\bibfnamefont {M.}~\bibnamefont {Marte}},\ and\
  \bibinfo {author} {\bibfnamefont {D.~F.}\ \bibnamefont {Walls}},\ }\href
  {https://doi.org/10.1103/PhysRevA.35.198} {\bibfield  {journal} {\bibinfo
  {journal} {Phys. Rev. A}\ }\textbf {\bibinfo {volume} {35}},\ \bibinfo
  {pages} {198} (\bibinfo {year} {1987})}\BibitemShut {NoStop}%
\bibitem [{\citenamefont {Gleyzes}\ \emph {et~al.}(2007)\citenamefont
  {Gleyzes}, \citenamefont {Kuhr}, \citenamefont {Guerlin}, \citenamefont
  {Bernu}, \citenamefont {Deléglise}, \citenamefont {Busk~Hoff}, \citenamefont
  {Brune}, \citenamefont {Raimond},\ and\ \citenamefont
  {Haroche}}]{gleyzes2007}%
  \BibitemOpen
  \bibfield  {author} {\bibinfo {author} {\bibfnamefont {S.}~\bibnamefont
  {Gleyzes}}, \bibinfo {author} {\bibfnamefont {S.}~\bibnamefont {Kuhr}},
  \bibinfo {author} {\bibfnamefont {C.}~\bibnamefont {Guerlin}}, \bibinfo
  {author} {\bibfnamefont {J.}~\bibnamefont {Bernu}}, \bibinfo {author}
  {\bibfnamefont {S.}~\bibnamefont {Deléglise}}, \bibinfo {author}
  {\bibfnamefont {U.}~\bibnamefont {Busk~Hoff}}, \bibinfo {author}
  {\bibfnamefont {M.}~\bibnamefont {Brune}}, \bibinfo {author} {\bibfnamefont
  {J.-M.}\ \bibnamefont {Raimond}},\ and\ \bibinfo {author} {\bibfnamefont
  {S.}~\bibnamefont {Haroche}},\ }\href {https://doi.org/10.1038/nature05589}
  {\bibfield  {journal} {\bibinfo  {journal} {Nature}\ }\textbf {\bibinfo
  {volume} {446}},\ \bibinfo {pages} {297–300} (\bibinfo {year}
  {2007})}\BibitemShut {NoStop}%
\bibitem [{\citenamefont {Minev}\ \emph {et~al.}(2019)\citenamefont {Minev},
  \citenamefont {Mundhada}, \citenamefont {Shankar}, \citenamefont {Reinhold},
  \citenamefont {Gutiérrez-Jáuregui}, \citenamefont {Schoelkopf},
  \citenamefont {Mirrahimi}, \citenamefont {Carmichael},\ and\ \citenamefont
  {Devoret}}]{minev2019}%
  \BibitemOpen
  \bibfield  {author} {\bibinfo {author} {\bibfnamefont {Z.}~\bibnamefont
  {Minev}}, \bibinfo {author} {\bibfnamefont {S.}~\bibnamefont {Mundhada}},
  \bibinfo {author} {\bibfnamefont {S.}~\bibnamefont {Shankar}}, \bibinfo
  {author} {\bibfnamefont {P.}~\bibnamefont {Reinhold}}, \bibinfo {author}
  {\bibfnamefont {R.}~\bibnamefont {Gutiérrez-Jáuregui}}, \bibinfo {author}
  {\bibfnamefont {R.}~\bibnamefont {Schoelkopf}}, \bibinfo {author}
  {\bibfnamefont {M.}~\bibnamefont {Mirrahimi}}, \bibinfo {author}
  {\bibfnamefont {H.}~\bibnamefont {Carmichael}},\ and\ \bibinfo {author}
  {\bibfnamefont {M.}~\bibnamefont {Devoret}},\ }\href
  {https://doi.org/10.1038/s41586-019-1287-z} {\bibfield  {journal} {\bibinfo
  {journal} {Nature}\ }\textbf {\bibinfo {volume} {570}},\ \bibinfo {pages}
  {200–204} (\bibinfo {year} {2019})}\BibitemShut {NoStop}%
\bibitem [{\citenamefont {Plenio}\ and\ \citenamefont
  {Knight}(1998)}]{plenio1998}%
  \BibitemOpen
  \bibfield  {author} {\bibinfo {author} {\bibfnamefont {M.~B.}\ \bibnamefont
  {Plenio}}\ and\ \bibinfo {author} {\bibfnamefont {P.~L.}\ \bibnamefont
  {Knight}},\ }\href {https://doi.org/10.1103/RevModPhys.70.101} {\bibfield
  {journal} {\bibinfo  {journal} {Rev. Mod. Phys.}\ }\textbf {\bibinfo {volume}
  {70}},\ \bibinfo {pages} {101} (\bibinfo {year} {1998})}\BibitemShut
  {NoStop}%
\bibitem [{\citenamefont {M{\o}lmer}\ \emph
  {et~al.}(1993{\natexlab{a}})\citenamefont {M{\o}lmer}, \citenamefont
  {Castin},\ and\ \citenamefont {Dalibard}}]{molmer93}%
  \BibitemOpen
  \bibfield  {author} {\bibinfo {author} {\bibfnamefont {K.}~\bibnamefont
  {M{\o}lmer}}, \bibinfo {author} {\bibfnamefont {Y.}~\bibnamefont {Castin}},\
  and\ \bibinfo {author} {\bibfnamefont {J.}~\bibnamefont {Dalibard}},\ }\href
  {https://doi.org/10.1364/JOSAB.10.000524} {\bibfield  {journal} {\bibinfo
  {journal} {J. Opt. Soc. Am. B}\ }\textbf {\bibinfo {volume} {10}},\ \bibinfo
  {pages} {524} (\bibinfo {year} {1993}{\natexlab{a}})}\BibitemShut {NoStop}%
\bibitem [{\citenamefont {Dalibard}\ \emph {et~al.}(1992)\citenamefont
  {Dalibard}, \citenamefont {Castin},\ and\ \citenamefont
  {M{\o}lmer}}]{dalibard1992}%
  \BibitemOpen
  \bibfield  {author} {\bibinfo {author} {\bibfnamefont {J.}~\bibnamefont
  {Dalibard}}, \bibinfo {author} {\bibfnamefont {Y.}~\bibnamefont {Castin}},\
  and\ \bibinfo {author} {\bibfnamefont {K.}~\bibnamefont {M{\o}lmer}},\
  }\href@noop {} {\bibfield  {journal} {\bibinfo  {journal} {Phys. Rev. Lett.}\
  }\textbf {\bibinfo {volume} {68}},\ \bibinfo {pages} {580} (\bibinfo {year}
  {1992})}\BibitemShut {NoStop}%
\bibitem [{\citenamefont {Dum}\ \emph {et~al.}(1992)\citenamefont {Dum},
  \citenamefont {Zoller},\ and\ \citenamefont {Ritsch}}]{dum1992}%
  \BibitemOpen
  \bibfield  {author} {\bibinfo {author} {\bibfnamefont {R.}~\bibnamefont
  {Dum}}, \bibinfo {author} {\bibfnamefont {P.}~\bibnamefont {Zoller}},\ and\
  \bibinfo {author} {\bibfnamefont {H.}~\bibnamefont {Ritsch}},\ }\href
  {https://doi.org/10.1103/PhysRevA.45.4879} {\bibfield  {journal} {\bibinfo
  {journal} {Phys. Rev. A}\ }\textbf {\bibinfo {volume} {45}},\ \bibinfo
  {pages} {4879} (\bibinfo {year} {1992})}\BibitemShut {NoStop}%
\bibitem [{\citenamefont {Daley}(2014)}]{daley2014}%
  \BibitemOpen
  \bibfield  {author} {\bibinfo {author} {\bibfnamefont {A.~J.}\ \bibnamefont
  {Daley}},\ }\href {https://doi.org/10.1080/00018732.2014.933502} {\bibfield
  {journal} {\bibinfo  {journal} {Advances in Physics}\ }\textbf {\bibinfo
  {volume} {63}},\ \bibinfo {pages} {77–149} (\bibinfo {year}
  {2014})}\BibitemShut {NoStop}%
\bibitem [{\citenamefont {Wiseman}\ and\ \citenamefont
  {Milburn}(1993)}]{wiseman1993}%
  \BibitemOpen
  \bibfield  {author} {\bibinfo {author} {\bibfnamefont {H.~M.}\ \bibnamefont
  {Wiseman}}\ and\ \bibinfo {author} {\bibfnamefont {G.~J.}\ \bibnamefont
  {Milburn}},\ }\href {https://doi.org/10.1103/PhysRevLett.70.548} {\bibfield
  {journal} {\bibinfo  {journal} {Phys. Rev. Lett.}\ }\textbf {\bibinfo
  {volume} {70}},\ \bibinfo {pages} {548} (\bibinfo {year} {1993})}\BibitemShut
  {NoStop}%
\bibitem [{\citenamefont {Collett}\ \emph {et~al.}(1987)\citenamefont
  {Collett}, \citenamefont {Loudon},\ and\ \citenamefont {and}}]{collett1987}%
  \BibitemOpen
  \bibfield  {author} {\bibinfo {author} {\bibfnamefont {M.}~\bibnamefont
  {Collett}}, \bibinfo {author} {\bibfnamefont {R.}~\bibnamefont {Loudon}},\
  and\ \bibinfo {author} {\bibfnamefont {C.~G.}\ \bibnamefont {and}},\ }\href
  {https://doi.org/10.1080/09500348714550811} {\bibfield  {journal} {\bibinfo
  {journal} {Journal of Modern Optics}\ }\textbf {\bibinfo {volume} {34}},\
  \bibinfo {pages} {881} (\bibinfo {year} {1987})},\ \Eprint
  {https://arxiv.org/abs/https://doi.org/10.1080/09500348714550811}
  {https://doi.org/10.1080/09500348714550811} \BibitemShut {NoStop}%
\bibitem [{\citenamefont {{Wiseman}}\ and\ \citenamefont
  {{Milburn}}(2014)}]{wiseman2014}%
  \BibitemOpen
  \bibfield  {author} {\bibinfo {author} {\bibfnamefont {H.~M.}\ \bibnamefont
  {{Wiseman}}}\ and\ \bibinfo {author} {\bibfnamefont {G.~J.}\ \bibnamefont
  {{Milburn}}},\ }\href@noop {} {\emph {\bibinfo {title} {{Quantum Measurement
  and Control}}}}\ (\bibinfo  {publisher} {Cambridge University Press},\
  \bibinfo {year} {2014})\BibitemShut {NoStop}%
\bibitem [{\citenamefont {Fuwa}\ \emph {et~al.}(2015)\citenamefont {Fuwa},
  \citenamefont {Takeda}, \citenamefont {Zwierz}, \citenamefont {Wiseman},\
  and\ \citenamefont {Furusawa}}]{fuwa2015}%
  \BibitemOpen
  \bibfield  {author} {\bibinfo {author} {\bibfnamefont {M.}~\bibnamefont
  {Fuwa}}, \bibinfo {author} {\bibfnamefont {S.}~\bibnamefont {Takeda}},
  \bibinfo {author} {\bibfnamefont {M.}~\bibnamefont {Zwierz}}, \bibinfo
  {author} {\bibfnamefont {H.~M.}\ \bibnamefont {Wiseman}},\ and\ \bibinfo
  {author} {\bibfnamefont {A.}~\bibnamefont {Furusawa}},\ }\bibfield  {journal}
  {\bibinfo  {journal} {Nature Communications}\ }\textbf {\bibinfo {volume}
  {6}},\ \href {https://doi.org/10.1038/ncomms7665} {10.1038/ncomms7665}
  (\bibinfo {year} {2015})\BibitemShut {NoStop}%
\bibitem [{\citenamefont {Cao}\ \emph {et~al.}(2019)\citenamefont {Cao},
  \citenamefont {Tilloy},\ and\ \citenamefont {{De~Luca}}}]{cao2019a}%
  \BibitemOpen
  \bibfield  {author} {\bibinfo {author} {\bibfnamefont {X.}~\bibnamefont
  {Cao}}, \bibinfo {author} {\bibfnamefont {A.}~\bibnamefont {Tilloy}},\ and\
  \bibinfo {author} {\bibfnamefont {A.}~\bibnamefont {{De~Luca}}},\ }\href
  {https://doi.org/10.21468/SciPostPhys.7.2.024} {\bibfield  {journal}
  {\bibinfo  {journal} {SciPost Phys.}\ }\textbf {\bibinfo {volume} {7}},\
  \bibinfo {pages} {024} (\bibinfo {year} {2019})}\BibitemShut {NoStop}%
\bibitem [{\citenamefont {Alberton}\ \emph {et~al.}(2021)\citenamefont
  {Alberton}, \citenamefont {Buchhold},\ and\ \citenamefont
  {Diehl}}]{alberton2021a}%
  \BibitemOpen
  \bibfield  {author} {\bibinfo {author} {\bibfnamefont {O.}~\bibnamefont
  {Alberton}}, \bibinfo {author} {\bibfnamefont {M.}~\bibnamefont {Buchhold}},\
  and\ \bibinfo {author} {\bibfnamefont {S.}~\bibnamefont {Diehl}},\ }\href
  {https://doi.org/10.1103/PhysRevLett.126.170602} {\bibfield  {journal}
  {\bibinfo  {journal} {Phys. Rev. Lett.}\ }\textbf {\bibinfo {volume} {126}},\
  \bibinfo {pages} {170602} (\bibinfo {year} {2021})}\BibitemShut {NoStop}%
\bibitem [{\citenamefont {Carisch}\ \emph {et~al.}(2023)\citenamefont
  {Carisch}, \citenamefont {Romito},\ and\ \citenamefont
  {Zilberberg}}]{carisch2023}%
  \BibitemOpen
  \bibfield  {author} {\bibinfo {author} {\bibfnamefont {C.}~\bibnamefont
  {Carisch}}, \bibinfo {author} {\bibfnamefont {A.}~\bibnamefont {Romito}},\
  and\ \bibinfo {author} {\bibfnamefont {O.}~\bibnamefont {Zilberberg}},\
  }\href {https://doi.org/10.1103/PhysRevResearch.5.L042031} {\bibfield
  {journal} {\bibinfo  {journal} {Phys. Rev. Res.}\ }\textbf {\bibinfo {volume}
  {5}},\ \bibinfo {pages} {L042031} (\bibinfo {year} {2023})}\BibitemShut
  {NoStop}%
\bibitem [{\citenamefont {Ladewig}\ \emph {et~al.}(2022)\citenamefont
  {Ladewig}, \citenamefont {Diehl},\ and\ \citenamefont
  {Buchhold}}]{ladewig2022}%
  \BibitemOpen
  \bibfield  {author} {\bibinfo {author} {\bibfnamefont {B.}~\bibnamefont
  {Ladewig}}, \bibinfo {author} {\bibfnamefont {S.}~\bibnamefont {Diehl}},\
  and\ \bibinfo {author} {\bibfnamefont {M.}~\bibnamefont {Buchhold}},\ }\href
  {https://doi.org/10.1103/PhysRevResearch.4.033001} {\bibfield  {journal}
  {\bibinfo  {journal} {Phys. Rev. Research}\ }\textbf {\bibinfo {volume}
  {4}},\ \bibinfo {pages} {033001} (\bibinfo {year} {2022})}\BibitemShut
  {NoStop}%
\bibitem [{\citenamefont {Misra}\ and\ \citenamefont
  {Sudarshan}(1977)}]{misra1977}%
  \BibitemOpen
  \bibfield  {author} {\bibinfo {author} {\bibfnamefont {B.}~\bibnamefont
  {Misra}}\ and\ \bibinfo {author} {\bibfnamefont {E.~C.~G.}\ \bibnamefont
  {Sudarshan}},\ }\href {https://doi.org/10.1063/1.523304} {\bibfield
  {journal} {\bibinfo  {journal} {Journal of Mathematical Physics}\ }\textbf
  {\bibinfo {volume} {18}},\ \bibinfo {pages} {756} (\bibinfo {year} {1977})},\
  \Eprint
  {https://arxiv.org/abs/https://pubs.aip.org/aip/jmp/article-pdf/18/4/756/19182345/756\_1\_online.pdf}
  {https://pubs.aip.org/aip/jmp/article-pdf/18/4/756/19182345/756\_1\_online.pdf}
  \BibitemShut {NoStop}%
\bibitem [{\citenamefont {Itano}\ \emph {et~al.}(1990)\citenamefont {Itano},
  \citenamefont {Heinzen}, \citenamefont {Bollinger},\ and\ \citenamefont
  {Wineland}}]{itano1990}%
  \BibitemOpen
  \bibfield  {author} {\bibinfo {author} {\bibfnamefont {W.~M.}\ \bibnamefont
  {Itano}}, \bibinfo {author} {\bibfnamefont {D.~J.}\ \bibnamefont {Heinzen}},
  \bibinfo {author} {\bibfnamefont {J.~J.}\ \bibnamefont {Bollinger}},\ and\
  \bibinfo {author} {\bibfnamefont {D.~J.}\ \bibnamefont {Wineland}},\ }\href
  {https://doi.org/10.1103/PhysRevA.41.2295} {\bibfield  {journal} {\bibinfo
  {journal} {Phys. Rev. A}\ }\textbf {\bibinfo {volume} {41}},\ \bibinfo
  {pages} {2295} (\bibinfo {year} {1990})}\BibitemShut {NoStop}%
\bibitem [{\citenamefont {Fischer}\ \emph {et~al.}(2001)\citenamefont
  {Fischer}, \citenamefont {Guti\'errez-Medina},\ and\ \citenamefont
  {Raizen}}]{raizen2001}%
  \BibitemOpen
  \bibfield  {author} {\bibinfo {author} {\bibfnamefont {M.~C.}\ \bibnamefont
  {Fischer}}, \bibinfo {author} {\bibfnamefont {B.}~\bibnamefont
  {Guti\'errez-Medina}},\ and\ \bibinfo {author} {\bibfnamefont {M.~G.}\
  \bibnamefont {Raizen}},\ }\href
  {https://doi.org/10.1103/PhysRevLett.87.040402} {\bibfield  {journal}
  {\bibinfo  {journal} {Phys. Rev. Lett.}\ }\textbf {\bibinfo {volume} {87}},\
  \bibinfo {pages} {040402} (\bibinfo {year} {2001})}\BibitemShut {NoStop}%
\bibitem [{\citenamefont {M{\o}lmer}\ \emph
  {et~al.}(1993{\natexlab{b}})\citenamefont {M{\o}lmer}, \citenamefont
  {Castin},\ and\ \citenamefont {Dalibard}}]{molmer1993}%
  \BibitemOpen
  \bibfield  {author} {\bibinfo {author} {\bibfnamefont {K.}~\bibnamefont
  {M{\o}lmer}}, \bibinfo {author} {\bibfnamefont {Y.}~\bibnamefont {Castin}},\
  and\ \bibinfo {author} {\bibfnamefont {J.}~\bibnamefont {Dalibard}},\
  }\href@noop {} {\bibfield  {journal} {\bibinfo  {journal} {J. Opt. Soc. Am.
  B}\ }\textbf {\bibinfo {volume} {10}},\ \bibinfo {pages} {524} (\bibinfo
  {year} {1993}{\natexlab{b}})}\BibitemShut {NoStop}%
\bibitem [{\citenamefont {Caldeira}\ and\ \citenamefont
  {Leggett}(1981)}]{caldeira1981}%
  \BibitemOpen
  \bibfield  {author} {\bibinfo {author} {\bibfnamefont {A.~O.}\ \bibnamefont
  {Caldeira}}\ and\ \bibinfo {author} {\bibfnamefont {A.~J.}\ \bibnamefont
  {Leggett}},\ }\href@noop {} {\bibfield  {journal} {\bibinfo  {journal} {Phys.
  Rev. Lett.}\ }\textbf {\bibinfo {volume} {46}},\ \bibinfo {pages} {211}
  (\bibinfo {year} {1981})}\BibitemShut {NoStop}%
\bibitem [{\citenamefont {Kitaev}(2015)}]{kitaev2015}%
  \BibitemOpen
  \bibfield  {author} {\bibinfo {author} {\bibfnamefont {A.}~\bibnamefont
  {Kitaev}},\ }\href@noop {} {\bibinfo {title} {A simple model of quantum
  holography}},\ \bibinfo {howpublished} {Talks at KITP: String seminar and
  Entanglement 2015 program, 12 Feb, 7 Apr, 27 May} (\bibinfo {year} {2015}),\
  \bibinfo {note}
  {\url{http://online.kitp.ucsb.edu/online/entangled15/}}\BibitemShut {NoStop}%
\bibitem [{\citenamefont {French}\ and\ \citenamefont
  {Wong}(1970)}]{french1970}%
  \BibitemOpen
  \bibfield  {author} {\bibinfo {author} {\bibfnamefont {J.~B.}\ \bibnamefont
  {French}}\ and\ \bibinfo {author} {\bibfnamefont {S.~S.~M.}\ \bibnamefont
  {Wong}},\ }\href@noop {} {\bibfield  {journal} {\bibinfo  {journal} {Phys.
  Lett. B}\ }\textbf {\bibinfo {volume} {33}},\ \bibinfo {pages} {449}
  (\bibinfo {year} {1970})}\BibitemShut {NoStop}%
\bibitem [{\citenamefont {Bohigas}\ and\ \citenamefont
  {Flores}(1971)}]{bohigas1971}%
  \BibitemOpen
  \bibfield  {author} {\bibinfo {author} {\bibfnamefont {O.}~\bibnamefont
  {Bohigas}}\ and\ \bibinfo {author} {\bibfnamefont {J.}~\bibnamefont
  {Flores}},\ }\href@noop {} {\bibfield  {journal} {\bibinfo  {journal} {Phys.
  Lett. B}\ }\textbf {\bibinfo {volume} {34}},\ \bibinfo {pages} {261}
  (\bibinfo {year} {1971})}\BibitemShut {NoStop}%
\bibitem [{\citenamefont {Sachdev}\ and\ \citenamefont
  {Ye}(1993)}]{sachdev1993}%
  \BibitemOpen
  \bibfield  {author} {\bibinfo {author} {\bibfnamefont {S.}~\bibnamefont
  {Sachdev}}\ and\ \bibinfo {author} {\bibfnamefont {J.}~\bibnamefont {Ye}},\
  }\href@noop {} {\bibfield  {journal} {\bibinfo  {journal} {Phys. Rev. Lett.}\
  }\textbf {\bibinfo {volume} {70}},\ \bibinfo {pages} {3339} (\bibinfo {year}
  {1993})}\BibitemShut {NoStop}%
\bibitem [{\citenamefont {Benet}\ \emph {et~al.}(2001)\citenamefont {Benet},
  \citenamefont {Rupp},\ and\ \citenamefont {Weidenm{\"u}ller}}]{benet2001}%
  \BibitemOpen
  \bibfield  {author} {\bibinfo {author} {\bibfnamefont {L.}~\bibnamefont
  {Benet}}, \bibinfo {author} {\bibfnamefont {T.}~\bibnamefont {Rupp}},\ and\
  \bibinfo {author} {\bibfnamefont {H.~A.}\ \bibnamefont {Weidenm{\"u}ller}},\
  }\href@noop {} {\bibfield  {journal} {\bibinfo  {journal} {Phys. Rev. Lett.}\
  }\textbf {\bibinfo {volume} {87}},\ \bibinfo {pages} {010601} (\bibinfo
  {year} {2001})}\BibitemShut {NoStop}%
\bibitem [{\citenamefont {Maldacena}\ and\ \citenamefont
  {Stanford}(2016)}]{maldacena2016}%
  \BibitemOpen
  \bibfield  {author} {\bibinfo {author} {\bibfnamefont {J.}~\bibnamefont
  {Maldacena}}\ and\ \bibinfo {author} {\bibfnamefont {D.}~\bibnamefont
  {Stanford}},\ }\href@noop {} {\bibfield  {journal} {\bibinfo  {journal}
  {Phys. Rev. D}\ }\textbf {\bibinfo {volume} {94}},\ \bibinfo {pages} {106002}
  (\bibinfo {year} {2016})}\BibitemShut {NoStop}%
\bibitem [{\citenamefont {Garc\'{\i}a-Garc\'{\i}a}\ \emph
  {et~al.}(2022)\citenamefont {Garc\'{\i}a-Garc\'{\i}a}, \citenamefont {S\'a},\
  and\ \citenamefont {Verbaarschot}}]{garcia2022d}%
  \BibitemOpen
  \bibfield  {author} {\bibinfo {author} {\bibfnamefont {A.~M.}\ \bibnamefont
  {Garc\'{\i}a-Garc\'{\i}a}}, \bibinfo {author} {\bibfnamefont
  {L.}~\bibnamefont {S\'a}},\ and\ \bibinfo {author} {\bibfnamefont {J.~J.~M.}\
  \bibnamefont {Verbaarschot}},\ }\href
  {https://doi.org/10.1103/PhysRevX.12.021040} {\bibfield  {journal} {\bibinfo
  {journal} {Phys. Rev. X}\ }\textbf {\bibinfo {volume} {12}},\ \bibinfo
  {pages} {021040} (\bibinfo {year} {2022})}\BibitemShut {NoStop}%
\bibitem [{\citenamefont {Chen}\ \emph {et~al.}(2017)\citenamefont {Chen},
  \citenamefont {Zhai},\ and\ \citenamefont {Zhang}}]{chen2017a}%
  \BibitemOpen
  \bibfield  {author} {\bibinfo {author} {\bibfnamefont {Y.}~\bibnamefont
  {Chen}}, \bibinfo {author} {\bibfnamefont {H.}~\bibnamefont {Zhai}},\ and\
  \bibinfo {author} {\bibfnamefont {P.}~\bibnamefont {Zhang}},\ }\href@noop {}
  {\bibfield  {journal} {\bibinfo  {journal} {J. High Energy Phys.}\ }\textbf
  {\bibinfo {volume} {2017}}\bibinfo  {number} { (7)}}\BibitemShut {NoStop}%
\bibitem [{\citenamefont {Zhang}(2019)}]{zhang2019}%
  \BibitemOpen
\bibfield  {number} {  }\bibfield  {author} {\bibinfo {author} {\bibfnamefont
  {P.}~\bibnamefont {Zhang}},\ }\href
  {https://doi.org/10.1103/PhysRevB.100.245104} {\bibfield  {journal} {\bibinfo
   {journal} {Phys. Rev. B}\ }\textbf {\bibinfo {volume} {100}},\ \bibinfo
  {pages} {245104} (\bibinfo {year} {2019})}\BibitemShut {NoStop}%
\bibitem [{\citenamefont {Eberlein}\ \emph {et~al.}(2017)\citenamefont
  {Eberlein}, \citenamefont {Kasper}, \citenamefont {Sachdev},\ and\
  \citenamefont {Steinberg}}]{eberlein2017}%
  \BibitemOpen
  \bibfield  {author} {\bibinfo {author} {\bibfnamefont {A.}~\bibnamefont
  {Eberlein}}, \bibinfo {author} {\bibfnamefont {V.}~\bibnamefont {Kasper}},
  \bibinfo {author} {\bibfnamefont {S.}~\bibnamefont {Sachdev}},\ and\ \bibinfo
  {author} {\bibfnamefont {J.}~\bibnamefont {Steinberg}},\ }\href
  {https://doi.org/10.1103/PhysRevB.96.205123} {\bibfield  {journal} {\bibinfo
  {journal} {Phys. Rev. B}\ }\textbf {\bibinfo {volume} {96}},\ \bibinfo
  {pages} {205123} (\bibinfo {year} {2017})},\ \Eprint
  {https://arxiv.org/abs/1706.07803} {arXiv:1706.07803 [cond-mat.str-el]}
  \BibitemShut {NoStop}%
\bibitem [{\citenamefont {Stefanucci}\ and\ \citenamefont
  {Van~Leeuwen}(2025)}]{stefanucci2025}%
  \BibitemOpen
  \bibfield  {author} {\bibinfo {author} {\bibfnamefont {G.}~\bibnamefont
  {Stefanucci}}\ and\ \bibinfo {author} {\bibfnamefont {R.}~\bibnamefont
  {Van~Leeuwen}},\ }\href {https://doi.org/10.1017/9781009536776} {\emph
  {\bibinfo {title} {Nonequilibrium {{Many-Body Theory}} of {{Quantum
  Systems}}: {{A Modern Introduction}}}}},\ \bibinfo {edition} {2nd}\ ed.\
  (\bibinfo  {publisher} {Cambridge University Press},\ \bibinfo {year}
  {2025})\BibitemShut {NoStop}%
\bibitem [{\citenamefont {Kamenev}(2023)}]{kamenev2023}%
  \BibitemOpen
  \bibfield  {author} {\bibinfo {author} {\bibfnamefont {A.}~\bibnamefont
  {Kamenev}},\ }\href
  {https://www.cambridge.org/ng/universitypress/subjects/physics/condensed-matter-physics-nanoscience-and-mesoscopic-physics/field-theory-non-equilibrium-systems-2nd-edition#about-the-authors}
  {\emph {\bibinfo {title} {Field Theory of Non-Equilibrium Systems}}},\
  \bibinfo {edition} {2nd}\ ed.\ (\bibinfo  {publisher} {Cambridge University
  Press},\ \bibinfo {year} {2023})\BibitemShut {NoStop}%
\bibitem [{\citenamefont {Almheiri}\ \emph {et~al.}(2024)\citenamefont
  {Almheiri}, \citenamefont {Milekhin},\ and\ \citenamefont
  {Swingle}}]{almheiri2019}%
  \BibitemOpen
  \bibfield  {author} {\bibinfo {author} {\bibfnamefont {A.}~\bibnamefont
  {Almheiri}}, \bibinfo {author} {\bibfnamefont {A.}~\bibnamefont {Milekhin}},\
  and\ \bibinfo {author} {\bibfnamefont {B.}~\bibnamefont {Swingle}},\ }\href
  {https://doi.org/10.1007/JHEP08(2024)034} {\bibfield  {journal} {\bibinfo
  {journal} {JHEP}\ }\textbf {\bibinfo {volume} {08}},\ \bibinfo {pages}
  {034}},\ \Eprint {https://arxiv.org/abs/1912.04912} {arXiv:1912.04912
  [hep-th]} \BibitemShut {NoStop}%
\bibitem [{\citenamefont {Maldacena}\ and\ \citenamefont
  {Milekhin}(2021)}]{milekhin2019}%
  \BibitemOpen
  \bibfield  {author} {\bibinfo {author} {\bibfnamefont {J.}~\bibnamefont
  {Maldacena}}\ and\ \bibinfo {author} {\bibfnamefont {A.}~\bibnamefont
  {Milekhin}},\ }\href@noop {} {\bibfield  {journal} {\bibinfo  {journal} {J.
  High Energy Phys.}\ }\textbf {\bibinfo {volume} {04}},\ \bibinfo {pages}
  {258}}\BibitemShut {NoStop}%
\bibitem [{\citenamefont {Milekhin}\ and\ \citenamefont
  {Popov}(2024)}]{milekhin2024}%
  \BibitemOpen
  \bibfield  {author} {\bibinfo {author} {\bibfnamefont {A.}~\bibnamefont
  {Milekhin}}\ and\ \bibinfo {author} {\bibfnamefont {F.~K.}\ \bibnamefont
  {Popov}},\ }\href {https://doi.org/10.21468/SciPostPhys.17.1.020} {\bibfield
  {journal} {\bibinfo  {journal} {SciPost Phys.}\ }\textbf {\bibinfo {volume}
  {17}},\ \bibinfo {pages} {020} (\bibinfo {year} {2024})}\BibitemShut
  {NoStop}%
\bibitem [{\citenamefont {Kuhlenkamp}\ and\ \citenamefont
  {Knap}(2020)}]{knap2020}%
  \BibitemOpen
  \bibfield  {author} {\bibinfo {author} {\bibfnamefont {C.}~\bibnamefont
  {Kuhlenkamp}}\ and\ \bibinfo {author} {\bibfnamefont {M.}~\bibnamefont
  {Knap}},\ }\href@noop {} {\bibfield  {journal} {\bibinfo  {journal} {Phys.
  Rev. Lett.}\ }\textbf {\bibinfo {volume} {124}},\ \bibinfo {pages} {106401}
  (\bibinfo {year} {2020})}\BibitemShut {NoStop}%
\bibitem [{\citenamefont {Berenguer}\ \emph {et~al.}(2024)\citenamefont
  {Berenguer}, \citenamefont {Dey}, \citenamefont {Mas}, \citenamefont
  {Santos-Su{\'a}rez},\ and\ \citenamefont {Ramallo}}]{berenguer2024}%
  \BibitemOpen
  \bibfield  {author} {\bibinfo {author} {\bibfnamefont {M.}~\bibnamefont
  {Berenguer}}, \bibinfo {author} {\bibfnamefont {A.}~\bibnamefont {Dey}},
  \bibinfo {author} {\bibfnamefont {J.}~\bibnamefont {Mas}}, \bibinfo {author}
  {\bibfnamefont {J.}~\bibnamefont {Santos-Su{\'a}rez}},\ and\ \bibinfo
  {author} {\bibfnamefont {A.~V.}\ \bibnamefont {Ramallo}},\ }\href@noop {}
  {\bibfield  {journal} {\bibinfo  {journal} {J. High Energy Phys.}\ }\textbf
  {\bibinfo {volume} {2024}}\bibinfo  {number} { (6)}}\BibitemShut {NoStop}%
\bibitem [{\citenamefont {Lindblad}(1976)}]{lindblad1976}%
  \BibitemOpen
\bibfield  {number} {  }\bibfield  {author} {\bibinfo {author} {\bibfnamefont
  {G.}~\bibnamefont {Lindblad}},\ }\href@noop {} {\bibfield  {journal}
  {\bibinfo  {journal} {Commun. Math. Phys.}\ }\textbf {\bibinfo {volume}
  {48}},\ \bibinfo {pages} {119} (\bibinfo {year} {1976})}\BibitemShut
  {NoStop}%
\bibitem [{\citenamefont {Gorini}\ \emph {et~al.}(1976)\citenamefont {Gorini},
  \citenamefont {Kossakowski},\ and\ \citenamefont {Sudarshan}}]{gorini1976}%
  \BibitemOpen
  \bibfield  {author} {\bibinfo {author} {\bibfnamefont {V.}~\bibnamefont
  {Gorini}}, \bibinfo {author} {\bibfnamefont {A.}~\bibnamefont
  {Kossakowski}},\ and\ \bibinfo {author} {\bibfnamefont {E.~C.~G.}\
  \bibnamefont {Sudarshan}},\ }\href@noop {} {\bibfield  {journal} {\bibinfo
  {journal} {J. Math. Phys.}\ }\textbf {\bibinfo {volume} {17}},\ \bibinfo
  {pages} {821} (\bibinfo {year} {1976})}\BibitemShut {NoStop}%
\bibitem [{\citenamefont {S{\'a}}\ \emph {et~al.}(2022)\citenamefont {S{\'a}},
  \citenamefont {Ribeiro},\ and\ \citenamefont {Prosen}}]{sa2022}%
  \BibitemOpen
  \bibfield  {author} {\bibinfo {author} {\bibfnamefont {L.}~\bibnamefont
  {S{\'a}}}, \bibinfo {author} {\bibfnamefont {P.}~\bibnamefont {Ribeiro}},\
  and\ \bibinfo {author} {\bibfnamefont {T.}~\bibnamefont {Prosen}},\
  }\href@noop {} {\bibfield  {journal} {\bibinfo  {journal} {Phys. Rev.
  Research}\ }\textbf {\bibinfo {volume} {4}},\ \bibinfo {pages} {L022068}
  (\bibinfo {year} {2022})}\BibitemShut {NoStop}%
\bibitem [{\citenamefont {Kulkarni}\ \emph {et~al.}(2022)\citenamefont
  {Kulkarni}, \citenamefont {Numasawa},\ and\ \citenamefont
  {Ryu}}]{kulkarni2022}%
  \BibitemOpen
  \bibfield  {author} {\bibinfo {author} {\bibfnamefont {A.}~\bibnamefont
  {Kulkarni}}, \bibinfo {author} {\bibfnamefont {T.}~\bibnamefont {Numasawa}},\
  and\ \bibinfo {author} {\bibfnamefont {S.}~\bibnamefont {Ryu}},\ }\href@noop
  {} {\bibfield  {journal} {\bibinfo  {journal} {Phys. Rev. B}\ }\textbf
  {\bibinfo {volume} {106}},\ \bibinfo {pages} {075138} (\bibinfo {year}
  {2022})}\BibitemShut {NoStop}%
\bibitem [{\citenamefont {Garc\'\i{}a-Garc\'\i{}a}\ \emph
  {et~al.}(2023)\citenamefont {Garc\'\i{}a-Garc\'\i{}a}, \citenamefont {S\'a},
  \citenamefont {Verbaarschot},\ and\ \citenamefont {Zheng}}]{garcia2022e}%
  \BibitemOpen
  \bibfield  {author} {\bibinfo {author} {\bibfnamefont {A.~M.}\ \bibnamefont
  {Garc\'\i{}a-Garc\'\i{}a}}, \bibinfo {author} {\bibfnamefont
  {L.}~\bibnamefont {S\'a}}, \bibinfo {author} {\bibfnamefont {J.~J.~M.}\
  \bibnamefont {Verbaarschot}},\ and\ \bibinfo {author} {\bibfnamefont {J.~P.}\
  \bibnamefont {Zheng}},\ }\href {https://doi.org/10.1103/PhysRevD.107.106006}
  {\bibfield  {journal} {\bibinfo  {journal} {Phys. Rev. D}\ }\textbf {\bibinfo
  {volume} {107}},\ \bibinfo {pages} {106006} (\bibinfo {year} {2023})},\
  \Eprint {https://arxiv.org/abs/2210.01695} {arXiv:2210.01695 [hep-th]}
  \BibitemShut {NoStop}%
\bibitem [{\citenamefont {Garc{\'i}a-Garc{\'i}a}\ \emph
  {et~al.}(2024)\citenamefont {Garc{\'i}a-Garc{\'i}a}, \citenamefont
  {Verbaarschot},\ and\ \citenamefont {Zheng}}]{garcia2024}%
  \BibitemOpen
  \bibfield  {author} {\bibinfo {author} {\bibfnamefont {A.~M.}\ \bibnamefont
  {Garc{\'i}a-Garc{\'i}a}}, \bibinfo {author} {\bibfnamefont {J.~J.~M.}\
  \bibnamefont {Verbaarschot}},\ and\ \bibinfo {author} {\bibfnamefont {J.-P.}\
  \bibnamefont {Zheng}},\ }\href@noop {} {\bibfield  {journal} {\bibinfo
  {journal} {Phys. Rev. D}\ }\textbf {\bibinfo {volume} {110}},\ \bibinfo
  {pages} {086010} (\bibinfo {year} {2024})}\BibitemShut {NoStop}%
\bibitem [{\citenamefont {Breuer}\ and\ \citenamefont
  {Petruccione}(2002)}]{breuer2002}%
  \BibitemOpen
  \bibfield  {author} {\bibinfo {author} {\bibfnamefont {H.-P.}\ \bibnamefont
  {Breuer}}\ and\ \bibinfo {author} {\bibfnamefont {F.}~\bibnamefont
  {Petruccione}},\ }\href@noop {} {\emph {\bibinfo {title} {The Theory of Open
  Quantum Systems}}}\ (\bibinfo  {publisher} {Oxford University Press},\
  \bibinfo {address} {Oxford},\ \bibinfo {year} {2002})\BibitemShut {NoStop}%
\bibitem [{\citenamefont {Garc{\'i}a-Garc{\'i}a}\ \emph
  {et~al.}(2025)\citenamefont {Garc{\'i}a-Garc{\'i}a}, \citenamefont {Liu},
  \citenamefont {S{\'a}}, \citenamefont {Verbaarschot},\ and\ \citenamefont
  {Zheng}}]{garcia2025}%
  \BibitemOpen
  \bibfield  {author} {\bibinfo {author} {\bibfnamefont {A.~M.}\ \bibnamefont
  {Garc{\'i}a-Garc{\'i}a}}, \bibinfo {author} {\bibfnamefont {C.}~\bibnamefont
  {Liu}}, \bibinfo {author} {\bibfnamefont {L.}~\bibnamefont {S{\'a}}},
  \bibinfo {author} {\bibfnamefont {J.~J.~M.}\ \bibnamefont {Verbaarschot}},\
  and\ \bibinfo {author} {\bibfnamefont {J.~P.}\ \bibnamefont {Zheng}},\
  }\href@noop {} {\bibinfo {title} {Anatomy of information scrambling and
  decoherence in the integrable sachdev-ye-kitaev model}} (\bibinfo {year}
  {2025}),\ \bibinfo {note} {preprint}\BibitemShut {NoStop}%
\bibitem [{\citenamefont {Garc\'{\i}a-Garc\'{\i}a}\ and\ \citenamefont
  {Verbaarschot}(2016)}]{garcia2016}%
  \BibitemOpen
  \bibfield  {author} {\bibinfo {author} {\bibfnamefont {A.~M.}\ \bibnamefont
  {Garc\'{\i}a-Garc\'{\i}a}}\ and\ \bibinfo {author} {\bibfnamefont {J.~J.~M.}\
  \bibnamefont {Verbaarschot}},\ }\href
  {https://doi.org/10.1103/PhysRevD.94.126010} {\bibfield  {journal} {\bibinfo
  {journal} {Phys. Rev. D}\ }\textbf {\bibinfo {volume} {94}},\ \bibinfo
  {pages} {126010} (\bibinfo {year} {2016})},\ \Eprint
  {https://arxiv.org/abs/1610.03816} {arXiv:1610.03816 [hep-th]} \BibitemShut
  {NoStop}%
\bibitem [{\citenamefont {Garc{\'i}a-Garc{\'i}a}\ \emph
  {et~al.}(2023)\citenamefont {Garc{\'i}a-Garc{\'i}a}, \citenamefont {S{\'a}},
  \citenamefont {Verbaarschot},\ and\ \citenamefont {Zheng}}]{garcia2023}%
  \BibitemOpen
  \bibfield  {author} {\bibinfo {author} {\bibfnamefont {A.~M.}\ \bibnamefont
  {Garc{\'i}a-Garc{\'i}a}}, \bibinfo {author} {\bibfnamefont {L.}~\bibnamefont
  {S{\'a}}}, \bibinfo {author} {\bibfnamefont {J.~J.~M.}\ \bibnamefont
  {Verbaarschot}},\ and\ \bibinfo {author} {\bibfnamefont {J.~P.}\ \bibnamefont
  {Zheng}},\ }\href@noop {} {\bibfield  {journal} {\bibinfo  {journal} {Phys.
  Rev. D}\ }\textbf {\bibinfo {volume} {107}},\ \bibinfo {pages} {106006}
  (\bibinfo {year} {2023})}\BibitemShut {NoStop}%
\bibitem [{\citenamefont {Larkin}\ and\ \citenamefont
  {Ovchinnikov}(1969)}]{larkin1969}%
  \BibitemOpen
  \bibfield  {author} {\bibinfo {author} {\bibfnamefont {A.~I.}\ \bibnamefont
  {Larkin}}\ and\ \bibinfo {author} {\bibfnamefont {Y.~N.}\ \bibnamefont
  {Ovchinnikov}},\ }\href
  {http://jetp.ras.ru/cgi-bin/e/index/e/28/6/p1200?a=list} {\bibfield
  {journal} {\bibinfo  {journal} {Sov. Phys. JETP}\ }\textbf {\bibinfo {volume}
  {28}},\ \bibinfo {pages} {1200} (\bibinfo {year} {1969})}\BibitemShut
  {NoStop}%
\bibitem [{\citenamefont {Mori}(2024)}]{mori2024}%
  \BibitemOpen
  \bibfield  {author} {\bibinfo {author} {\bibfnamefont {T.}~\bibnamefont
  {Mori}},\ }\href {https://doi.org/10.1103/PhysRevB.109.064311} {\bibfield
  {journal} {\bibinfo  {journal} {Phys. Rev. B}\ }\textbf {\bibinfo {volume}
  {109}},\ \bibinfo {pages} {064311} (\bibinfo {year} {2024})}\BibitemShut
  {NoStop}%
\bibitem [{\citenamefont {ping Zheng}\ \emph {et~al.}(2025)\citenamefont {ping
  Zheng}, \citenamefont {Dukelsky}, \citenamefont {Molina},\ and\ \citenamefont
  {García-García}}]{garcia2025a}%
  \BibitemOpen
  \bibfield  {author} {\bibinfo {author} {\bibfnamefont {J.}~\bibnamefont {ping
  Zheng}}, \bibinfo {author} {\bibfnamefont {J.}~\bibnamefont {Dukelsky}},
  \bibinfo {author} {\bibfnamefont {R.~A.}\ \bibnamefont {Molina}},\ and\
  \bibinfo {author} {\bibfnamefont {A.~M.}\ \bibnamefont {García-García}},\
  }\href {https://arxiv.org/abs/2510.15793} {\bibinfo {title} {Role of
  exceptional points in the dynamics of the lindblad sachdev-ye-kitaev model}}
  (\bibinfo {year} {2025}),\ \Eprint {https://arxiv.org/abs/2510.15793}
  {arXiv:2510.15793 [quant-ph]} \BibitemShut {NoStop}%
\bibitem [{\citenamefont {Maldacena}\ \emph {et~al.}(2016)\citenamefont
  {Maldacena}, \citenamefont {Shenker},\ and\ \citenamefont
  {Stanford}}]{maldacena2015}%
  \BibitemOpen
  \bibfield  {author} {\bibinfo {author} {\bibfnamefont {J.}~\bibnamefont
  {Maldacena}}, \bibinfo {author} {\bibfnamefont {S.~H.}\ \bibnamefont
  {Shenker}},\ and\ \bibinfo {author} {\bibfnamefont {D.}~\bibnamefont
  {Stanford}},\ }\href {http://dx.doi.org/10.1007/JHEP08(2016)106} {\bibfield
  {journal} {\bibinfo  {journal} {J. High Energy Phys.}\ }\textbf {\bibinfo
  {volume} {2016}}\bibfield  {number} {\bibinfo  {number} { (106)}},\ }\Eprint
  {https://arxiv.org/abs/1503.01409} {arXiv:1503.01409 [hep-th]} \BibitemShut
  {NoStop}%
\bibitem [{\citenamefont {Romero-Bermúdez}\ \emph {et~al.}(2019)\citenamefont
  {Romero-Bermúdez}, \citenamefont {Schalm},\ and\ \citenamefont
  {Scopelliti}}]{bermudez2019}%
  \BibitemOpen
  \bibfield  {author} {\bibinfo {author} {\bibfnamefont {A.}~\bibnamefont
  {Romero-Bermúdez}}, \bibinfo {author} {\bibfnamefont {K.}~\bibnamefont
  {Schalm}},\ and\ \bibinfo {author} {\bibfnamefont {V.}~\bibnamefont
  {Scopelliti}},\ }\bibfield  {journal} {\bibinfo  {journal} {Journal of High
  Energy Physics}\ }\textbf {\bibinfo {volume} {2019}},\ \href
  {https://doi.org/10.1007/jhep07(2019)107} {10.1007/jhep07(2019)107} (\bibinfo
  {year} {2019})\BibitemShut {NoStop}%
\bibitem [{\citenamefont {Ferrari}\ and\ \citenamefont
  {Schaposnik~Massolo}(2019)}]{Ferrari:2019ogc}%
  \BibitemOpen
  \bibfield  {author} {\bibinfo {author} {\bibfnamefont {F.}~\bibnamefont
  {Ferrari}}\ and\ \bibinfo {author} {\bibfnamefont {F.~I.}\ \bibnamefont
  {Schaposnik~Massolo}},\ }\href {https://doi.org/10.1103/PhysRevD.100.026007}
  {\bibfield  {journal} {\bibinfo  {journal} {Phys. Rev. D}\ }\textbf {\bibinfo
  {volume} {100}},\ \bibinfo {pages} {026007} (\bibinfo {year} {2019})},\
  \Eprint {https://arxiv.org/abs/1903.06633} {arXiv:1903.06633 [hep-th]}
  \BibitemShut {NoStop}%
\end{thebibliography}%

\clearpage
\onecolumngrid

\setcounter{table}{0}
\renewcommand{\thetable}{S\arabic{table}}%
\setcounter{figure}{0}
\renewcommand{\thefigure}{SM\arabic{figure}}%
\setcounter{equation}{0}
\renewcommand{\theequation}{S\arabic{equation}}%
\setcounter{page}{1}
\renewcommand{\thepage}{SM-\arabic{page}}%
\setcounter{secnumdepth}{3}
\setcounter{section}{0}
\renewcommand{\thesection}{S\arabic{section}}%
\setcounter{subsection}{0}
\renewcommand{\thesubsection}{\arabic{section}.\arabic{subsection}}%

\begin{center}
		\textbf{Supplemental Material for } 
		\textbf{``Inducing, and enhancing, many-body quantum chaos by continuous monitoring''} \\
		\vspace{10pt}
        Xianlong Liu ({\begin{CJK}{UTF8}{gbsn}刘显龙\end{CJK}}), 
        Jie-ping Zheng ({\begin{CJK*}{UTF8}{gbsn}郑杰平\end{CJK*}}),
        and Antonio M. Garc\'ia-Garc\'ia
\end{center}

\section{Action and Kadanoff-Baym equations}

In this section we present the details for the Schwinger-Keldysh path integral and the Kadanoff-Baym equations. Given the Liouvillian of the SYK monitored dissipative dynamics, we first perform the disorder averaging over random couplings $J_{j_1, \dots j_q}$, $V_{j_1 \dots j_{f_{\sys}}}^{i_1 \dots i_{f_{\bath}}}$, resulting in an action that is bi-local in the time arguments. Next, introducing the bi-local collective fields for the system and the bath Majorana fermions along the Schwinger-Keldysh (SK) contour $z \in \mathcal{C} = \mathcal{C}^+ \cup \mathcal{C}^-$
\begin{align}
    G(z_1, z_2) & = - \frac{\ii}{N} \sum_{j=1}^{N} \psi_j(z_1) \psi_j(z_2) \, , \\
    G_{\bath}(z_1, z_2) & = - \frac{\ii}{N_{\bath}} \sum_{j=1}^{N_{\bath}} \chi_j(z_1) \chi_j(z_2) \, ,
\end{align}
and the Lagrange multipliers $\Sigma$, $\Sigma'$ that enforce these relations, we can express the interaction terms in terms of $G$ and $G_{\bath}$, leaving a Gaussian integral for the Majorana fermions which can be directly integrated out. Finally we are left with the partition function given by
\begin{equation}
    Z = \int \mathcal{D} G \mathcal{D} G_{\bath} \mathcal{D} \Sigma \mathcal{D} \Sigma' \, \ee^{\ii S[G, G_{\bath}, \Sigma, \Sigma']} \, ,
\end{equation}
where the action consists of four parts
\begin{equation}
    \ii S[G, G_{\bath}, \Sigma, \Sigma'] =  \ii S_{\sys}[G, \Sigma] + \ii S_{\bath}[G_{\bath}, \Sigma'] + \ii S_{\sysb}[G, G_{\bath}] + \ii S_{\lind}[G] \, .
\end{equation}
Explicitly,
\begin{align}
    \frac{\ii}{N} S_{\sys}[G, \Sigma] & = \frac{1}{2}\Tr_{\mathcal{C}} \log(\ii \partial - \Sigma) - \frac{1}{2} \int_{\mathcal{C}} \Sigma(z_1, z_2) G(z_1, z_2) \dd{z_1} \dd{z_2} - \frac{\ii^{q} J^2}{2 q} \int_{\mathcal{C}} [G(z_1, z_2)]^q \dd{z_1} \dd{z_2} \, , \\[2pt]
    \frac{\ii}{N_\bath} S_{\bath}[G_{\bath}, \Sigma'] & = \frac{1}{2}\Tr_{\mathcal{C}} \log(\ii \partial - \Sigma') - \frac{1}{2} \int_{\mathcal{C}} \Sigma'(z_1, z_2) G_{\bath}(z_1, z_2) \dd{z_1} \dd{z_2} - \frac{\ii^{q_\bath} J_\bath^2}{2 q_\bath} \int_{\mathcal{C}} [G_{\bath}(z_1, z_2)]^{q_\bath} \dd{z_1} \dd{z_2} \, , \\[2pt]
    \frac{\ii}{N} S_{\sysb}[G, G_{\bath}] & = - \frac{\ii^{f_{\sys} + f_{\bath}} V^2}{2 f_{\sys}} \int_{\mathcal{C}} [G(z_1, z_2)]^{f_{\sys}} [G_{\bath}(z_1, z_2)]^{f_{\bath}} \theta(t_1) \theta(t_2) \dd{z_1} \dd{z_2} \, , \\[2pt]
    \frac{\ii}{N} S_{\lind}[G] & = \frac{\ii \kappa}{2} \int_{\mathcal{C}} K(z_1, z_2) G(z_1, z_2) \dd{z_1} \dd{z_2} \, ,
\end{align}
where the kernel $K(z_1, z_2)$ is defined as \cite{sa2022,garcia2023}
\begin{equation}
    \int_{\mathcal{C}} K(z_1, z_2) G(z_1, z_2) \dd{z_1} \dd{z_2} = \int_{-\infty}^{\infty} [G(t^+, t^-) - G(t^-, t^+)] \dd{t} \, .
\end{equation}
We note that the right hand side is an integration over bi-local fields whereby time difference equals $0$, reflecting the fact that the Lindblad dynamics is Markovian, in the sense that it does not depend on status of the system at earlier times. For convenience we also drop out the constant term $-N \kappa /2$ in the Lindblad action arising from the anti-commutator term in the Lindbladian, which does not affect the equations of motion.

We have the Schwinger-Dyson (SD) equations for the system Green's function $G$
\begin{equation}
    (\ii \partial - \Sigma) * G = 1_{\mathcal{C}} \, ,
\end{equation}
where `$*$' denotes the convolution along the SK contour. Restricting the SD equation to the `$-+$' and `$+-$' components leads to the  Kadanoff-Baym (KB) equations:
\begin{align}
\ii \partial_{t_1} G^{\gtrless}(t_1, t_2) & = \int_{-\infty}^{\infty} \dd t \Bigl(\Sigma^\gR(t_1, t) G^{\gtrless}(t, t_2) + \Sigma^{\gtrless}(t_1, t) G^\gA(t, t_2) \Bigr) \, , \\
- \ii \partial_{t_2} G^{\gtrless}(t_1, t_2) & = \int_{-\infty}^{\infty} \dd t \Bigl(G^\gR(t_1, t) \Sigma^{\gtrless}(t, t_2) + G^{\gtrless}(t_1, t)\Sigma^\gA(t, t_2) \Bigr) \, ,
\end{align}
where $G^> \equiv G^{-+}$ and $G^< \equiv G^{+-}$ are the greater and lesser Green's functions, respectively, and $G^{\gR}$ ($G^{\gA}$) denotes the retarded (advanced) Green's function. The self-energies $\Sigma^{\gtrless}$ consist of three parts:
\begin{equation}
    \Sigma^{\gtrless} = \Sigma^{\gtrless}_{\sys} + \Sigma^{\gtrless}_{\bath} + \Sigma^{\gtrless}_{\lind} \, , 
\end{equation}
and each term is given by
\begin{align}
\Sigma^{\gtrless}_{\sys}(t_1, t_2) & = - \ii^{q} J^2 \big[G^{\gtrless}(t_1, t_2)\big]^{q-1} \, , \\
\Sigma^{\gtrless}_{\bath}(t_1, t_2) & = - \ii^{f_{\sys} + f_{\bath}} V^2 \theta(t_1) \theta(t_2) \big[G^{\gtrless}(t_1, t_2)\big]^{f_{\sys} - 1} \big[G_{\bath}^{\gtrless}(t_1, t_2)\big]^{f_{\bath}} \, , \\
\Sigma^{\gtrless}_{\lind}(t_1, t_2) & = \mp \ii \kappa \, \delta(t_1 - t_2) \, .
\end{align}
The retarded, advanced and Keldysh Green's functions (RAK) are defined as
\begin{align}
    G^{\gR}(t_1, t_2) & = \theta(t_1 - t_2) \big[G^>(t_1, t_2) - G^<(t_1, t_2) \big] \, , \\
    G^{\gA}(t_1, t_2) & = \theta(t_2 - t_1) \big[G^<(t_1, t_2) - G^>(t_1, t_2)] \, , \\
    G^{\gK}(t_1, t_2) & = G^>(t_1, t_2) + G^<(t_1, t_2) \, ,
\end{align}
and similarly for the self-energies. On the other hand, due to the relation $N_{\bath} \gg N$, the back-reaction is negligible, hence the bath Green's function $G_{\bath}$ obeys the same SD equation as a closed Hermitian SYK model, which is at a thermal state with inverse temperature $\beta_{\bath}$.

\begin{figure}[t]
\begin{center}
    \includegraphics[width=0.4\linewidth]{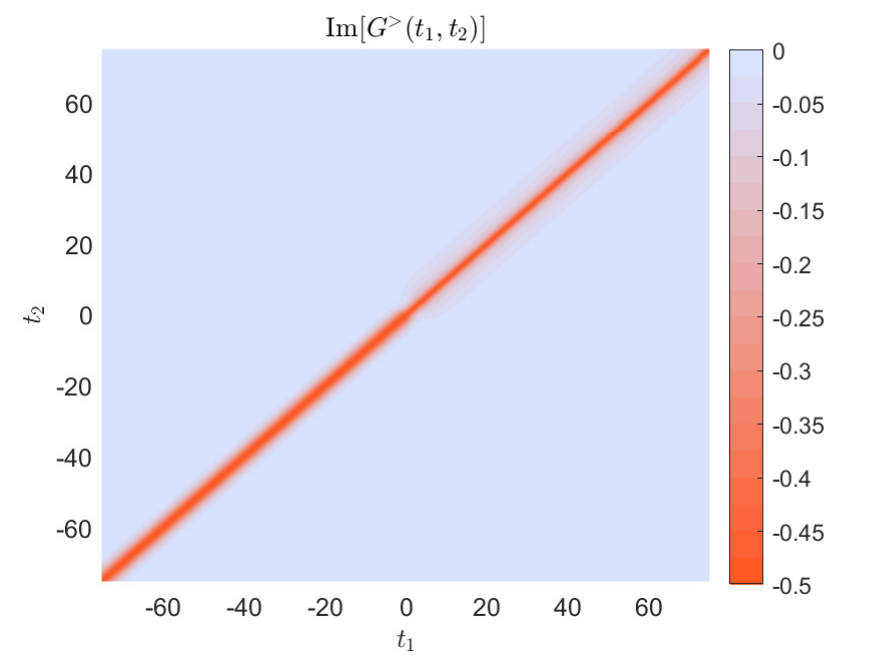}
    \includegraphics[width=0.4\linewidth]{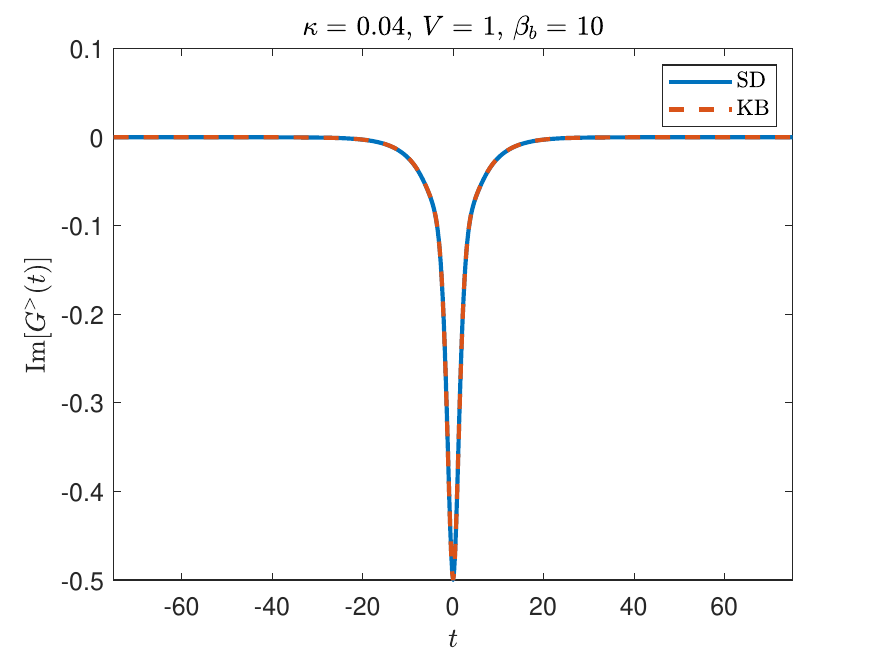}
    \includegraphics[width=0.4\linewidth]{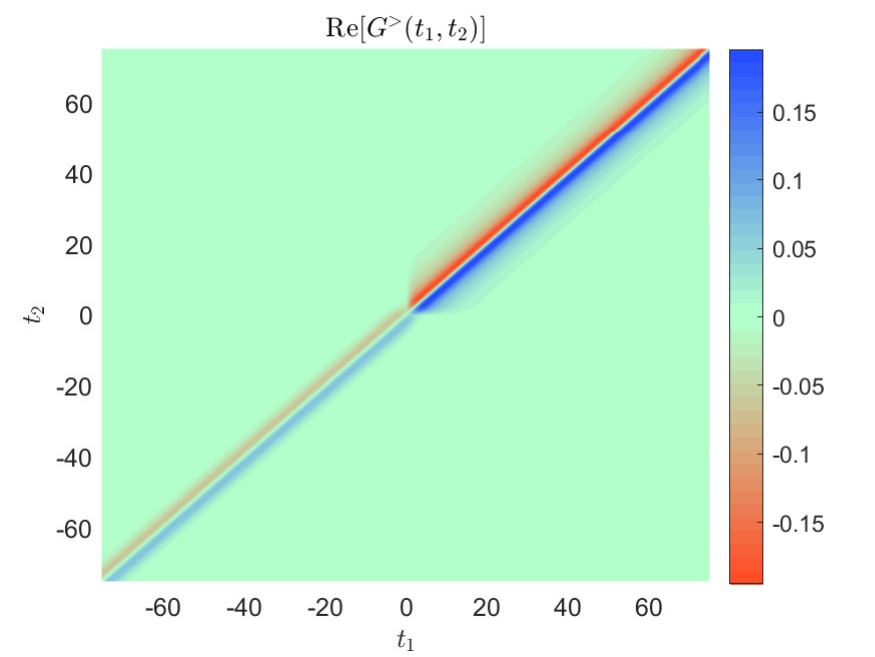}
    \includegraphics[width=0.4\linewidth]{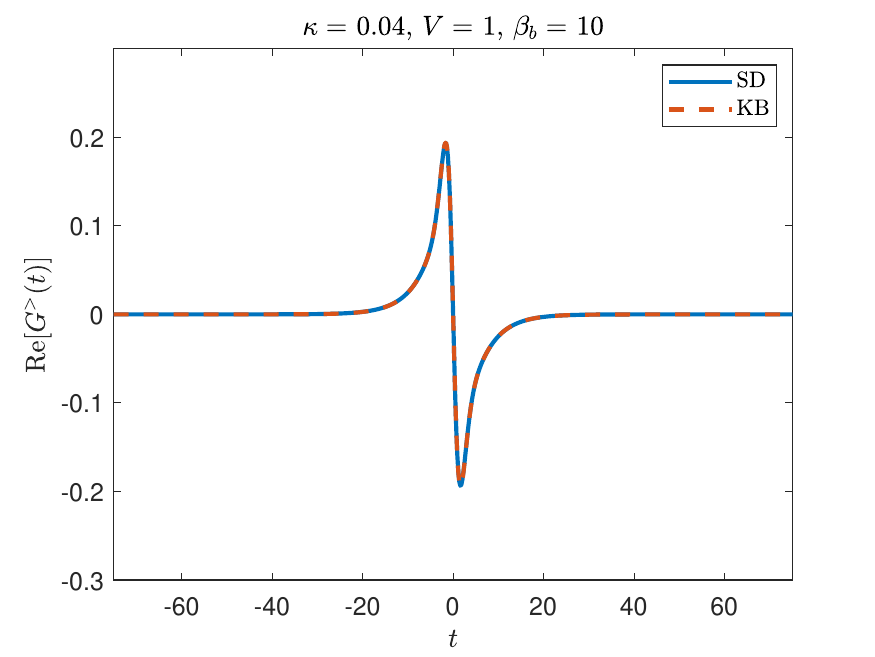}
    \caption{Left column: $\mathrm{Im}[G^>(t_1,t_2)]$ (Top) and $\mathrm{Re}[G^>(t_1, t_2)]$ (Bottom) solved from the Kadanoff-Baym (KB) equations, for a system starting from a thermal initial state with $\beta=1$ suffering from monitoring with strength $\kappa=0.04$ and cooling by a thermal bath with $V=1$, $\beta_{\mathrm{B}}=10$. Right column: Comparison of Green's functions solved from KB equations when $\mathcal{T}=60$ and the solution from the SD equation at the steady state, exhibiting excellent agreement.}
    \label{fig:gfgrmtx_ld_04_V_1_beta_10}
\end{center}
\end{figure}

In the long time limit $\mathcal{T} \equiv (t_1+t_2)/2 \rightarrow \infty$, the KB equations reduce to time translational invariant SD equations that involving only the time different $t \equiv t_1 - t_2$,
\begin{equation}
    \ii \partial_t G^{\alpha \beta}(t) =  s_{\alpha} \delta_{\alpha \beta} \delta(t) + \sum_{\gamma = \pm} s_{\gamma} \int_{-\infty}^{\infty} \Sigma^{\alpha \gamma}(t - t^\prime) G^{\gamma \beta}(t^\prime) \dd t^\prime  \, , 
\end{equation}
with $s_{\pm} = \pm 1$. The self-energies then obey 
\begin{equation}
    \Sigma^{\gtrless}(t) = -\ii^q J^2 \big[G^{\gtrless}(t)\big]^{q-1} - \ii^{f_{\sys}+f_{\bath}} V^2 \big[G^{\gtrless}(t)\big]^{f_{\sys}-1} \big[G^{\gtrless}_{\bath}(t)\big]^{f_{\bath}} \mp \ii \kappa \delta(t) \, .
\end{equation}
Performing Fourier transform,
\begin{equation}
G^{\alpha \beta}(t) = \int_{-\infty}^{\infty} G^{\alpha \beta}(\omega) \ee^{-\ii \omega t} \frac{\dd\omega}{2\pi} \, , 
\end{equation}
we have the time translational invariant SD equations in frequency space in the RAK basis:
\begin{align}
    \big[\omega - \Sigma^{\gR}(\omega)\big] G^{\gR}(\omega) & = 1 \, , \\
    \big[\omega - \Sigma^{\gA}(\omega)\big] G^{\gA}(\omega) & = 1 \, , \\
    \big[\omega - \Sigma^{\gR}(\omega)\big] G^{\gK}(\omega) & = \Sigma^{\gK}(\omega) G^{\gA}(\omega) \, .
\end{align}
These equations can be solved numerically, in a similar manner as the time-translational invariant SD equations for SYK model at thermal equilibrium state \cite{maldacena2016}. The only difference is that in that case, one can impose the KMS condition $G^>(\omega) = - \ee^{\beta \omega} G^<(\omega)$ so that in frequency space one only needs to solve for the retarded Green's function. In this case, however, we need also solve for the Keldysh Green's function $G^{\gK}$ due to the violation of the KMS condition. Nevertheless, we are able to obtain unique solutions once the parameters are fixed, and find precise agreements with the long time results of the KB equations, as shown in Fig. \ref{fig:gfgrmtx_ld_04_V_1_beta_10}.

\section{Long time decay rate}

\begin{figure}[htb!]
\begin{center}
\includegraphics[width=0.4\textwidth]{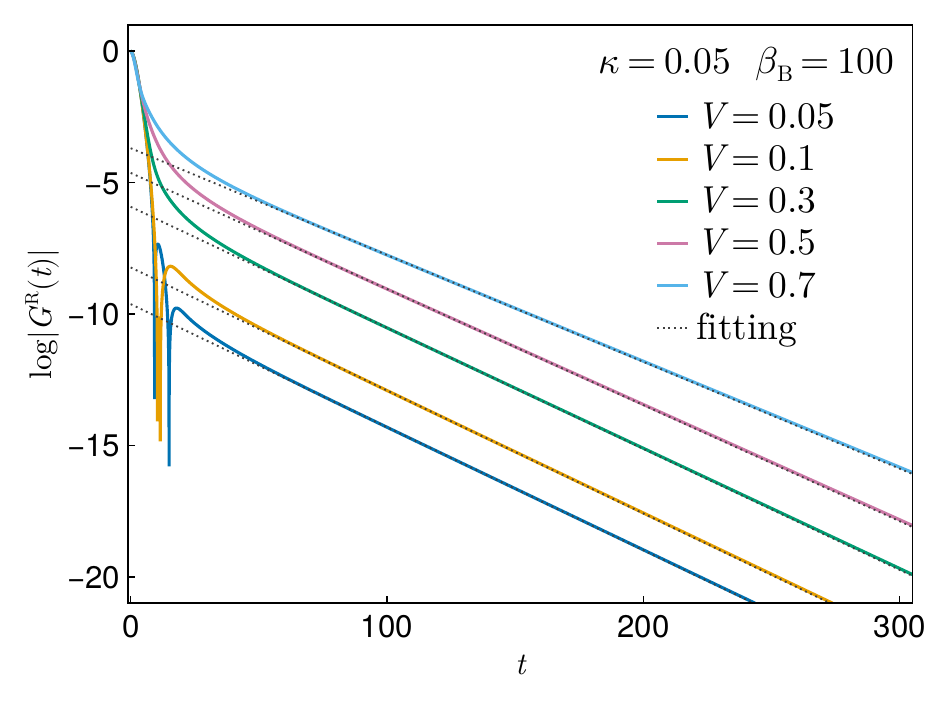}
\caption{Long time decay rate for $q=4$.}
\label{fig:log_absGRt_plot}
\end{center}
\end{figure}

In this section we summarize the long time decay rate $\gamma$ for SYK monitored dissipative dynamics. A generic feature for non-zero $\kappa$ and small but non zero $V$, the long time decay rate $\gamma = (q-1) \gamma_{\rm B} + \delta \gamma$, where $\gamma_{\rm B} = 2\pi \Delta_{\rm B} / \beta_{\rm B}$ with $\Delta_{\rm B} = 1/q_{\rm B}$ the conformal dimension of the bath, and $\delta \gamma$ is some small corrections. In Fig.~\ref{fig:log_absGRt_plot} we present the results for $\kappa = 0.05$ and $\beta_{\rm B}=100$ for different values of $V$. In the main text we present the ratio $\gamma/\gamma_{\rm B}$ versus $V$ for different values of $\kappa$ and $\beta_{\rm B}$. We see the correction $\delta \gamma$ is suppressed as the bath temperature is lowered.

We then provide an analytic explanation for this results. Since the system-bath interaction is of type $(1, q-1)$, the system-bath self energy $\Sigma_{\rm B}^{\rm R}$ then has a decay rate $(q-1) \gamma_{\rm{B}}$. In frequency space, we may expand it near $\omega_0 \equiv -\ii (q-1) \gamma_{\rm{B}}$, with $\omega = \omega_0 + \delta \omega$
\begin{equation}
    \Sigma_{\rm B}^{\rm R}(\omega) = \frac{A}{\omega - \omega_0} +  B + \mathcal{O}(\delta\omega) \, ,
\end{equation}
where 
\begin{align}
    A & \equiv - 2 V^2 b^{q-1} \cos(\pi \Delta) \left(\frac{2\pi}{\beta_{\rm B}}\right)^{2(1-\Delta)} \, , \\ 
    B & \equiv - 2 \ii V^2 b^{q-1} \cos(\pi \Delta) \left(\frac{2\pi}{\beta_{\rm B}}\right)^{1-2\Delta} [\gamma_{\rm E} + \psi(2\Delta - 1)] \, , \\
    \pi J_{\rm B}^2 b^q & = (\frac{1}{2} - \Delta) \tan(\pi \Delta) \, .
\end{align}
Here $\gamma_{\rm E} \approx 0.577$ is Euler's constant, and $\psi(z)$ is the digamma function. $\gamma_{\rm E} + \psi(2\Delta - 1) = H_{2\Delta-2}$, where $H_n$ is Harmonic number defined by $H_n=\int_0^1 \frac{1 - x^n}{1-x}\dd{x}$.

\begin{figure}[t]
\begin{center}
    \includegraphics[width=0.4\linewidth]{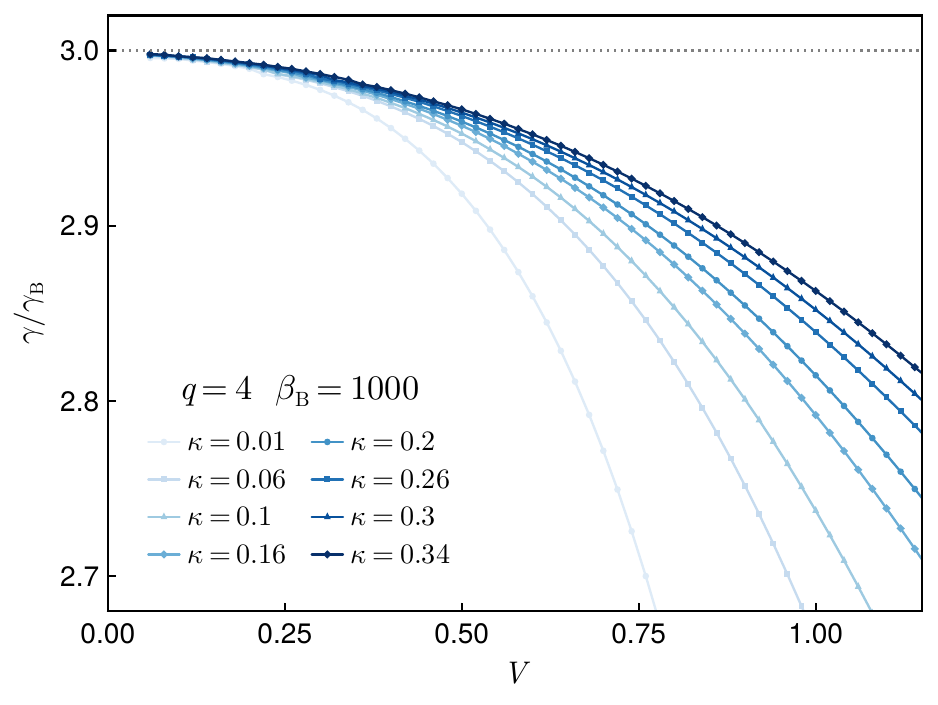}
    \includegraphics[width=0.4\linewidth]{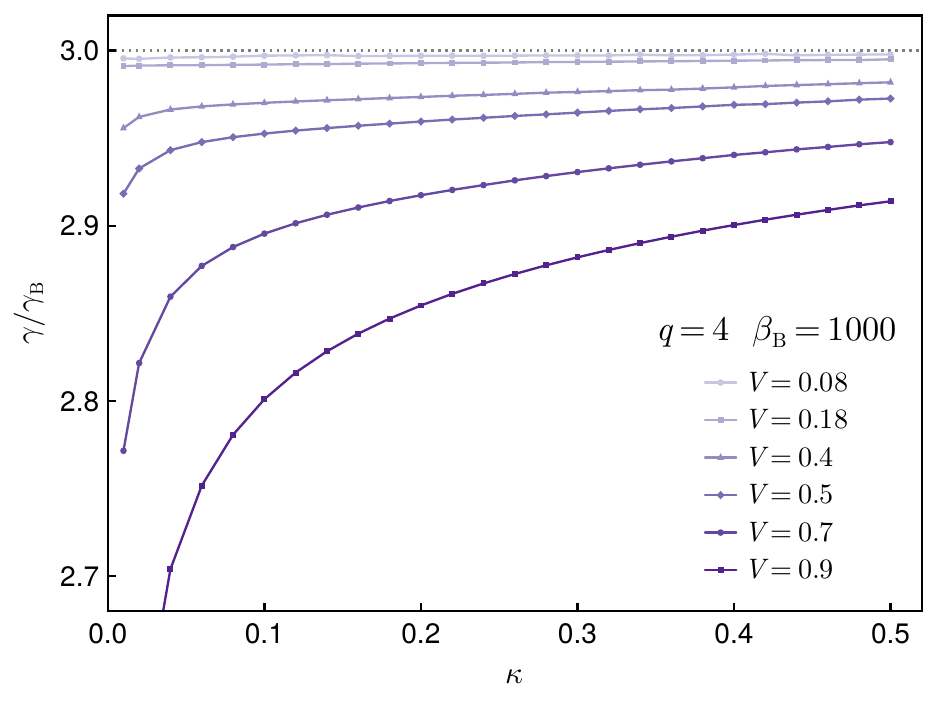}
    \caption{$\gamma / \gamma_{{\rm B}}$ versus $V$ and $\kappa$ with for different values for $\beta_{\rm B}=1000$. We take $q=4$ and $J=1$.}
    \label{fig:gamma_ratio_vs_V_beta_1000}
\end{center}
\end{figure}

From the Schwinger-Dyson equations we have
\begin{equation}
    G^{\rm R}(\omega) = \frac{1}{D(\omega)} \, , 
    \quad
    D(\omega) \equiv \omega - \Sigma_{\rm S}^{\rm R}(\omega) - \Sigma_{\rm B}^{\rm R}(\omega) + \ii \kappa \, .
\end{equation}
The long time decay rate is given by the pole located in the lower half plane that is closet to the real axis, satisfying the necessary condition $D(\omega_*) = 0$ and $D'(\omega_*) \neq 0$. Then this gives a simple pole at $\omega_*$ and we can expand the retarded Green's function as 
\begin{equation}
    G^{\rm R}(\omega) = \frac{1/D'(\omega_*)}{\omega - \omega_*} + \dots \, .
\end{equation}
We may expand the simple near $\omega_0$, $\omega_* = \omega_0 + \delta \omega$, where $\delta \omega$ is small compared to $\omega_0$. Next, we may expand $D(\omega)$ around $\omega_0$ to solve for $\delta \omega$. Substituting the expansion of $\Sigma_{\rm B}$ into the expression for $D(\omega)$, and keep up to order $\mathcal{O}(\delta \omega)$ we obtain the correction and the residue
\begin{align}
    \delta \omega & = \frac{A}{D_{\rm reg}(\omega_0)} \, , \\[5pt]
    \frac{1}{D'(\omega_*)} & = \frac{A}{[D_{\rm reg}(\omega_0)]^2} + \mathcal{O}(\delta \omega^4)\, , \\[5pt]
    D_{\rm reg}(\omega_0) & \equiv \omega_0 - \Sigma_{\rm S}^{\rm R}(\omega_0) - B + \ii \kappa \, ,
\end{align}
In computing the residue we plug into the solution for $\delta \omega$. The correction to the long time decay rate therefore is given by $\delta \gamma = - \Im \delta \omega$. Due to the non-negligible system self-energy $\Sigma_{\rm S}^{\rm R}(\omega_0)$, one cannot explicitly compute the correction. However, this gives us qualitative explanations for numerical results. Firstly, it explains that $\delta \gamma$ is negative, such that the corrections decrease the decay rate. Secondly, Since $A$ is proportional to $V^2$, as $V$ is increased, the magnitude of $\delta \gamma$ increases, and therefore $\gamma$ decreases with $V$. Thirdly, for larger $\beta_{\rm B}$ (lower temperature) we have smaller $A$, consequently the corrections are smaller than higher temperature results, as is verified numerically. Finally, since the Lindblad term appears in $D_{\rm reg}$, as $\kappa$ increases, $\gamma$ also increases, which can also be seen from figure in the main text. For a complementary comparisons, in Fig.~\ref{fig:gamma_ratio_vs_V_beta_1000} we present the long-time fitting results for $\beta_{\mathrm{B}}=1000$ and $q=4$, in which the case the sub-leading order correction is smaller than $\beta_{\mathrm{B}}=100$ presented in the main text.

\begin{figure}[th]
\begin{center}
    \includegraphics[width=0.4\linewidth]{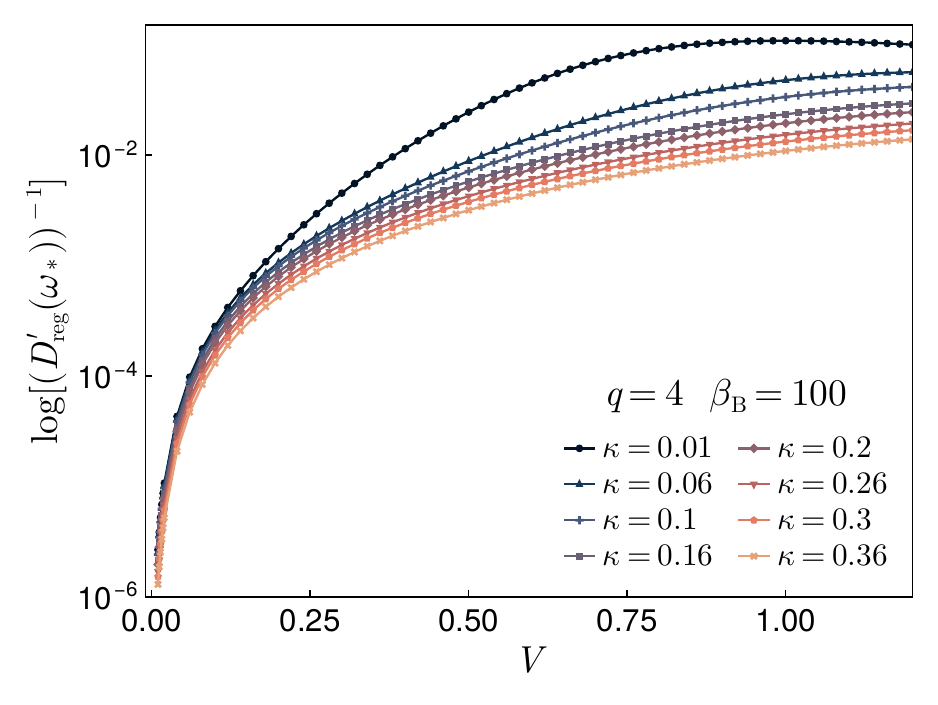}
    \includegraphics[width=0.4\linewidth]{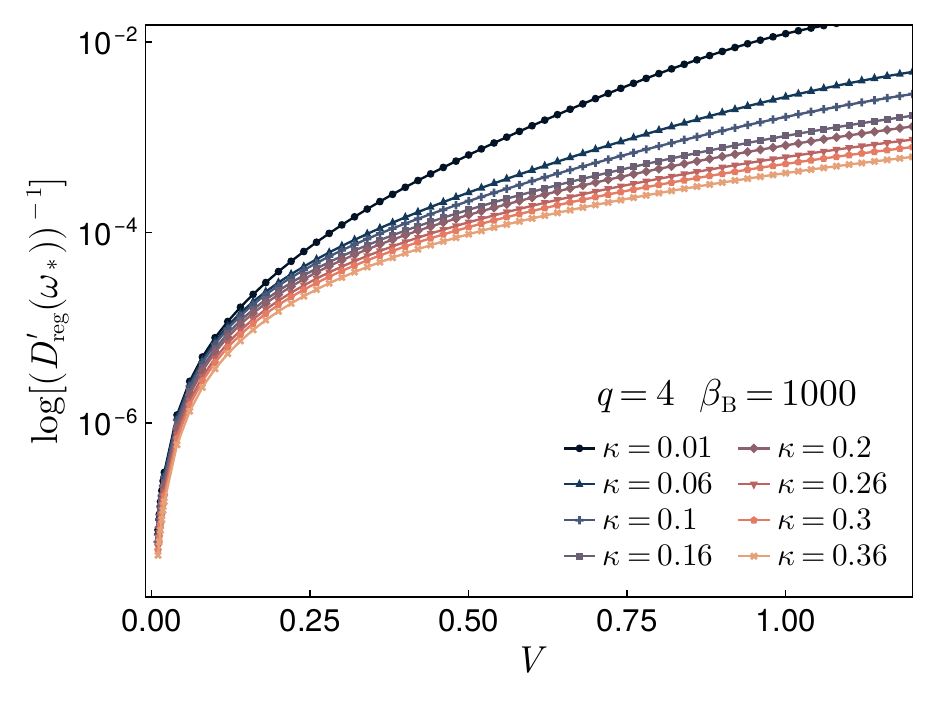}
    \caption{Numerical fitting results for the residue $1/D'(\omega_*)$ that corresponds the amplitude for the long time decay.}
    \label{fig:residue_vs_V}
\end{center}
\end{figure}

Even though the long time behavior is controlled by the simple pole at $\omega_0$, we also note that the residue $1/D'(\omega_*)$ is proportional to $\delta \omega^2$, which is very small. This means it is almost invisible from the spectral function $\rho^- = -\Im(G^{\rm R})/\pi$. In Fig.~\ref{fig:residue_vs_V} we exhibit the results for the residue from numerical fittings, and indeed they are small and do not give significant contribution to the spectrum.

\section{Out of time ordered correlation functions and Lyapunov exponents}

\begin{figure}[!t]
\begin{center}
    \includegraphics[width=0.9\textwidth]{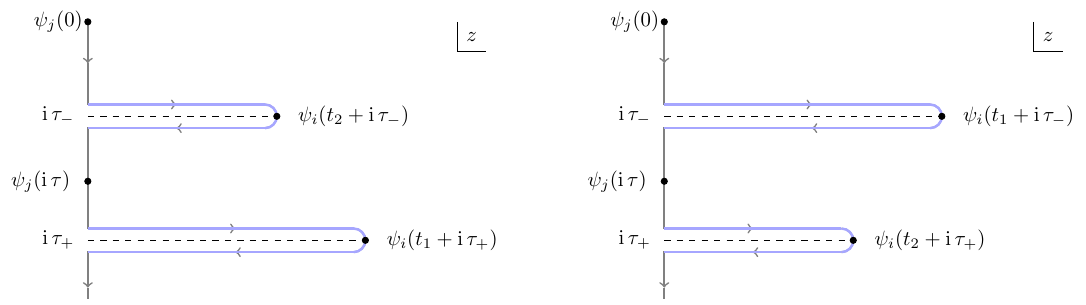}
    \caption{Keldysh contours for two possible OTOCs.}
    \label{fig:OTOC_contours_supp}
\end{center}
\end{figure}

In this section, we discuss the computation of the out-of-time-ordered correlators (OTOCs) in the SYK monitored dissipative dynamics. Consider the time argument $z = t + \ii \tau$ on the Keldysh contour $\mathcal{C}$, let $t = \mathrm{Re}(z)$ denote the real time, and $\tau = \mathrm{Im}(z) \leq 0$ denote the imaginary time. The OTOCs are defined as the four-point functions on the Keldysh contour $\mathcal{C}$, with specific contour-ordering of the operators, where the time of the last operator can be set to 0 due to time translational invariance:
\begin{equation}
    \langle \operatorname{T}_{\mathcal{C}} G(z_1, z_2) G(\ii \tau, 0) \rangle \equiv \left(-\frac{\ii}{N}\right)^2 \sum_{i,j=1}^{N} \langle \operatorname{T}_{\mathcal{C}} \psi_i(z_1) \psi_i(z_2) \psi_j(\ii \tau) \psi_j(0) \rangle \, .
\end{equation}
As illustrated in Fig.~\ref{fig:OTOC_contours_supp}, there are two possible time orderings relevant for the OTOCs:
\begin{equation}
    \Im z_1 < \tau < \Im z_2 < 0 \quad \text{(Case I)} \, , \quad \text{or} \quad \Im z_2 < \tau < \Im z_1 < 0 \quad \text{(Case II)} \, .
\end{equation}
We denote the connected part of these two OTOCs as $C_1(t_1, t_2)$ and $C_2(t_1, t_2)$, respectively, 
\begin{align}
C_1(t_1, t_2) & \equiv \mathcal{F}(t_1 + \ii \tau_+, t_2 + \ii \tau_-, \ii \tau, 0) \, , \\
C_2(t_1, t_2) & \equiv \mathcal{F}(t_1 + \ii \tau_-, t_2 + \ii \tau_+, \ii \tau, 0) \, ,
\end{align}
where $\mathcal{F}$ denotes the connected four-point function. They obey recursion relations \cite{Ferrari:2019ogc}
\begin{align}
\label{eq:OTOC_recursion_relations_C1}
C_1(t_1, t_2) & = - 2 G_0(t_1 + \ii \tau_+, \ii \tau) G_0(t_2 + \ii \tau_-, 0) + \int_{\mathcal{C}} \mathcal{K}(t_1+\ii\tau_+, t_2 + \ii\tau_-, z, z') \mathcal{F}(z, z', \ii \tau, 0) \dd z \dd z' \, , \\
\label{eq:OTOC_recursion_relations_C2}
C_2(t_1, t_2) & = - 2 G_0(t_1 + \ii \tau_-, \ii \tau) G_0(t_2 + \ii \tau_+, 0) + \int_{\mathcal{C}} \mathcal{K}(t_1+\ii\tau_-, t_2+\ii\tau_+, z, z') \mathcal{F}(z, z', \ii \tau, 0) \dd z \dd z' \, ,
\end{align}
The kernel for our setting with $q=4$ is given by
\begin{equation}
   \mathcal{K}(z_1, z_2, z_3, z_4) = 3 J^2 G_0(z_1, z_3) G_0(z_4, z_2) (G_0(z_3, z_4))^{2} \, ,
\end{equation}
representing a building block for the ladder diagrams.

\begin{figure}[!t]
\begin{center}
    \includegraphics[width=0.8\textwidth]{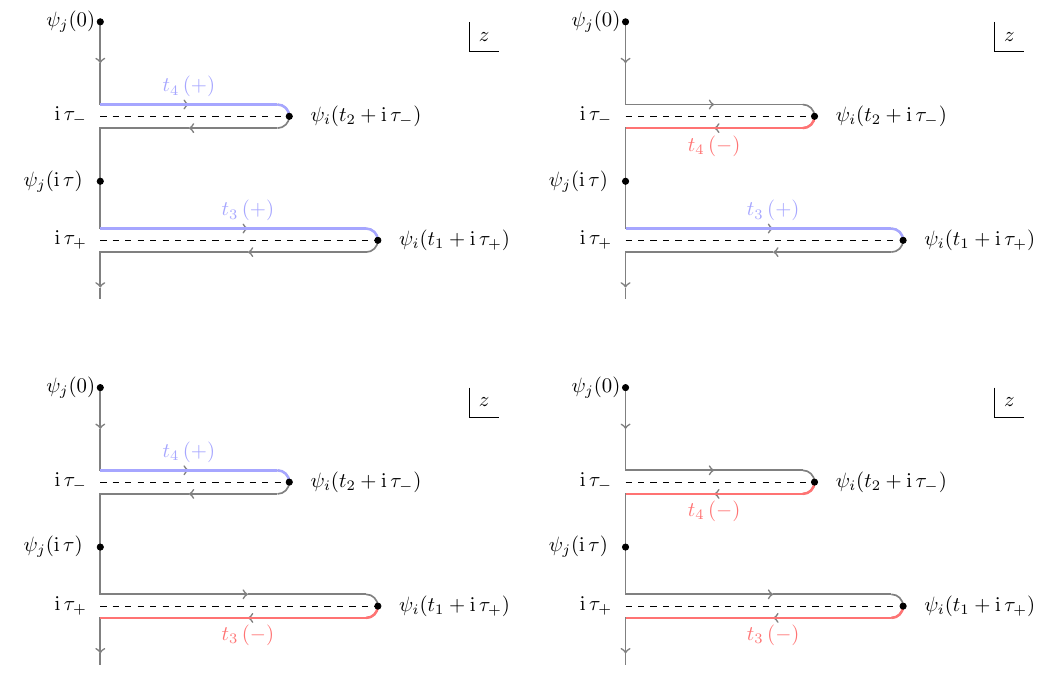}
    \caption{Contour integrations that contribute to the kernel $\mathcal{K}_{1,1}$.}
    \label{fig:K_11_kernel}
\end{center}
\end{figure}

The contour integrations on the right hand sides of Eq.~\eqref{eq:OTOC_recursion_relations_C1} and \eqref{eq:OTOC_recursion_relations_C2} should in principle travel all the segments of the Keldysh contour that are relevant to the OTOCs, leading to a set of un-closed recursion equations, which involve $\mathcal{F}(t_1 + \ii \tau_+, t_2 + \ii \tau_+, \ii \tau, 0)$, for example. However, to capture the long time limit ($t_1, t_2 \gg 0$) responsible for the OTOCs, $z$ and $z'$ should be in the different folds of the contour \cite{Ferrari:2019ogc}. Therefore, we can make the equations self-closed by the approximation:
\begin{align}
C_1(t_1, t_2) & \approx - 2 G_0(t_1+i\tau_+, \ii\tau) G_0(t_2+\ii\tau_-, 0) + \sum_{i=1}^{2} \int_0^{\infty} \mathcal{K}_{1,i}(t_1, t_2, t_3, t_4) C_i(t_3, t_4) \dd t_3 \dd t_4 \, , \\
C_2(t_1, t_2) & \approx - 2 G_0(t_1+\ii\tau_-, \ii\tau) G_0(t_2+\ii\tau_+, 0) + \sum_{i=1}^{2} \int_0^{\infty} \mathcal{K}_{2,i}(t_1, t_2, t_3, t_4) C_i(t_3, t_4) \dd t_3 \dd t_4 \, .
\end{align} 
The kernels are found to be
\begin{align}
\mathcal{K}_{1,1}(t_1, t_2, t_3, t_4) & = 3 J^2 G_0^{\gR}(t_1 - t_3) G_0^{\gA}(t_4 - t_2) \big(G_0^{\gW+}(t_3 - t_4)\big)^2 \, , \\
\mathcal{K}_{1,2}(t_1, t_2, t_3, t_4) & = 0 \, ,\\
\mathcal{K}_{2,1}(t_1, t_2, t_3, t_4) & = 0 \, , \\
\mathcal{K}_{2,2}(t_1, t_2, t_3, t_4) & = 3 J^2 G_0^{\gR}(t_1 - t_3) G_0^{\gA}(t_4 - t_2) \big(G_0^{\gW-}(t_3 - t_4)\big)^2 \, .
\end{align}
Here $G^{\gR}$ and $G^{\gA}$ denote the retarded and advanced Green's function, respectively, and the Wightman functions $G^{\gW\pm}$ are defined as
\begin{align}
G_0^{\gW+}(t_1, t_2) & \equiv G_0(t_1 - t_2 + \ii (\tau_+ - \tau_-)) \, , \\
G_0^{\gW-}(t_1, t_2) & \equiv G_0(t_1 - t_2 + \ii (\tau_- - \tau_+)) \, .
\end{align}
To give a concrete example for the derivation of the kernels, let us consider $\mathcal{K}_{1,1}$, which consists of the contour integrations illustrated in Fig.~\ref{fig:K_11_kernel}. For the upper segments of the fold (indicated by light blue color) the time runs forward, giving a `$+$' sign. While for the lower segments of the fold (indicated by light red color) the time runs backward, giving a `$-$' sign. Taking into account the contour ordering, we have
\begin{align*}
    \mathcal{K}_{1,1}(t_1, t_2, t_3, t_4) = & 3 J^2 \theta(t_{13}) \theta(t_{24}) \big(G_0(t_1 - t_3^+) - G_0(t_1 - t_3^-) \big) \big(G_0(t_4^+ - t_2) - G_0(t_4^- - t_2)\big) \\ 
    & \quad \times \big(G_0(t_3 - t_4 + \ii (\tau_+ - \tau_-)) \big)^2 \\[4pt]
    = & 3 J^2 \theta(t_{13}) \theta(t_{24}) \big(G_0^>(t_1 - t_3)  - G_0^<(t_1 - t_3) \big) \big(G_0^<(t_4 - t_2) - G_0^>(t_4 - t_2) \big) \\
    & \quad \times \big(G_0^{\gW+}(t_3 - t_4) \big)^2 \\[4pt]
    = & 3 J^2 G_0^{\gR}(t_1 - t_3) G_0^{\gA}(t_4 - t_2) \big(G_0^{\gW+}(t_3 - t_4)\big)^2 \, .
\end{align*}

In the SYK literature, in order to compute the out-of-time-ordered correlators (OTOCs), one usually considers the ``regularized'' OTOCs which involve the Wightman functions $G_0(t \pm \ii \frac{\beta}{2})$ \cite{maldacena2015,maldacena2016}, and consequently the OTOCs can be written as a norm in the thermofield double state. The regularization inherently requires a well-defined thermal equilibrium state and hence the KMS symmetry. In our case, however, the steady state is not thermal due to the Lindblad bath. Therefore, to compute the OTOCs in this model, we have to go back to the un-regularized OTOCs by taking all imaginary times to be zero while keeping the contour ordering fixed. Specifically, we are interested in the limit
\begin{equation}
    \tau_+ \rightarrow 0 \, , \quad \tau \rightarrow 0 \, , \quad \tau_- \rightarrow 0 \, , \quad 
\text{with fixed relations } \tau_+ < \tau < \tau_- < 0 \, .
\end{equation}
In this limit the Wightman functions reduces to greater and lesser Green's functions, respectively, 
\begin{equation}
\lim_{\tau_{-} \rightarrow 0^-} \lim_{\tau_{+} \rightarrow \tau_{-}^{-}} G_0^{\gW \pm}(t_1, t_2) = G_0^{\gtrless}(t_1 - t_2) \, .
\end{equation}
Collecting the results together we obtain two decoupled integral equations for the un-regularized OTOCs:
\begin{align}
    C_i(t_1, t_2) & = - 2 G_0^{\gtrless}(t_1) G_0^>(t_2) + \int_0^{\infty} \mathcal{K}_{i}(t_1, t_2, t_3, t_4) C_i(t_3, t_4) \dd t_3 \dd t_4 \, , \quad i = 1, 2 \, , \\
    \mathcal{K}_{i}(t_1, t_2, t_3, t_4) & = - 3 J^2 G_0^{\gR}(t_1 - t_3) G_0^{\gR}(t_2 - t_4) \big(G_0^{\gtrless}(t_3 - t_4)\big)^2 \, ,
\end{align}
where we use the relation $G^{\gA}(-t) = - G^{\gR}(t)$.

\begin{figure}[!t]
\begin{center}
    \includegraphics[width=0.4\textwidth]{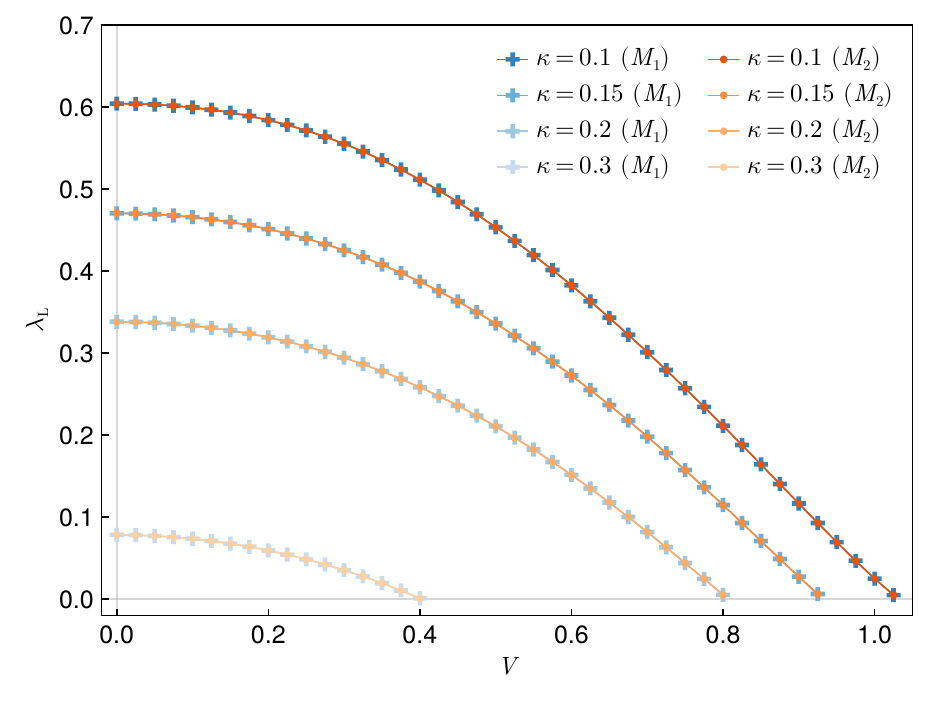}
    \caption{Consistency check for the Lyapunov exponents of quenched Lindblad SYK$_4$ with $\beta_{\bath}=100$: $M_1$ and $M_2$ results comparison.}
    \label{fig:Lyapunov_M1_and_M2_quenched_Lindblad_SYK}
\end{center}
\end{figure}

\begin{figure}[!t]
\begin{center}
    \includegraphics[width=0.4\textwidth]{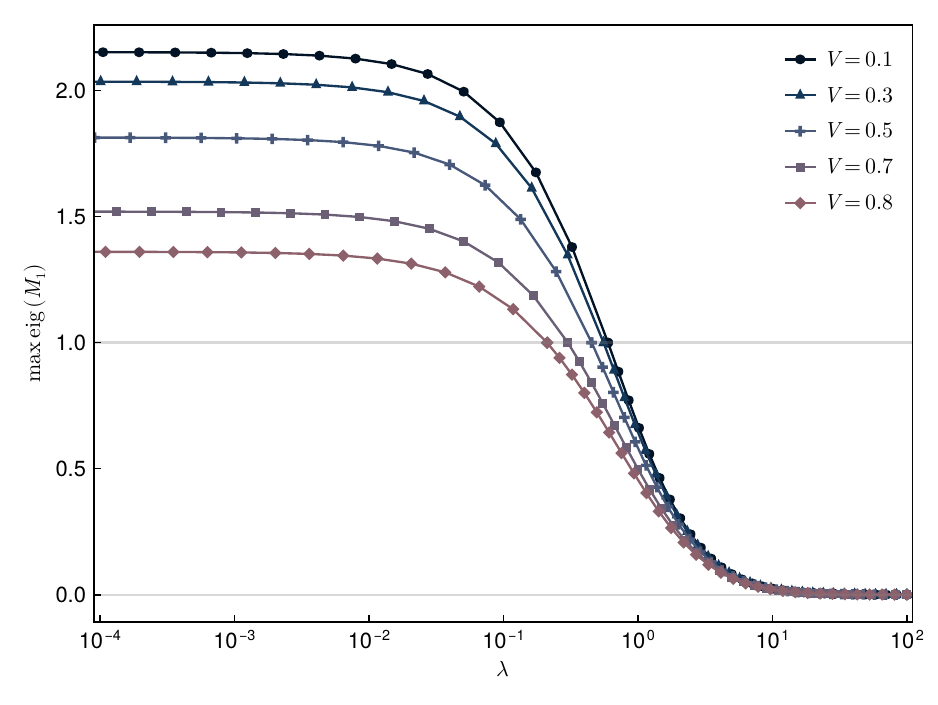}
    \includegraphics[width=0.4\textwidth]{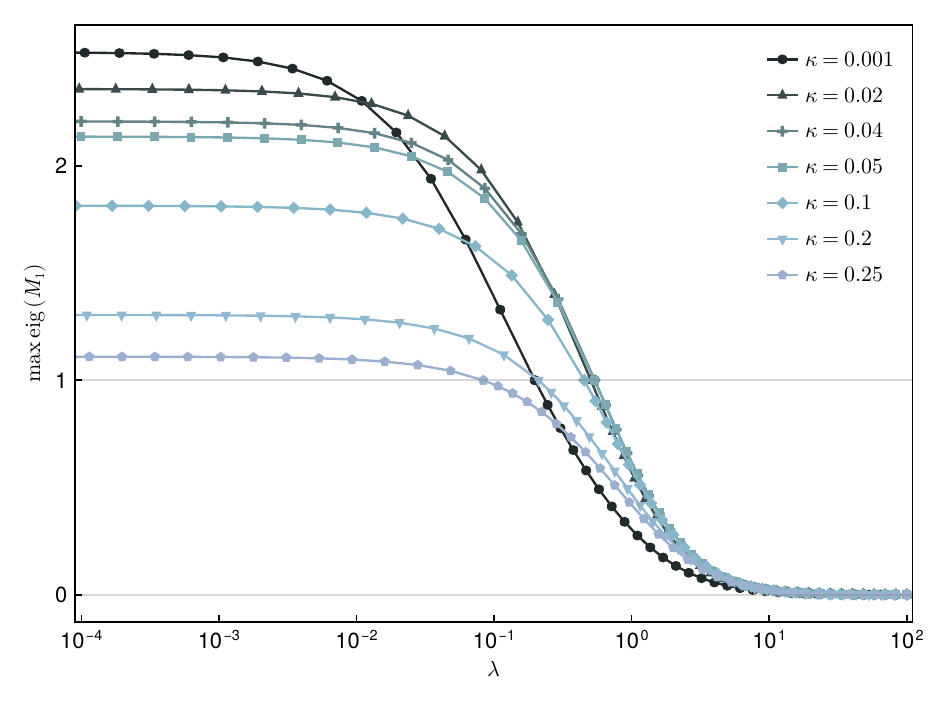}
    \caption{Uniqueness check for the Lyapunov exponents of quenched Lindblad SYK$_4$ with $\beta_{\bath}=100$.}
    \label{fig:uniqueness_check}
\end{center}
\end{figure}

We are now at the stage on extracting the Lyapunov exponent from the above recursion relations for the connected four-point functions that are relevant to OTOCs. In particular, we focus on the un-regularized OTOCs, in which case the Wightman functions that involve time arguments on two separate folds reduce to the greater and lesser Green's functions. Besides, the Lyapunov exponent is expected to be independent of the regularization scheme. We will consider the case that $t_1, t_2 \gg 0$, such that the first terms on the recursion formula decay to 0, leaving
\begin{equation}
    C_i(t_1, t_2) = \int_0^{\infty} \mathcal{K}_{i}(t_1, t_2, t_3, t_4) C_i(t_3, t_4) \dd t_3 \dd t_4 \, , \quad i = 1, 2 \, .
\end{equation}
In the center-of-mass frame, 
\begin{equation}
    \mathcal{T} \equiv \frac{t_1 + t_2}{2} \, , \quad t \equiv t_1 - t_2 \, ,
\end{equation}
we can make an ansatz
\begin{equation} \label{eq:OTOC_ansatz_supp}
    C_i(\mathcal{T}, t) = \ee^{\lambda_{\mathrm{L}} \mathcal{T}} f_i(t) \, .
\end{equation}
Then the center-of-mass time $\mathcal{T}$ cancels out from the recursion relation, giving an eigenvalue problem for $f_i(t)$:
\begin{equation}
    f_i(t) = \int_{-\infty}^{\infty} M_i(t, t'; \lambda_\mathrm{L}) f_i(t') \dd t' \, ,
\end{equation}
with a reduced kernel
\begin{equation} \label{eq:reduced_kernel_M_i_supp}
    M_i(t, t' ; \lambda_\mathrm{L}) = 3 J^2 \big(G^{\gtrless}(t')\big)^{2} \int_{-\infty}^{\infty} G^{\gR}\left(u + \frac{1}{2}(t-t')\right) G^{\gR}\left(u - \frac{1}{2}(t - t')\right)
    \ee^{-\lambda_{\mathrm{L}} u} \dd u \, .
\end{equation}
In order to be consistent with the exponential growth ansatz \eqref{eq:OTOC_ansatz_supp}, the eigenvalue problem indicates that, the Lyapunov exponent $\lambda_\mathrm{L}$ corresponds to the matrix $M_i(t, t' ; \lambda_{\mathrm{L}})$ that should have the largest eigenvalue $1$. This can be done by discretizing the time $t_a = a \Delta t$ ($a \in \mathbb{Z}$) such that $M$ becomes a discretized matrix $M_{ab} \equiv M(t_a, t_b)$. It will be a highly sparse matrix and one can use the Krylov method (\texttt{KrylovKit.jl} package in \texttt{Julia} programming language, for example) to solve for the largest eigenvalue, and then perform a binary search for the Lyapunov exponent $\lambda_\mathrm{L}$ such that the largest eigenvalue of $M_i$ equals $1$. To speed up the calculation of the kernel matrix $M$, the above integral can be expressed as a Fourier transform~\cite{Ferrari:2019ogc}. Specifically, let
\begin{equation}
A(t ;\lambda_{\mathrm{L}}) = \int \frac{\dd \omega}{2\pi} g(\omega) g(-\omega) \ee^{-\ii \omega t} \, , \quad 
g(t) \equiv G^{\gR}(t) \ee^{- \frac{1}{2} \lambda_{\mathrm{L}} t} \, ,
\end{equation}
then we can write the kernel as
\begin{equation}
M_i(t, t' ; \lambda_{\mathrm{L}}) = 3 J^2 \big(G^{\gtrless}(t')\big)^2 A(t - t' ; \lambda_\mathrm{L}) \, .
\end{equation}
The advantage of this expression is that we can compute the kernel matrix $M_i$ column by column instead of element by element, hence significantly reduces the computational cost.

\begin{figure}[!t]
	\begin{center}
        \includegraphics[width=0.4\textwidth]{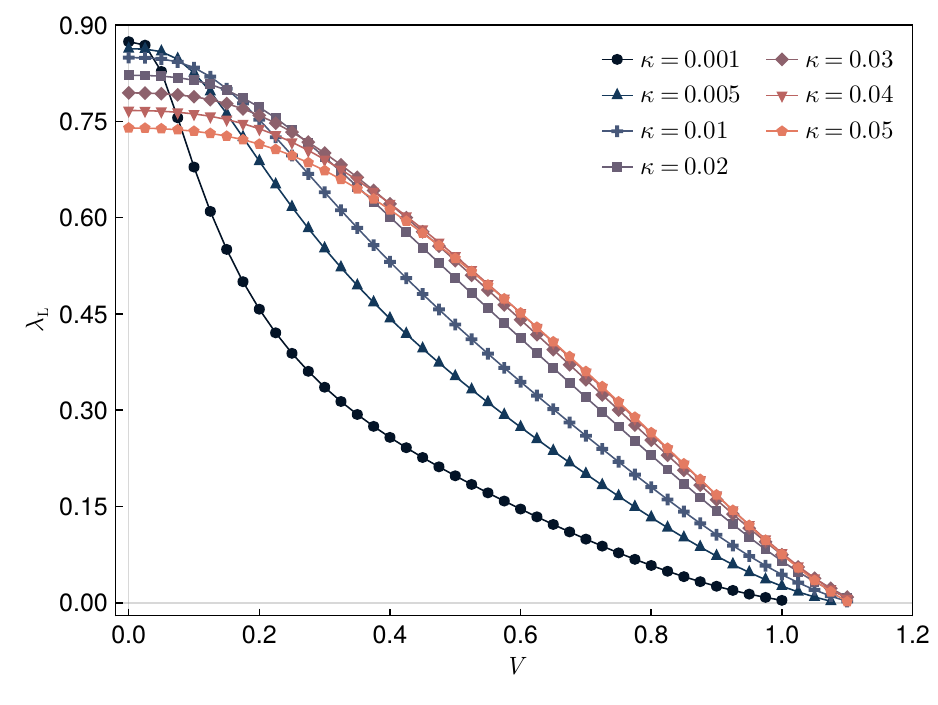}
        \label{fig:small_kappa_region}
        \includegraphics[width=0.4\textwidth]{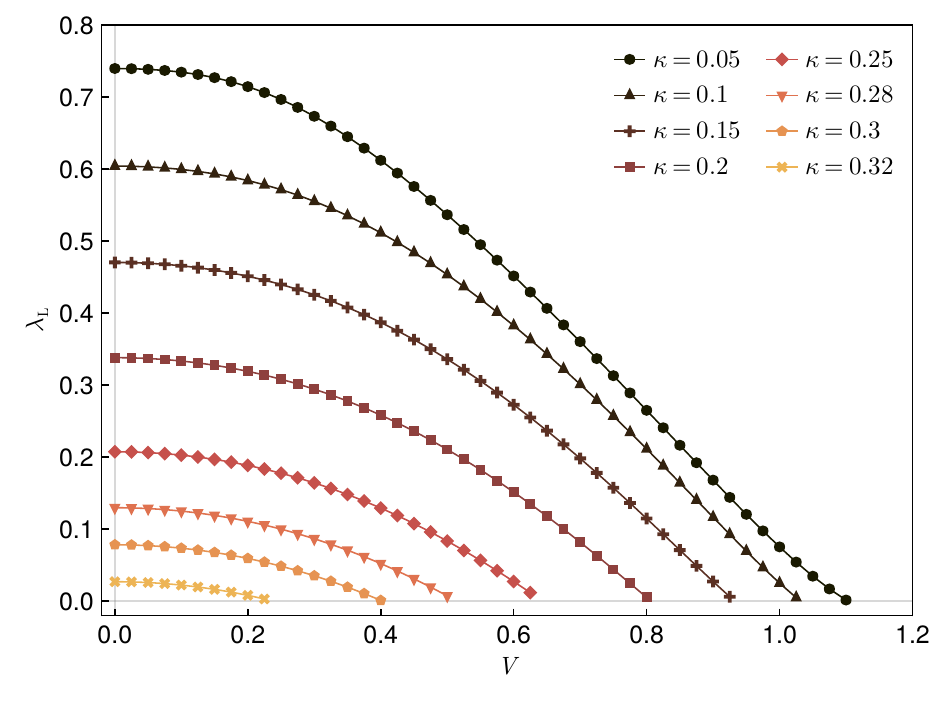}
        \label{fig:large_kappa_region}
		\caption{The Lyapunov exponents versus $V$ for various $\kappa$ with $\beta_{\bath}=100$.}
		\label{fig:Lyapunov_vs_V_quenched_Lindblad_SYK}
	\end{center}
\end{figure}

\begin{figure}[!t]
\begin{center}
    \includegraphics[width=0.45\textwidth]{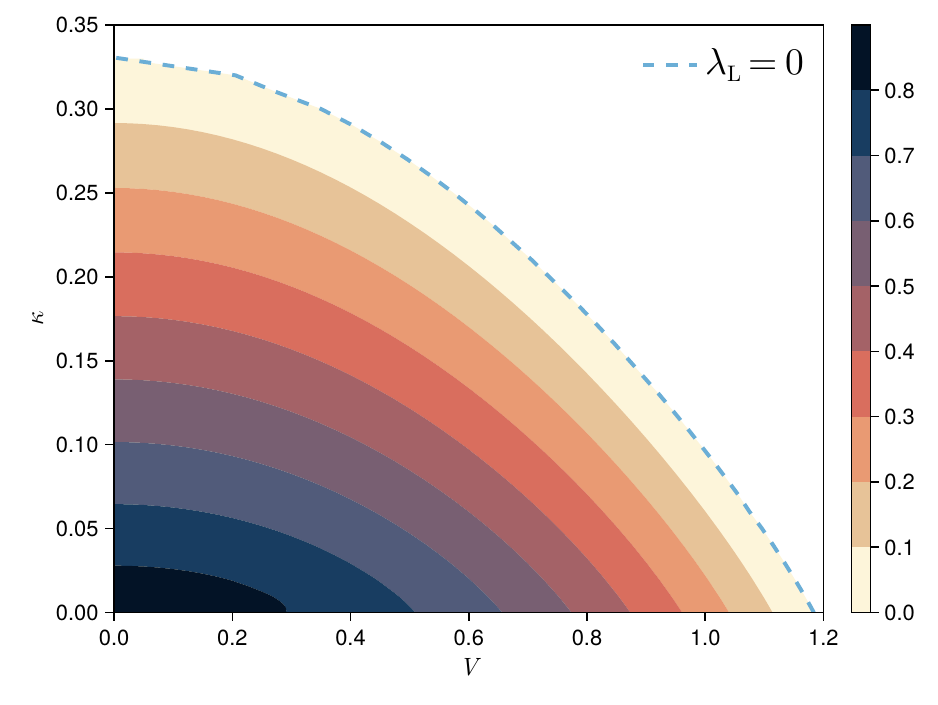}
    %
    \includegraphics[width=0.45\textwidth]{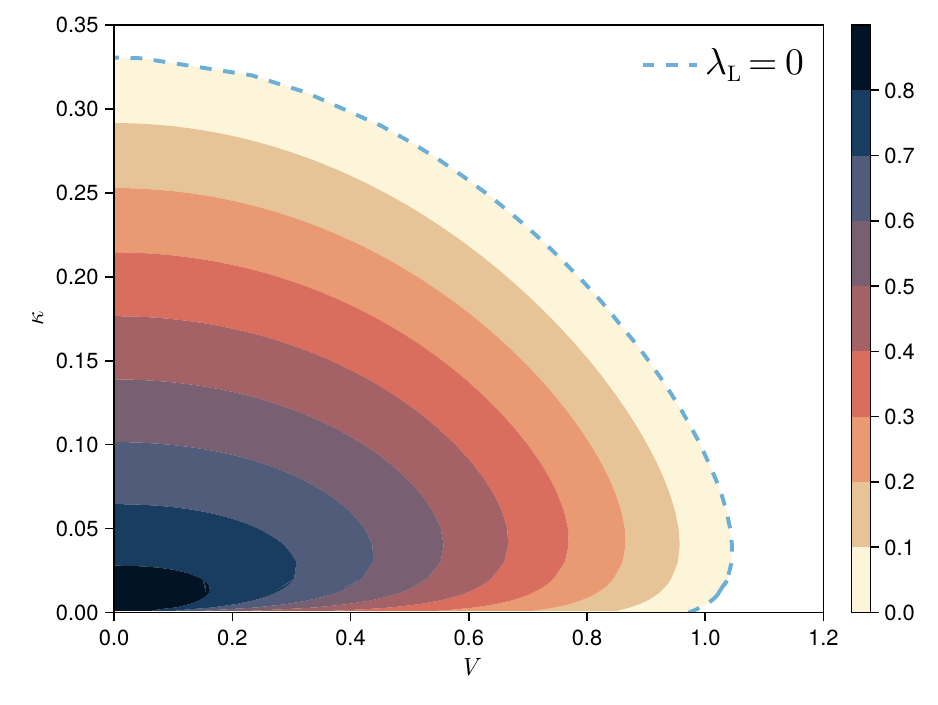}
    %
    \includegraphics[width=0.45\textwidth]{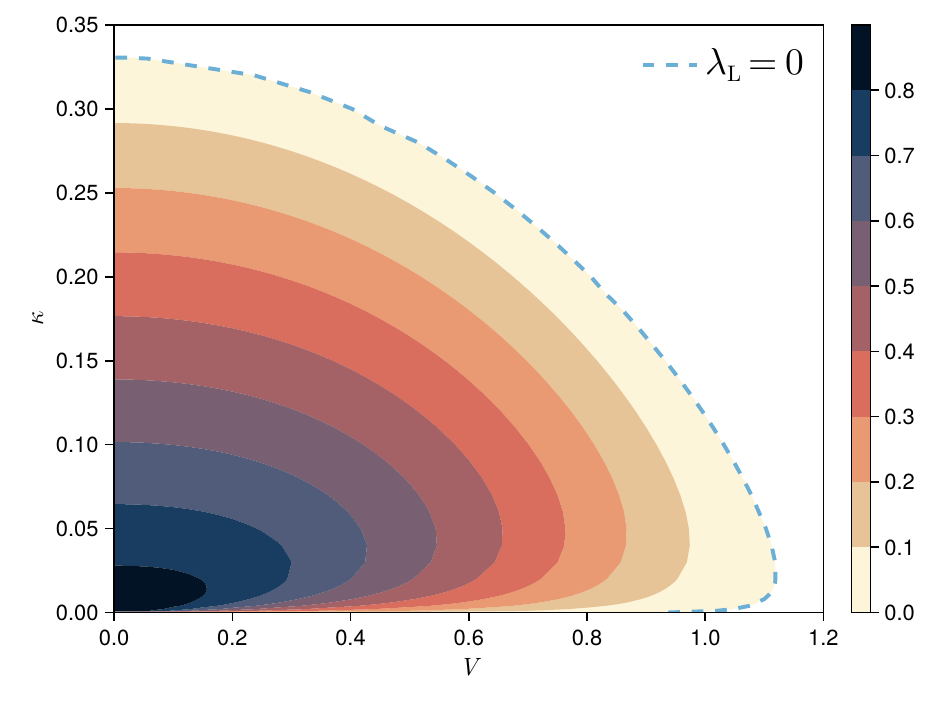}
    \includegraphics[width=0.45\textwidth]{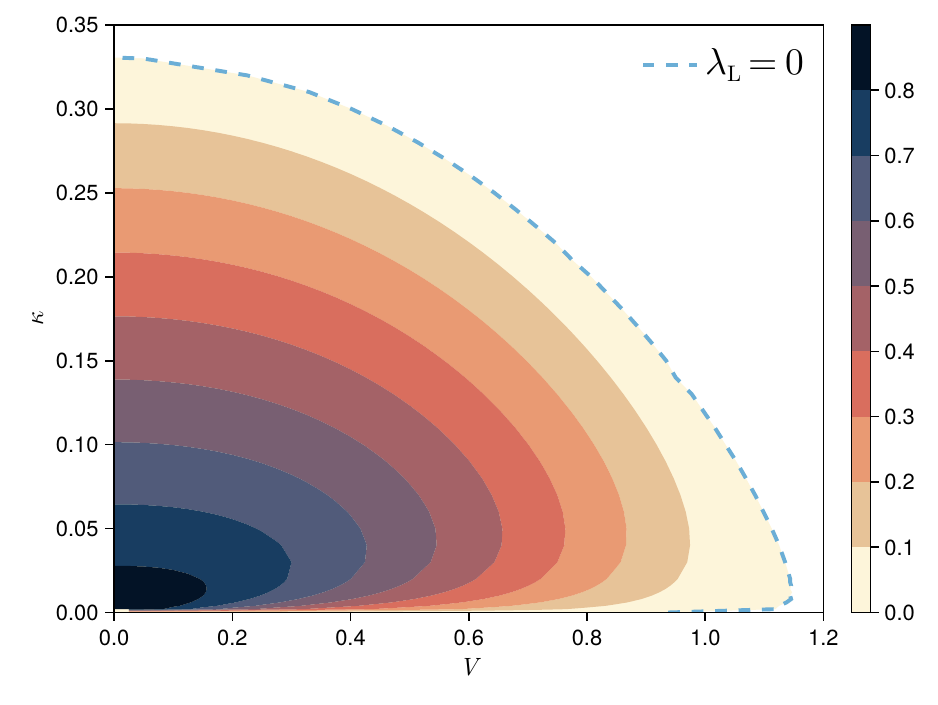}
    \caption{Contour plots of the Lyapunov exponents for the quenched Lindblad SYK$_4$ model with $\beta_{\bath}=1$ (Top Left), 10 (Top Right), 100 (Bottom Left), 1000 (Bottom Right) on the ($\kappa$, $V$)-plane.}
    \label{fig:Lyapunov_kappa_V_plane_quenched_Lindblad_SYK}
\end{center}
\end{figure}

The two kernels $M_1$ and $M_2$ correspond to different time orderings in the four-point function, but they should give precisely the same Lyapunov exponent for consistency. In Fig.~\ref{fig:Lyapunov_M1_and_M2_quenched_Lindblad_SYK}, we present the Lyapunov exponents $\lambda_{\mathrm{L}}$ versus $V$ for several $\kappa$, obtained from the two kernels $M_1$ and $M_2$. Precise agreements of are found, indicating consistency results for $\lambda_{\mathrm{L}}$. We also checked that different choices of the discretization of time $\Delta t$ and different cutoffs of the time give the same results, hence verified the convergence of the numerical results. Besides, we also numerically checked uniqueness of the solutions, as is illustrated in Fig.~\ref{fig:uniqueness_check}: the largest eigenvalues of $M_1(\lambda)$ decrease monotonically as $\lambda$ increases, and cross 1 at a unique point, giving the Lyapunov exponent $\lambda_{\mathrm{L}}$. With these sanity checks, in the main text, we therefore only present the results obtained from $M_1$ numerics. In Fig.~\ref{fig:Lyapunov_kappa_V_plane_quenched_Lindblad_SYK} we present the Lyapunov exponents on the $(V, \kappa)$ coupling space for various $\beta_{\rm B}$, as complementary plots for the main text. Taking the horizontal slices, we present in Fig.~\ref{fig:Lyapunov_vs_V_quenched_Lindblad_SYK} the relation of the Lyapunov exponents versus $V$ for fixed $\kappa$.

\end{document}